\documentclass[fleqn,usenatbib,amsmath,amssymb]{emulatemnras}

\usepackage{epsf}
\usepackage{verbatim} 
\usepackage{graphicx}
\usepackage{rotating}
\usepackage{tablefootnote}
\usepackage{threeparttable}

\usepackage{dcolumn}
\usepackage{bm}
\usepackage{epsf}
\usepackage{amsmath}
\usepackage{amssymb}
\usepackage{amsfonts}
\usepackage{appendix}
\usepackage{xcolor}
\usepackage{soul}
\usepackage[normalem]{ulem} 
\usepackage{pifont}
\usepackage{multirow}
\usepackage{adjustbox}

\usepackage{epstopdf}
\usepackage[percent]{overpic}
\usepackage{footnote}
\usepackage{tikz}
\usepackage{dblfloatfix}

\makesavenoteenv{tabular}

\defcitealias{ReissKeshet18}{RK18}
\newcommand{\RK}{{\citetalias{ReissKeshet18}}}
\definecolor{darkgreen}{rgb}{0.0,0.5,0.0}

\newcommand{\ie}{\emph{i.e.} }
\newcommand{\eg}{\emph{e.g.,} }

\newcommand{\be}{\begin{equation}}
\newcommand{\ee}{\end{equation}}
\newcommand{\bea}{\begin{equation*}}
\newcommand{\eea}{\end{equation*}}
\newcommand{\beqr}{\begin{eqnarray} \nonumber}
\newcommand{\eeqr}{\end{eqnarray}}
\newcommand{\beqrb}{\begin{eqnarray}}
\newcommand{\eeqrb}{\nonumber \end{eqnarray}}
\newcommand{\fin}{\mbox{ .}}
\newcommand{\coma}{\mbox{ ,}}

\newcommand{\cm}{\mbox{ cm}}
\newcommand{\sr}{\mbox{ sr}}
\newcommand{\se}{\mbox{ s}}

\newcommand{\erg}{\mbox{ erg}}

\newcommand{\MHz}{\mbox{ MHz}}

\newcommand{\km}{\mbox{ km}}

\newcommand{\Mpc}{\mbox{ Mpc}}

\newcommand{\keV}{\mbox{ keV}}

\newcommand{\GeV}{\mbox{ GeV}}

\newcommand{\muG}{\mbox{ $\mu$G}}
\newcommand{\Jy}{\mbox{ Jy}}

\newcommand{\gama}{$\gamma$}

\newcommand{\mynewcommand}[2]{\ifdefined #1 \else \newcommand{#1}{#2} \fi}
\newcommand{\myR}{R_{500}}

\mynewcommand{\apj}{ApJ}     
\mynewcommand{\apjl}{ApJL}     
\mynewcommand{\apjs}{ApJS}    
\mynewcommand{\aap}{A\&A}    
\mynewcommand{\nat}{Nature}  

\newcommand{\DrawFig}[1]{{#1}}
\graphicspath{{./figure/}}

\newcommand{\dgr}{^{\circ}}
\newcommand{\dgrdot}{{\overset{^\circ}{.}}}

\newcommand{\pxN}{{\mathsf{n}}}

\newcommand{\Msun}{{M_\odot}}

\newcommand{\MyT}{{T_b}}
\newcommand{\MyTzero}{{T_{b,0}}}
\newcommand{\Myc}{{\mathsf{c}}}
\newcommand{\MyH}{{\mathsf{h}}}
\newcommand{\Myfr}{{\nu}}
\newcommand{\MySig}{{S}}

\newcommand{\MyChiCorr}{{\mathcal{C}}}

\newcommand{\oridata}{{restored data}}
\newcommand{\cleandata}{{cleared data}}
\newcommand{\orig}{{restored}}
\newcommand{\clean}{{cleared}}

\newcommand{\planar}{{planar}}
\newcommand{\sh}{{s}}

\newcommand{\taumax}{{\tau_{max}}}

\newcommand{\ext}{{e}}
\newcommand{\pnt}{{p}}
\newcommand{\myK}{{C}}
\newcommand{\gmin}{{\gamma_{1}}}
\newcommand{\gmax}{{\gamma_{2}}}
\newcommand{\gminP}[1]{{\gamma_{1}^{#1}}}
\newcommand{\gmaxP}[1]{{\gamma_{2}^{#1}}}

\newcommand{\DF}{{\mathcal{N}}}
\newcommand{\TS}{{\mbox{TS}}}

\newcommand{\append}[1]{{#1}}

\usepackage{xstring}

\newcommand{\MakeDouble}{0}
\newcommand{\MyTitle}{{Synchrotron emission from virial shocks around stacked OVRO-LWA galaxy clusters}}
\newcommand{\MySTitle}{{Stacked LWA virial ring}}

\title[\MySTitle]{\MyTitle}

\author[Hou et al.]{
Kuan-Chou Hou,$^{1}$\thanks{E-mail: hou@post.bgu.ac.il}
Gregg Hallinan,$^{2}$
and
Uri Keshet$^{1}$\thanks{E-mail: ukeshet@bgu.ac.il}
\\
$^{1}$ Physics Department, Ben-Gurion University of the Negev, POB 653, Be'er-Sheva 84105, Israel\\
$^{2}$Department of Astronomy, California Institute of Technology, 1200 E California Blvd, Pasadena, CA 91125\\
}

\newcommand{\MyPlace}{h}

\begin{document}
\pubyear{2022}
\label{firstpage}
\pagerange{\pageref{firstpage}--\pageref{lastpage}}
\maketitle

\renewcommand*\thesection{\arabic{section}}

\if \MakeDouble 1
\setlength{\columnsep}{15pt}
\mytwocolumn
\fontsize{11pt}{8pt}\selectfont
\parindent 0.2in
\sloppy
\fi

\begin{abstract}
Galaxy clusters accrete mass through large scale, strong, structure-formation shocks. Such a virial shock is thought to deposit fractions $\xi_e$ and $\xi_B$ of the thermal energy in cosmic-ray electrons (CREs) and magnetic fields, respectively, thus generating a leptonic virial ring.
However, the expected synchrotron signal was not convincingly established until now.
We stack low-frequency radio data from the OVRO-LWA around the 44 most massive, high latitude, extended MCXC clusters, enhancing the ring sensitivity by rescaling clusters to their characteristic, $R_{500}$ radii.
Both high (73 MHz) and co-added low ($36\text{--}68\text{ MHz}$) frequency channels separately indicate a significant ($4$--$5\sigma$) excess peaked at
$(2.4 \mbox{--} 2.6) R_{500}$, coincident with a previously stacked \emph{Fermi} $\gamma$-ray signal interpreted as inverse-Compton emission from virial-shock CREs.
The stacked radio signal is well fit (TS-test: $4$--$6\sigma$ at high frequency, $4$--$8\sigma$ at low frequencies, and $8$--$10\sigma$ joint)
by virial-shock synchrotron emission from the more massive clusters, with $\dot{m}\xi_e\xi_B\simeq (1\mbox{--}4)\times 10^{-4}$, where $\dot{m}\equiv \dot{M}/(MH)$ is the dimensionless accretion rate for a cluster of mass $M$ and a Hubble constant $H$.
The inferred CRE spectral index is flat, $p \simeq 2.0 \pm 0.2$, consistent with acceleration in a strong shock.
Assuming equipartition or using $\dot{m}\xi_e\sim0.6\%$ inferred from the \emph{Fermi} signal yields $\xi_B\simeq (2\mbox{--}9)\%$, corresponding to $B \simeq (0.1\text{--}0.3)~\mu\text{G}$ magnetic fields downstream of typical virial shocks.
Preliminary evidence suggests non-spherical shocks, with factor $2$--$3$ elongations.
\end{abstract}

\date{Accepted ---. Received ---; in original ---}
\begin{keywords}
galaxies: clusters: general - galaxies: clusters: intracluster medium - intergalactic medium - magnetic fields - radio continuum: general
\end{keywords}

\section{Introduction}
\label{sec:Intro}

Galaxy clusters and filaments are the largest gravitational bound objects in the Universe, with clusters located at the nodes of the cosmic web of large-scale structure (LSS).
A cluster of mass $M$ is thought to grow by accreting surrounding matter at a rate $\dot{M}=\dot{m}MH$, where $H$ is the Hubble parameter and $\dot{m}$ is a dimensionless parameter of order unity.
As the accreted gas is violently decelerated, strong, collisionless, so-called virial or structure formation shocks form near the virial radius of the cluster.
These virial shocks define the edges of clusters, so observing them can provide useful information about the formation of LSS, in particular the local accretion rate.
As the accreted gas is pristine and weakly magnetized, virial shocks also provide a useful laboratory for studying collisionless shock physics.
A virial shock is distinct from the weak shocks found in its heated downstream, known as the intracluster medium (ICM); such weak shocks, usually observed at $r\lesssim 1\Mpc$ radii following a merger, are more difficult to model as they propagate into a pre-heated, magnetised plasma, already enriched by metals and cosmic-rays.

Virial shocks are thought to accelerate charged particles to highly relativistic, $\gtrsim$ 10 TeV energies, in resemblance of strong supernova remnant shocks that share similar, $\sim10^3\km\se^{-1}$ velocities.
The cosmic ray (CR) electrons (CREs) and ions (CRIs) accelerated by such strong shocks are thought to develop a nearly flat, $dN/dE\propto E^{-p}$, energy spectrum with $p\simeq2$ (equal energy per logarithmic CR energy bin), as expected in diffusive shock acceleration theory \citep[for a review, see][]{Blandford_Eichler_87} when scattering is not too anisotropic \citep{2020ApJ...891..117K}.
A fraction $\xi_e$ of the downstream thermal energy, deposited in CREs, is subsequently radiated away, forming a distinctive non-thermal signature, which should surface at the extreme ends of the electromagnetic spectrum \citep{LoebWaxman00,TotaniKitayama00,WaxmanLoeb00, KeshetEtAl03, Miniati02, KeshetEtAl04}.

The high energy CREs accelerated by the virial shock cool rapidly, primarily by inverse-Compton scattering off cosmic microwave background (CMB) photons,
resulting in a thin shell of energetic CREs.
These CREs produce a nonthermal, detectable leptonic ring around the cluster, as predicted analytically \citep{LoebWaxman00, WaxmanLoeb00, TotaniKitayama00} and calibrated using cosmological simulations \citep{KeshetEtAl03,Miniati02, KeshetEtAl04}.
The rings are expected to be fairly spherical, but somewhat elongated toward the main filament feeding the cluster \citep{KeshetEtAl03}.

Evidence for the virial inverse-Compton signal first surfaced in the Coma cluster: an ideal virial-shock target, thanks to its high mass, proximity (redshift $z\simeq0.023$), and location in a low-foreground region near the north Galactic pole.
Consistent signals were reported in $\sim220\GeV$ data from VERITAS \citep{2017ApJ...845...24K}, $\sim\GeV$ data from \emph{Fermi}-LAT, and $\sim0.1\keV$ data from \emph{ROSAT} \citep{2018ApJ...869...53K}.
The inferred virial ring is elongated toward the large-scale filament between Coma and the nearby cluster A1367, with a semi-minor axis $\sim 2.1R_{500}$ coinciding with the virial radius, and is thin; searches for a thick or spherical LAT signal did not recover the signal.
Here, $R_\delta$ is a characteristic cluster radius, where subscript $\delta$ designates an enclosed mass density $\delta$ times above the critical mass density of the universe.
The three aforementioned signals are all consistent with a normalisation $\dot{m}\xi_e\simeq 0.3\%$ (up to an uncertainty factor $\sim3$) and with a $2.0\lesssim p\lesssim2.2$ spectrum \citep{2018ApJ...869...53K}.

Galaxy cluster signals can often be amplified by stacking data around multiple clusters, utilising the self-similarity of the latter.
By co-adding \emph{Fermi}-LAT data around 112 massive, high latitude, extended clusters, rescaled by their $R_{500}$ radii to a dimensionless, $\tau\equiv r/R_{500}$ scale, binning the outcome radially, and utilising the anticipated flat spectrum, a high significance ($>5\sigma$) stacked \gama-ray ring was identified 
\citep[][henceforth \RK]{ReissEtAl17, ReissKeshet18}.
The stacked signal is best fit by a normalised shock (subscript $s$) radius $\tau_s\equiv r_s/R_{500}=2.3\pm0.1$, a CRE injection rate $\dot{m}\xi_e=(0.6\pm0.1)\%$, and a spectral index $p=2.1\pm0.2$, all consistent with predictions and with the Coma signal.
While stacking increases the sensitivity, it carries an elevated systematic uncertainty (here admitting best-fitting peak radii in the range $2.2\lesssim\tau_s\lesssim2.5$, and an additional uncertainty factor $\sim2$ in $\dot{m}\xi_e$) and washes away non-spherical components and variations among clusters, due to the co-addition and radial binning.
Moreover, the projected shock signal of an individual cluster should not be precisely circular, so stacking effectively biases $\dot{m}$ towards lower values.

A direct tracer of virial shocks, independent of any particle acceleration, is their anticipated \citep{KocsisEtAl05} imprint on the thermal Sunyaev-Zel’dovich \citep[SZ;][]{1972CoASP...4..173S} signal.
A spatial correlation was identified between the \gama-ray ring in Coma and an outward drop in the $y$-parameter inferred from \emph{WMAP} data, supporting the association of both signals with the virial shock \citep{2017ApJ...845...24K}.
A highly significant, $8.6\sigma$ drop in $y$-parameter was later identified in the cluster A2319, which has the highest signal-to-noise detection in the \emph{Planck} SZ catalogueues, indicating a strong shock around $\tau\simeq 3$ with a Mach number $\Upsilon>3.25$ \citep[at the $95\%$ confidence level;][]{HurierEtAl19, 2020ApJ...895...72K}.
A joint search for SZ and \emph{Fermi} virial signals in the massive clusters Coma, A2319, and A2142 showed a significant drop in $y$ coincident with a weak \gama-ray excess in all three clusters, consistent with $0.2\%\lesssim\dot{m}\xi_e\lesssim0.7\%$  \citep{2020ApJ...895...72K}.
Combining the SZ signal with galaxy counts gives a measure of the accretion rate \citep{HurierEtAl19}, allowing independent measurements of $\xi_e\sim 0.5\%$ and $\dot{m}\simeq 1.1$ in A2319 \citep{2020ApJ...895...72K}.
More recently, stacking the \emph{Planck} $y$-parameter around 10 galaxy groups \citep{Pratt2021}
indicated a significant drop around $2.0\lesssim\tau\lesssim2.6$, and stacking the $y$-parameter among 500 clusters using South Pole Telescope data \citep{Anbajagane2022} indicated a projected drop starting at $\tau\simeq2$, both consistent with a virial shock at the stacked $\gamma$-ray radius of {\RK}. 
Interestingly, stacking the 500 clusters also indicated a second drop starting at $\tau\simeq 6$; if verified, this would suggest elongated virial shocks with a $\sim2.5$ ellipticity ratio, as indicated by the virial ring found in Coma by  \citet{2017ApJ...845...24K, 2018ApJ...869...53K}.

In addition to the inverse-Compton signal, the CREs accelerated by the virial shock should also emit a synchrotron radio signal, as they gyrate in the post-shock magnetic fields \citep{WaxmanLoeb00, KeshetEtAl04}.
Assuming that the shock deposits a fraction $\xi_B\simeq 1\%$ of the thermal energy in magnetic fields \cite[as inferred from cluster halo observations; see][and references therein]{WaxmanLoeb00, KeshetEtAl04}, the cumulative signal from all clusters should constitute a considerable fraction ($\sim 30\%$ for $\xi_e\xi_B=10^{-4}$) of the diffuse extragalactic radio background below 500 MHz, where it dominates fluctuations on $1\arcmin\lesssim\theta\lesssim1\dgr$ scales \citep{KeshetEtAl04}.
The signal, similar in morphology but somewhat more patchy than its inverse-Compton counterpart, has proven difficult to detect, although a correlation found between the \emph{WMAP} synchrotron signal and the VERITAS \gama-ray ring in Coma corresponds to $\xi_B\sim1\%$ \citep{2017ApJ...845...24K}.

The flat, $\alpha\simeq p/2\simeq 1$ radio spectral index of the anticipated virial shock signal from cooled CREs is somewhat softer than that of the Galactic foreground, giving low-frequency interferometers an advantage in searching for this elusive signal \citep{KeshetEtAl04}.
The Owens Valley Radio Observatory Long Wavelength Array (OVRO-LWA; hereafter LWA) is particularly suited for such a search, providing clean sky maps of the northern sky in the (36--73) MHz frequency range.
Searching the low-frequency sky for the virial shock signature is complicated by the weakness of the signal, an abundance of unrelated point and diffuse sources, including radio halos and relics associated with the galaxy clusters themselves \citep{2008Natur.455..944B, Keshet10, 2014MNRAS.444L..44B, 2018MNRAS.478.2927B, 2019ApJ...881L..18C, 2019A&A...622A..20H, 2020arXiv201108249O}, and instrumental difficulties associated in particular with beam sidelobes of sources in the centres of clusters. Such studies therefore require careful modelling and extensive sensitivity tests.

We select the 44 most massive, nearby, extended clusters in the northern sky from the Meta-catalogue of X-ray Clusters \citep[MCXC;][]{PiffarettiEtAl11}, after avoiding regions of strong Galactic contamination or a suboptimal point spread function (PSF).
LWA data are excised around these clusters, cleaned from point sources and from smooth background and foreground signals, rescaled to the $R_{500}$ extent of each cluster, co-added over all clusters, and radially binned, to test if a residual virial-shock signal can be identified.
A robust (in particular, independent of PSF residuals) and highly-significant signal, found around the virial radius and coincident with the previously stacked \gama-ray signal, is then modelled as the cumulative synchrotron emission from cluster virial shocks.

The paper is organised as follows.
We outline the LWA data and the cluster sample in \S\ref{sec:Data}.
The stacking analysis, including background modelling and Monte-Carlo simulations of control cluster samples, is described in \S\ref{sec:analysis}.
In \S\ref{sec:result}, we outline the results of the stacking, in particular a strong central signal and a significant virial excess.
Simple models for the central emission are presented and fit to the data in \S\ref{sec:centralemission}.
Models for virial-shock emission are fit to the peripheral signal, alone and jointly with the central signal, in \S\ref{sec:virialemission}.
Finally, the results are summarised and discussed in \S\ref{sec:summary}.

The Supplementary Material provides sky maps (\S\ref{append:sky_maps}), technical details concerning different aspects of the analysis,
including cluster co-addition (\append{\S\ref{append:intensity}}), background and foreground removal (\append{\S\ref{append:backgroundremove}}), control sample analyses (\append{\S\ref{append:describe_controlsample}}), noise and correlation corrections (\append{\S\ref{append:controlsample}}), modelling the central emission from clusters (\append{\S\ref{append:CentralResults}}), additional tests constraining residual sidelobe artefacts (\append{\S\ref{append:psf}}), sensitivity and consistency tests (\append{\S\ref{append:sensi}}), the theory of synchrotron emission from virial shocks (\append{\S\ref{append:model}}), cross contamination by  central and virial modelling residuals (\append{\S\ref{append:mutual_effects}}), methods of inferring the peripheral magnetic fields (\append{\S\ref{append:decouple_xib}}), the list of clusters in our sample (\S\ref{append:MCXC_clusters_list}), and extended 2D significance images (\S\ref{append:2D_sig_maps}).

We adopt a flat $\Lambda$CDM cosmological model with a Hubble constant $H_0=70\km\se^{-1}\Mpc^{-1}$, a mass fraction $\Omega_{m}$ = 0.3, and a cosmic baryon fraction $f_b = 0.17$.
Assuming a $76\%$ hydrogen mass fraction, we use a mean particle mass $\bar{m} \simeq 0.59 m_p$, where $m_p$ is the proton mass.
Error bars designate 68\% containment projected for a single parameter.

\section{Data and data preparation}
\label{sec:Data}

\begin{table*}
 	\caption{LWA stacked-data properties.
 	} 
	\centering 
\begin{tabular}{cccccccccc}
$\nu$ ($\Delta \nu$) & $T_0$ & FWHM & Restored          & \multicolumn{3}{l}{Cleared} & \multicolumn{3}{l}{Restored} \\
{[MHz]} & {[K]} & ($\delta = 45\dgr$)  & FWHM & $\eta_s$    & $\bar{\eta}$ & $\psi$ & $\eta_s$    & $\bar{\eta}$ & $\psi$                       \\
(1)                          & (2)                                 & (3)                        & (4)                                             & (5)                         & (6)         & (7)          & (8)                          & (9)         & (10) \\ \hline
36.528                       (0.024)                               & 595                        &  $20\arcmin.2 \times 16\arcmin.9$                & $16\arcmin.1$ & 1.24 (1.20) & 1.17 (1.14)  & 0.70 (0.71) & 1.91 (1.86) & 1.90 (1.83)  & 0.85 (0.85) \\
41.760 (0.024)                               & 541                        & $18\arcmin.5 \times 16\arcmin.0$                & $14\arcmin.5$ & 0.99 (0.95) & 0.93 (0.90)  &                              & 1.54 (1.49) & 1.51 (1.45)  &                              \\
46.992 (0.024)                               & 417                        & $17\arcmin.4 \times 15\arcmin.2$                & $14\arcmin.0$ & 1.06 (1.03) & 1.02 (0.99)  &                              & 1.63 (1.57) & 1.61 (1.55)  &                              \\
52.224 (0.024)                               & 418                        & $16\arcmin.2 \times 15\arcmin.0$                & $13\arcmin.1$ & 0.89 (0.86) & 0.87 (0.84)  &                              & 1.33 (1.28) & 1.30 (1.25)  &                              \\
57.456 (0.024)                               & 354                        & $15\arcmin.9 \times 15\arcmin.0$                & $12\arcmin.9$ & 0.84 (0.80) & 0.81 (0.78)  &                              & 1.18 (1.13) & 1.17 (1.13)  &                              \\
62.688 (0.024)                               & 309                        & $15\arcmin.8 \times 14\arcmin.9$                & $12\arcmin.7$ & 0.77 (0.74) & 0.75 (0.72)  &                              & 1.05 (1.02) & 1.05 (1.01)  &                              \\
67.920 (0.024)                               & 281                        & $15\arcmin.9 \times 14\arcmin.7$                & $12\arcmin.6$ & 0.71 (0.68) & 0.69 (0.67)  &                              & 0.94 (0.91) & 0.93 (0.90)  &                              \\
73.152 (0.024)                               & 154 $^{\dagger}$                       & $12\arcmin.2 \times 10\arcmin.0\,\, ^{\dagger}$ & $9\arcmin.3$ & 0.97 (0.94) & 0.94 (0.91)  & ---                          & 1.26 (1.22) & 1.24 (1.20)  & ---                          \\
\end{tabular}
\label{tab:maps_sum}
\begin{tablenotes}
\item
    {\bf
    Columns:} (1) Central frequency (and bandwidth) (MHz) of each channel;
    (2) Noise temperature \citep[in K, from][]{2018AJ....156...32E};
    (3) FWHM major axis $\times$ minor axis of the synthesized beam at a typical declination $\delta\sim 45\dgr$;
    (4) Effective FWHM of the restored data in the $20\dgr<\delta<60\dgr$ range (see \S\ref{sec:CentralModels});
    (5) Correction factor $\eta_s\equiv\eta(2.25 \leq \tau < 2.5)$ for the product $T_0\cdot\mbox{FWHM}$ in the virial shock region, inferred from control samples for the beam- (cluster-)stacked \cleandata;
    (6) Correction factor $\bar{\eta}\equiv\eta(0 < \tau < 10)$ uniformly averaged over the full control range for the \cleandata;
    (7) Correlation parameter of low-frequency channels uniformly averaged over the range $0<\tau<10$,
    for the \cleandata;
    (8) $\eta_s$ for the \oridata;
    (9) $\bar{\eta}$ for the \oridata;
    (10) $\psi$ for the \oridata.\\
    $^{\dagger}$ --- In the high-frequency channel we use a dedicated  high-resolution data set, provided by the OVRO-LWA collaboration, rather than the nominal-resolution $16\arcmin.8 \times 14\arcmin.6$ data set of \citet{2018AJ....156...32E}. The $T_0=154$ K of the nominal-resolution map is shown to approximately hold at high-resolution. The  $\delta=45^\circ$ FWHM provided here is extracted from the tabulated high-resolution PSF.
\end{tablenotes}
\end{table*}

\subsection{LWA sky maps}

The LWA is a 288-antenna interferometer with a maximum baseline of $\sim 1.5$ km
\citep{2019ApJ...886..123A,2019AJ....158...84E}.
We use data produced from 28 hours of observation during 2017--2018.
The data, in the form of eight
low-frequency maps of the sky north of declination $\delta = -30^\circ$, were constructed with Tikhonov-regularised \emph{m}-mode analysis imaging \citep{2018AJ....156...32E}.
The maps were taken at frequencies 36.53, 41.76, 46.99, 52.22, 57.46, 62.69, 67.92, and 73.15 MHz, with a 24 kHz bandwidth, a $15'$--$20'$ FWHM angular resolution, and $\sim 800$ mJy/beam thermal noise.
More precise parameters, and the thermal noise level $T_0(\Myfr)$ in each frequency $\Myfr$, are discussed in \citet{2018AJ....156...32E} and summarised in Table \ref{tab:maps_sum}.

For the high-frequency, $\nu_h=73.152\MHz$ channel, the sky map was prepared with a substantially better, $\sim 11'$ FWHM resolution.
Therefore, we separate the study of data in this channel (henceforth referred to as the high-frequency channel) from the analysis of the other, lower-frequency data sets (referred to as the low-frequency channels).

\begin{figure}
    \centering
    \DrawFig{\includegraphics[width=0.5\textwidth,trim={0 0.7cm 0 0},clip]{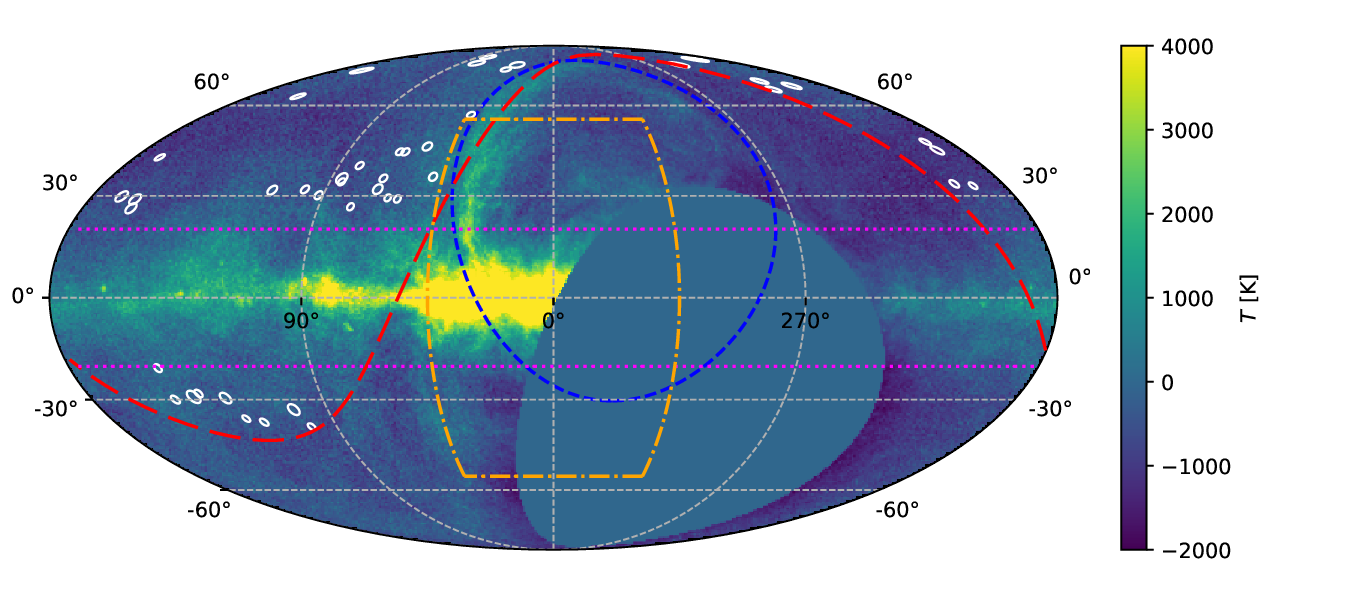}}
    \caption{
    Brightness temperature map of the
    cleared LWA data in the high-frequency channel using a Mollweide projection, in the full (declination $\delta>-30\dgr$) field of view.
    White circles of radius $5R_{500}$ are shown around the 44 clusters in our sample.
    The cuts discussed in \S\ref{subsec:cluster} are shown as curves:
    around the Galactic plane (dotted magenta lines), Loop-I (dashed blue circle), Fermi bubbles (dash-dotted yellow curve), and $\delta = 20\dgr$ (long-dashed red curve). 
    A larger version and a map based on restored data may be found in \S\ref{append:sky_maps}.
    }
    \label{fig:allsky}
\end{figure}

Point sources were removed from the maps using a version of the CLEAN algorithm, adapted to deconvolve the PSF for \emph{m}-mode analysis imaging 
by drift-scanning telescopes like the OVRO-LWA \citep[see algorithm details in][]{2018AJ....156...32E}.
To better understand the point-source contribution and the performance of CLEAN, we utilise two sets of maps in each channel, denoted '\clean' and '\orig', provided by the OVRO-LWA collaboration.
The {\clean} maps are the usual output of CLEAN, after the flux associated with each source was removed by convolving a point source with the PSF.
In the {\orig} maps, the flux of each source thus removed was subsequently restored using a simpler, Gaussian beam.
Figure \ref{fig:allsky} shows the resulting sky map of the
cleared data in the high-frequency channel.

Our nominal analyses are based on the cleared data.
To further avoid contamination from bright source residuals,
we mask all HEALPix pixels within a radius $0\dgrdot3$ of $F_{\rm 1.4GHz} > 1$ Jy sources in the NRAO VLA Sky Survey \citep[NVSS;][]{1998AJ....115.1693C}.
Our results are not sensitive to small variations in the masking radius or in the flux cut, as illustrated in \append{\S\ref{append:sensi}}.

The PSF of the LWA varies with declination. The PSF sidelobes become increasingly less circularly-symmetric toward lower declination, in particular for $\delta < 20\dgr$ \citep{2018AJ....156...32E}.
This gradual distortion of the PSF renders sidelobe artefacts increasingly difficult to test and correct.
Therefore, we impose a cut on the cluster sample, and include only clusters with $\delta > 20\dgr$ in our sample, thus selecting for a well-behaved, approximately circular PSF.
The basic information and properties of the maps are summarised in Table \ref{tab:maps_sum}.

The central regions of galaxy clusters are often radio-bright, due to both point sources and diffuse emission.
Searching for an annular signal around such a cluster is challenged by the possible presence of sidelobes that were not fully accounted for.
This difficulty is addressed and tested throughout the text, in particular in \S\ref{subsec:sidelobe}, \S\ref{subsec:LowFrequencyVirial}, \S\ref{append:centre_restored},
and \append{\S\ref{append:psf}}, as summarised in \S\ref{sec:summary}.
Another difficulty is that if the virial ring is sufficiently clumpy, it may in part be misidentified by CLEAN as a collection of point sources and thus removed; we show in \S\ref{subsub:sig_prof} that this is not the case.

The sky maps are discretised using a HEALPix scheme \citep{GorskiEtAl05} of order 11 ($\text{nside}=2048$),
involving $3\times 2^{24}\simeq 5\times 10^7$ pixels on the sky, so each pixel spans a solid angle
\begin{equation}
    \delta \Omega\simeq 2.5\times 10^{-7}\sr\fin
\end{equation}
The mean, $\sim 0\dgrdot03$ separation between pixels is sufficiently small with respect to the angular resolution of the LWA to avoid discretisation errors.
Our modelling utilises tabulated PSF maps, provided for the full declination range by the OVRO-LWA collaboration.

\subsection{Cluster sample}
\label{subsec:cluster}

We select groups and clusters of galaxies (henceforth, clusters) for radio stacking from the MCXC catalogue.
This catalogue, based on the \emph{ROSAT} all sky-survey, was homogenised to an overdensity of 500, thus providing the characteristic length and mass scales, required for our analysis, for 1743 clusters.
In addition to the position of each cluster on the sky and the cluster mass $M_{500}$ enclosed within $\myR$, the catalogue specifies the redshift $z$ and radius $R_{500}$ of each cluster, so the corresponding angular radius $\theta_{500}$ can be computed.

To construct an optimal sample of clusters for the virial shock search, we select only clusters that satisfy all of the following criteria.
\begin{enumerate}
\item
Massive clusters: the anticipated synchrotron signal arises from massive clusters,  so we select clusters with a mass $M_{500} > 10^{13}M_\odot$.
Note that the mass dependence of synchrotron emission is stronger than its inverse-Compton counterpart, due to expected scaling of the magnetic field.
\item
Resolvable $R_{500}$ core: an angular radius $\theta_{500} > 0\dgrdot2$, chosen such that $R_{500}$ is marginally resolved at the $\sim 12\arcmin$ PSF scale of our nominal, high frequency channel.
\item
Avoiding excessively extended, bright clusters of $\theta_{500}>0\dgrdot4$, which are too bright and extended for our analysis;  in particular, such clusters complicate the background removal procedure.
\item
Avoiding contamination from the Galactic plane by excluding clusters located at latitudes $|b|<20\dgr$.
\item
Avoiding possible contamination from Loop-I, by excluding a circle on the sky centred on Galactic coordinates $\{l, b\} = \{337.2, 24.6\}$ with a $55\dgr$ radius. This region is based on LWA data, and is largely consistent with previous findings \citep[e.g.][]{Haslam1970,Dickinson2018}.
\item
Avoiding possible contamination from the Fermi bubbles, by excluding the regions defined as $|b| < 55\dgr$ and either $l < 45\dgr$ or $l > 315\dgr$,
\citep[approximating the bubbles structure found in][]{2017ApJ...840....7K}.
\item
Optimising for PSF: as in the cut on the data,
we select only clusters with $\delta > 20\dgr$ to avoid the PSF deterioration at lower declination.
\end{enumerate}

Out of the 1743 clusters in the MCXC catalogue, the above cuts leave us with a sample consisting of $N_c=44$ high-quality clusters.
These are nearby clusters, found at redshifts $z \lesssim 0.07$.  
The clusters in our sample and their basic properties are listed in the supplementary material \S\ref{append:MCXC_clusters_list}. 
Their median values are $z \simeq 0.03$,  $\theta_{500} \simeq 0\dgrdot24 $, and $M_{500} \simeq 0.56 \times 10^{14} \Msun$.
Variations in the above cuts are discussed in \append{\S\ref{append:sensi}}; 
our results are shown to be robust to reasonable changes in all cluster selection criteria.

\section{Stacking method}
\label{sec:analysis}

Along with the stacking of multiple rescaled maps around clusters, detecting the virial shock signal requires careful removal of backgrounds and foregrounds (henceforth background, for brevity), angular data-folding or radial binning, and Monte-Carlo simulations to calibrate the noise and correlation
parameters and validate the relevant sky statistics.

The procedures of scaling, co-adding, and binning the data are outlined in \S\ref{sec:StackingBinning}.
The stacked quantities and their computation are detailed in \S\ref{sec:StackedQuantities}.
Different background removal methods are discussed in \S\ref{sec:T0}.
Finally, we describe the Monte-Carlo simulations of control cluster samples in \S\ref{subsec:controlsamples}.
These control samples, consisting of random position in the sky analysed as if they contain clusters analogous to the real sample, are used for calibration purposes throughout   \S\ref{sec:StackingBinning}--\S\ref{sec:T0}, where we refer to them as 'control clusters'; in later sections, where we inject mock signals into these random sky locations for modelling purposes, we refer to them as 'mock clusters'.

\subsection{Stacking and binning procedure}
\label{sec:StackingBinning}

In terms of (proper) spatial scales, virial shocks span a wide range of radii, due to the diversity in cluster parameters.
The dispersion in terms of angular scales is even larger, due to the $0.017 \lesssim z \lesssim 0.072$ range of cluster redshifts in our sample.
Therefore, before stacking cluster maps, we normalise the sky coordinates around each cluster as  $\{\tau_x,\tau_y\}\equiv\{\theta_x,\theta_y\}/\theta_{500}$, and define the normalised radius $\tau\equiv\theta/\theta_{500}$ or equivalently $r/\myR$.
Here, angles $\bm{\theta}=\{\theta_x,\theta_y\}$ are measured with respect to the X-ray peak of the cluster.
The data around each cluster
are rotated randomly in the $\{\tau_x,\tau_y\}$ plane to minimize background effects, although this is inconsequential for most parts of the analysis,
where radial binning is employed.
With a background model, the excess flux and brightness above the background can now be co-added over clusters, and the significance of the excess can be estimated directly
from the sky statistics or
from Monte-Carlo simulations.

The virial shock models and detected signals reviewed in \S\ref{sec:Intro} indicate that the anticipated signal, after stacking the scaled maps, should show a ring of enhanced brightness peaked
near the $2.2\lesssim\tau\lesssim2.5$ radius of its similarly stacked, \gama-ray counterpart.
The radio ring might be somewhat offset from this radius by the different weighting of clusters --- which are not precisely spherical --- in the present sample, and by the effect of magnetic fields on the synchrotron signature of virial shocks.
The anticipated signal is weak, so it is beneficial to bin it radially, although we show in \S\ref{subsec:virial_excess} that the signal can be identified even without such binning.
The corresponding scaled, stacked, binned, radial plot --- based on bins in the form of scaled concentric rings about the centres of the clusters in the sample --- should then show a peak in, or close to, the above $\tau$ range.
As projected virial shock signals are not expected to be circular, some information and flux are lost when randomly rotated clusters are stacked or when the data are radially binned; see discussion in \S\ref{sec:summary}.

We choose $\Delta\tau=0.25$ as our nominal radial bin size, and $2.25\leq\tau<2.5$ as the a-priori most-likely virial shock bin, based (only) on the \gama-ray stacking of {\RK}.
Although such a resolution may seem finer than the beam FWHM, it exceeds the limiting resolution near the virial radius after integrating over the $2\pi\tau$ circumference of the radial bin.
Both larger and smaller choices of bin size are examined throughout the text; the robustness of the results to resolution is demonstrated in \append{\S\ref{append:sensi}}.
While co-adding the flux is straightforward, there is more than one way to co-add the brightness and determine the significance of the excess, as we next discuss.

\subsection{Stacked quantities}
\label{sec:StackedQuantities}

In each frequency channel
$\Myfr$, each cluster $\Myc$, and each HEALPix pixel $\MyH$,
denote the measured
brightness temperature as $T(\Myfr,\Myc,\MyH)$, and the brightness temperature of the background as $\MyT(\Myfr,\Myc,\MyH)$.
Estimates of the background $\MyT$ are discussed in \S\ref{sec:T0} and in \append{\S\ref{append:backgroundremove}}.
We may now define the local brightness temperature excess as
\begin{equation}
\Delta T(\Myfr,\Myc,\MyH) \equiv T-\MyT \fin
\end{equation}

We anticipate a spectrally flat synchrotron signal from virial-shocks,
with a specific flux $F_\Myfr\propto \Myfr^{-1}$, \ie with a photon index $\alpha_{vir}\simeq p/2\simeq 1$ corresponding to rapidly cooled CREs injected with an initially flat, $p\simeq 2$ spectrum.
Therefore, it is useful to estimate the excess
flux
over the background per e-fold in photon energy (flux excess; henceforth),
\begin{equation} \label{eq:FjDef}
\Delta F(\Myfr,\Myc,\MyH) \equiv \nu \,\Delta F_{\nu} \simeq 2\frac{\Myfr^3}{c^2}k_B \Delta T\,\delta\Omega \coma
\end{equation}
approximately independent of frequency for the expected
signal.
Here, $k_B$ is the Boltzmann constant and $c$ is the speed of light.
We use $\Delta F\sim \nu F_\nu$ instead of $F_\nu$
for convenience; this of course does not bias the measurements of the spectrum, carried out in  \S\ref{sec:centralemission} and \S\ref{sec:virialemission}.

We may now stack the flux excess \eqref{eq:FjDef} around the clusters
in our sample to obtain an image
of the mean flux excess per cluster at a given channel $\Myfr$,
\begin{equation}
\Delta F\left(\Myfr, \bm{\tau}\equiv\{ \tau_x,\tau_y \}\right) = \frac{1}{N_c} \sum_{\Myc=1}^{N_c} \sum_{\MyH =1}^{N_p (\bm{\tau},\Myc)} \Delta F(\Myfr,\Myc,\MyH)  \coma
\label{eq:excess_b_pixel}
\end{equation}
where the second sum runs over the
$N_p (\bm{\tau},\Myc)$ HEALPix pixels mapped from cluster $\Myc$ onto the desired 
$(\tau_x,\tau_y)$ pixel.
Radial binning,
used in the remainder of this subsection,
is obtained by formally identical summations,
\begin{equation}
\Delta F(\Myfr, \tau) = \frac{1}{N_c} \sum_{\Myc=1}^{N_c} \sum_{\MyH =1}^{N_p (\tau,\Myc)} \Delta F(\Myfr,\Myc,\MyH)  \coma
\label{eq:excess_b}
\end{equation}
where
the second sum here runs over the
$N_p (\tau,\Myc)$ HEALPix pixels that fall in the radial $\tau$ bin of cluster $\Myc$.
We henceforth focus mostly on the radially binned Eq.~\eqref{eq:excess_b}.

The significance $S$ of the excess intensity over the background
in some bin $\tau$ of the cluster $\Myc$ may be estimated as
\begin{align} \label{eq:SingleBinSignificance}
\MySig(\Myfr,\tau,\Myc)
& = \frac{\sum_{\MyH=1}^{N_p (\tau,\Myc)} \Delta T(\Myfr,\Myc,\MyH)}{\eta(\Myfr) T_0(\Myfr)\sqrt{N_p (\tau,\Myc) \pxN}} \coma
\end{align}
where $\pxN$ is the number of (correlated, HEALPix) pixels in the beam,
\begin{equation}
\pxN
\simeq \frac{\pi\left( \mbox{FWHM}/2\right)^2}{\delta\Omega}
\simeq 60\left(\frac{\mbox{FWHM}}{15'}\right)^2 \fin
\label{eq:LWA_PixPerBeam}
\end{equation}
Here, we assumed that the noise is predominantly thermal, at the $T_0(\Myfr)$ antenna noise temperature, and given by Poisson statistics among the
$\sim N_p/\pxN$
independent beams mapped onto
the $\tau$ bin.
The product $T_0\cdot\mbox{FWHM}$ varies with the PSF across the sky, so its effective, co-added value could deviate from its nominal, tabulated value.
Therefore, in Eq.~\eqref{eq:SingleBinSignificance} we introduced order-unity correction factors
$\eta$.
These factors vary among channels, but in principle should be fairly independent across the sky, as shown below, and so roughly independent of $\tau$ and $\Myc$.

The factors $\eta$
are calibrated by sampling the sky with control clusters, as discussed in \S\ref{subsec:controlsamples} and in \append{\S\ref{append:controlsample}}.
In the cleared data, \ie after CLEAN, $\eta$ varies between $\sim 0.7$ and $\sim 1.3$ as a function of frequency. In the restored data, $\eta$ tends to be larger and varies between $\sim 0.9$ and $\sim 1.9$ for different frequencies.
In both cleared and restored data,
$\eta$ shows small, $\lesssim 5\%$ spatial variations
across the stacked control maps.
Hence, we define mean factors
$\bar{\eta}$, by averaging
$\eta$ in the domain $0<\tau<10$ uniformly in $\tau$.
The resulting averaged factors are used throughout the following analysis.
Table \ref{tab:maps_sum} provides the averaged factors, as well as the local factors
$\eta_s$
estimated near the anticipated shock position.
The factors $\eta$ and $\eta_s$ are consistent within $\lesssim 6\%$.

There are different ways to stack the signal over clusters.
We consider two opposed methods for estimating the significance of an excess signal stacked over multiple clusters, for given $\Myfr$ and $\tau$.
We refer to the first, more standard method, as beam co-addition (equivalent to photon co-addition in \RK).
Here,
we compute the excess
brightness
summed over all clusters, and estimate its significance over random fluctuations of the summed background, leading to
\begin{equation}
\MySig^{(bm)}(\Myfr,\tau) = \frac
{\sum_{\Myc=1}^{N_c} \sum_{\MyH=1}^{N_p (\tau,\Myc)} \Delta T(\Myfr,\Myc,\MyH)}
{\eta(\Myfr) T_0(\Myfr) \sqrt{\pxN\sum_{\Myc=1}^{N_c} N_p(\tau,\Myc)}} \coma
\label{eq:photon_add}
\end{equation}
directly generalizing Eq.~\eqref{eq:SingleBinSignificance} for multiple clusters.
As this method weighs the
stacked excess against the stacked solid angle,
it has the advantage of being less
sensitive to individual clusters
with a small $\theta_{500}$
that leaves fewer independent beams for determining $\MySig(\Myfr,\Myc,\MyH)$.

In the second method, referred to as cluster co-addition (also used in \RK), we co-add the significance of the coincident excess $\MySig(\Myfr,\Myc,\MyH)$ in individual clusters as unit-normal
random variables of equal weights, leading to
\begin{equation}
\MySig^{(cl)}(\Myfr,\tau) = \frac
{\sum_{\Myc=1}^{N_c} \MySig(\Myfr,\tau,\Myc) }
{
\sqrt{N_c}} \fin
\label{eq:cluster_add}
\end{equation}
As each cluster contributes equally, regardless of its angular extent, this method
avoids the large weights that beam co-addition effectively assigns to both the excess and the estimated fluctuations of the most extended clusters.

The radially-binned Eqs.~\eqref{eq:photon_add} and \eqref{eq:cluster_add} are readily generalised for a $S(\nu,\tau_x,\tau_y)$ map,
by replacing the $\tau$ bins by  $(\tau_x, \tau_y)$ pixels, as in Eq.~\eqref{eq:excess_b_pixel}.
The beam and cluster co-addition methods, analysed separately below,
are generally found to yield results in good agreement with each other, provided that
bright NVSS sources are masked.
The correction factors $\eta$ are calibrated separately for beam and cluster co-addition
after masking NVSS sources, showing small, $\lesssim 5\%$ differences between the two stacking methods for both cleared and restored data.

To demonstrate the agreement between beam and cluster co-addition, consider for example the high-frequency channel.
Near the virial radius, we find $\lesssim0.03$ ($\lesssim0.07$) changes in $S$, which correspond to $\lesssim 3\%$ ($\lesssim 6\%$) fractional differences, for the
cleared (restored) data (see \S\ref{subsub:sig_prof}).
The largest deviations in $S$ between the two methods, $\lesssim 0.4$ ($\lesssim 0.3$), are found in the central, $\tau\lesssim 1$ regions of the clusters.
Substantial differences between beam and cluster co-addition can be seen in the restored data when bright NVSS point sources are not masked (see \append{\S\ref{append:centre_restored})}.

We use control samples to verify (see \S\ref{subsec:controlsamples} and \append{\S\ref{append:controlsample}})
that the distribution of the excess emission above the background, stacked around mock clusters randomly placed on the sky,
is approximately normal in each channel when using the cleared data.
In the restored data, this distribution is slightly skew-normal but, within a $\sim 0.1$ accuracy in $S$, can be approximated as normal.
The control samples are used to calibrate the noise parameters $\eta$, as summarised in Table
\ref{tab:maps_sum}.
The distribution of the calibrated significance $S$ attributed to the excess is then found to approximately follow a unit-normal
(\ie zero mean and unit variance) distribution.
Namely, the distribution of control samples is found to be consistent with the calibrated Eqs.~(\ref{eq:photon_add}) and (\ref{eq:cluster_add}), out to at least the $\pm3\sigma$ confidence level of a normal distribution.

Our analysis is largely based on the above, single-channel expressions, applied to
the high-frequency channel, where the angular resolution is superior.
The low-frequency channels are mostly used to measure the spectrum and verify the signal.
However, we also co-add the different channels, adopting the
significance estimate
\begin{equation}
\MySig(\tau) = \frac
{\sum_{\Myfr=1}^{N_\Myfr} \MySig(\Myfr,\tau)}
{N_\Myfr^{(1+\psi)/2}} \coma
\label{eq:band_stacking}
\end{equation}
where $N_\Myfr=7$ is the number of co-added, low-frequency channels.
Anticipating some degree of inherent correlation between the different channels, here we introduced a correlation parameter $\psi$, which can in principle range from $1$ (full correlations) to $0$ (no correlations) or even become negative (for anti-correlations), and is determined empirically below based on control samples.

These control samples indicate strong positive inter-correlations among the seven low-frequency channels.
The simple correction factor in Eq.~(\ref{eq:band_stacking}) is motivated by the approximately unit-normal
$S$ distribution it produces once $\psi$ is calibrated; we thus find
that $\psi\sim 0.70$ ($\psi\sim 0.85$) for the cleared (restored) data.
The strong correlations appear to emerge primarily from the similar small-scale structures seen across the radio sky in the different channels, rather than from any instrumentation effects, because point source restoration significantly  strengthens the correlations and because the correlations among different channels roughly scale with their PSF.
As such, Eq.~(\ref{eq:band_stacking}) provides a reasonable and simple alternative to the determination of the full covariance matrices.
Notice that
introducing a positive $\psi$ provides a conservative estimate of $S$: it effectively lowers the significance attributed to the appearance of the virial excess emission in multiple channels.

The channel co-addition Eq.~(\ref{eq:band_stacking}) is applied to both beam and cluster stacking methods, with typically similar results.
For example, the two methods show
$\lesssim 0.2$ ($\lesssim 0.3$) differences in $S$ corresponding to
$\lesssim 7\%$ ($\lesssim 19\%$) fractional changes, for the cleared (restored) data around the virial radius (see \S\ref{subsub:lowf_sigprof}).
In terms of $\psi$, we find $\lesssim 1\%$ differences between beam and cluster co-addition.

It is instructive to compute the stacked intensity excess profile, $\Delta I \equiv \nu \, \Delta I_\nu(\nu,\tau)$, and relate it to the significance $S$ of the local excess.
The procedure is similar to the above stacking of $S$, as discussed in \append{\S\ref{append:intensity}}.

\subsection{Background modelling}
\label{sec:T0}

The virial signal is typically weaker than the Galactic foreground, the extragalactic background, and the emission from the centre of the cluster, so a careful removal of the background is essential.
We consider several different models for the $\MyT(\Myfr,\Myc,\MyH)$ background field.

The simplest model is a uniform background around each cluster, which we define as the mean brightness temperature $\MyTzero(\Myfr,\Myc)$ measured within an angular separation $\theta<\theta_b$ of the centre of the cluster.
We find that $\theta_b\simeq 5\dgr$ is sufficiently large to establish the virial ring, and not too large to be offset by the large-scale curvature of the Galactic foreground.
Our results do not vary substantially with small changes in $\theta_b$ around this value
(see \append{\S\ref{append:backgroundremove}}).
This pertains both to the above uniform background and to the polynomial backgrounds discussed next.
Therefore, we adopt $\theta_b = 5 \dgr$ as our nominal value.

More accurate models allow spatial variations, to better capture the local background.
As our cluster selection avoids extended Galactic structures, namely the Galactic plane, Loop-I, and the Fermi bubbles (see selection criteria (iv) -- (vi) in \S\ref{subsec:cluster}), we apply identical masks on the data, with a negligible effect on the results.
After bright point sources and these Galactic structures are masked,
the remaining background varies mainly on scales larger than the anticipated,
$\lesssim 1\dgr$ extent of the virial signal.
This remaining background can be fairly well approximated using a polynomial fit on larger scales. 
For each cluster, we thus consider the region $\theta < \theta_b$, and fit the enclosed data in each band by an order $N_b$ polynomial in
$\{\tau_x,\tau_y\}$.

In \append{\S\ref{append:backgroundremove}}, we consider different choices of $N_b$ and $\theta_b$, quantifying their implications using the control samples presented in \S\ref{subsec:controlsamples} below.
Overall, we find that the statistical distribution of $S(\tau)$ is approximately symmetric, normal and normalised, as constructed
above, provided that $N_b\geq 2$.
In contrast, a low, $N_b\leq 1$ order polynomial is not sufficient to capture the curvature in the background, and leads to $S(\tau)$ distributions that are neither symmetric, nor normal, nor normalised.
However, higher orders $N_b$ cause increasingly more of the virial excess itself to be fitted as part of the background, thus spuriously diminishing the estimated signal.

These considerations indicate that $N_b =2$ or $N_b = 3$ best balance a fitting power
sufficiently high to render $S(\tau)$ normal and normalised,
but sufficiently low
to avoid over-fitting much of the local excess due to the cluster as part of the background.
Odd-order terms in a polynomial are integrated out by the radial binning, so for any odd $N_b$, the results are very similar, although not identical, to those of $N_b-1$.
Therefore, and for better accuracy in the $\{\tau_x,\tau_y\}$ stacked maps, we adopt $N_b=3$ as the nominal polynomial order.
Interestingly, similar background properties were found by {\RK} when fitting the stacked brightness of \emph{Fermi} clusters.
For our nominal, $N_b=3$, the results are
not sensitive
to the fitting range for $\theta_b \geq 5 \dgr$.
For sensitivity tests and a discussion of alternative choices of $N_b$, see \append{\S\ref{append:backgroundremove}}.

\subsection{Control sample tests}
\label{subsec:controlsamples}

In order to estimate the statistics of the stacked quantities, calibrate the parameters $\eta$ and $\psi$, test Eqs.~(\ref{eq:photon_add})--(\ref{eq:band_stacking}) for $S$ (and their counterparts \eqref{eq:err_Intensity_bm}--\eqref{eq:err_Intensity_cl}, and \eqref{eq:err_Intensity_sum} for $I$),
examine different background
models, and investigate possible systematic biases, we generate control samples of mock clusters and stack the real LWA data around them with the same pipelines applied to the MCXC cluster sample.
For each test, we thus generate $N_{\rm mock} = 5000$ control samples, each chosen to mimic the real cluster sample.
This value of $N_{\rm mock}$
samples the analysed sky many ($\gtrsim 40$) times, leading to sufficiently converged estimates
($<\{2\%,2\%,3\%\}$
in $\{1,2,3\}\sigma$ cleared-data containment around $\tau\sim 2.4$; $<1\%$ in $\bar{\eta}$ and $\xi$).

Each control sample
consists of 44 control clusters with random positions on the sky, chosen under the same geometric constraints as the real sample, and in addition, constrained to lie at least $1\dgrdot5$ away from
any MCXC catalogue source
and any bright (1.4 GHz flux exceeding 1 Jy) point source
from the NVSS catalogue.
We verify that the latter constraint, avoiding bright NVSS sources for more efficient control samples, has no significant effect on our nominal results, as pixels are already masked within $0\dgrdot3$ from such bright NVSS point sources.
The control clusters are assigned with the same relevant parameters (angular scale and, for modelling the virial signal, also mass and redshift) as the real clusters, and are identically analysed.
The significance $S(\nu,\tau)$ of the mock stacked excess is then estimated for each cluster sample.

\begin{figure*}
    \centering
    \includegraphics[width=\textwidth,trim={0 0.5cm 0 0},clip]{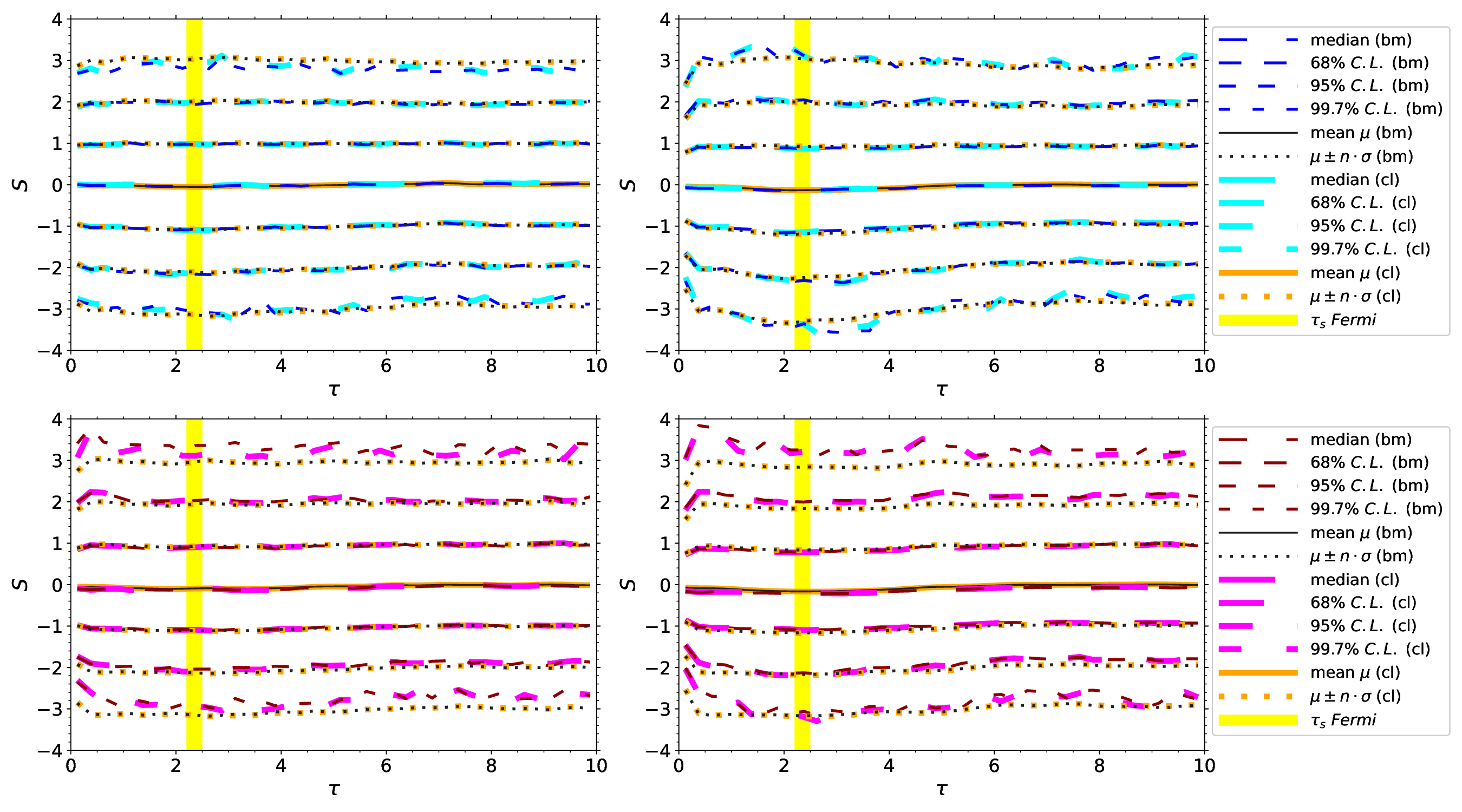}
    \caption{
    Statistics of data rescaled, stacked, and radially binned around the control clusters, for the high-frequency [left panels, using Eqs.~(\ref{eq:photon_add}) and (\ref{eq:cluster_add}) including uniform
    $\bar{\eta}$
    noise corrections] and the co-added low-frequency [right panels, using Eq.~(\ref{eq:band_stacking}) including the $\psi$ correlation correction] channels, shown (see legend) for both cleared (top panels)
    and restored (bottom panels)
    data, with both beam (abbrev. ``bm''; thin curves) and cluster (abbrev. ``cl''; thick curves) co-addition.
    Each panel shows (long-dashed to short-dashed curves) the median and the 68\%, 95\% and 99.7\% confidence intervals
    based on the control sample distribution.
    Confidence intervals $\mu\pm n\sigma$ of a corresponding normal distribution are shown (dotted curves) by offsetting the mean ($\mu$, solid curves)
    by integer multiples $n=1$, $2$, and $3$ of the standard deviation $\sigma$.
    Results are shown for nominal parameters (see text),
    as a function of $\tau$, highlighting (vertical yellow band) the $2.2<\tau_s<2.5$ virial shock radius inferred (\RK) from stacking \emph{Fermi} data.
    }
    \label{fig:mock}
\end{figure*}

Figure \ref{fig:mock} shows the resulting distribution of $S$ around the control cluster sample, for nominal analysis parameters, in particular $N_b=3$ and $\theta_b=5$ for background removal, and a mask of radius $0\dgrdot3$ around $F_{\rm 1.4GHz} > 1\Jy$ NVSS sources.
The figure shows results for both the high-frequency (left panels) and the co-added low-frequency (right) channels, using either \clean\ (top row) or  \orig\ (bottom row) data.
The results are stacked using both beam (thin lines) and cluster (thick lines) co-addition, found to be in good agreement with each other.

The results are shown after calibrating the single-channel noise correction factors $\bar{\eta}$,
and the multi-channel correlation factor $\psi$.
These parameters are tuned such that $S$ is normalised to have unit variance, $\sigma(S)=1$, when averaged over $\tau$, in each channel separately and when the low-frequency channels are co-added; the resulting $\bar{\eta}$ and $\psi$ values are not far from unity (see Table \ref{tab:maps_sum}).
The figure then contrasts, for all relevant $\tau$ bins separately, the nominal values of $S$ (y-axis labels) inferred from Eqs.~(\ref{eq:photon_add})--(\ref{eq:band_stacking}), against the mock confidence intervals (dashed curves) extracted from the control samples up to the maximal, $\pm3\sigma$ levels available with $N_{\rm mock}$, and against the normal-distribution intervals
$\mu\pm n\sigma$ (dot-dashed curves) anticipated from the locally measured mean $\mu$ and standard deviation $\sigma$ of the control samples for
$n=1$, $2$, and $3$.
The respective curves should therefore coincide with each other and with integer values of $S$ only when the underlying distribution of $S$ is locally symmetric, normal and normalised; see \append{\S\ref{append:describe_controlsample}} and  \append{\S\ref{append:controlsample}} for details.

Overall, we find that the stacking equations for $S$
[Eqs.~\eqref{eq:photon_add}--\eqref{eq:band_stacking}] and
$I$ (\append{\S\ref{append:intensity}})
can be applied to the cleared data,
with minor errors with respect to the control sample distribution,
even when the latter is extrapolated as a normal distribution to high-confidence levels.
Our tests indicate that in the radial range $0.5<\tau<10$, we may then slightly underestimate the significance of an excess signal by $-0.06\lesssim|\Delta S/S| <0$ in the high-frequency channel, and overestimate it by $0<|\Delta S/S| \lesssim 0.02$ in the co-added low-frequency channels.
In particular, around the virial radius,
the nominal
confidence levels are conservative (slightly exaggerated) and consistent with the extrapolated control sample distributions within $\gtrsim -4\%$ ($\lesssim +3\%$),
for the high (co-added low) frequency data.
For clarity, when displaying the $S$ profiles measured from the real sample, we overplot the extrapolated significance levels of the corresponding control sample.

\section{Stacking results}
\label{sec:result}

We begin in \S\ref{subsec:73MHz} by presenting our nominal results, based on the high-frequency channel.
To test these results and constrain the spectrum, in \S\ref{subsec:7lowerbands}, we present the co-added results for the seven low-frequency channels.
We focus on the {\cleandata}, showing \oridata\ results mainly to clarify the role of point sources.
The correction factors, namely $\bar{\eta}$ for channel noise and $\psi$ for cross-channel correlations, are used throughout, as inferred above from the control cluster catalogues.

\subsection{Nominal analysis: high-frequency channel}
\label{subsec:73MHz}

In our nominal analysis, we focus on the high-frequency, $73.15\MHz$ channel.
Here, the spatial resolution is considerably better than in other channels, complications due to PSF sidelobes are minor, and the control samples are well behaved even without any noise corrections.
We inspect the brightness profile in \S\ref{subsubsec:HighFreqIntensity}, and the significance of the implied excess in  \S\ref{subsub:sig_prof}; a pronounced excess near the virial is shown to be robust in \S\ref{subsec:virial_excess}, and argued to be genuine and not associated with beam sidelobes in \S\ref{subsec:sidelobe}.

Nominal parameters are used (henceforth, unless otherwise stated), including the standard cuts giving our nominal, 44 cluster sample, a radial binning with $\Delta\tau=0.25$ resolution, polynomial background modelling of order $N_b=3$
within $\theta_b=5\dgr$, masking data within $0\dgrdot3$ of NVSS point sources with a flux exceeding 1 Jy, and applying the (small) $\bar{\eta}$ correction.

\subsubsection{Brightness profile}
\label{subsubsec:HighFreqIntensity}

\begin{figure} 
    \centering
    \includegraphics[width=0.45\textwidth,trim={0 0.5cm 0 0},clip]{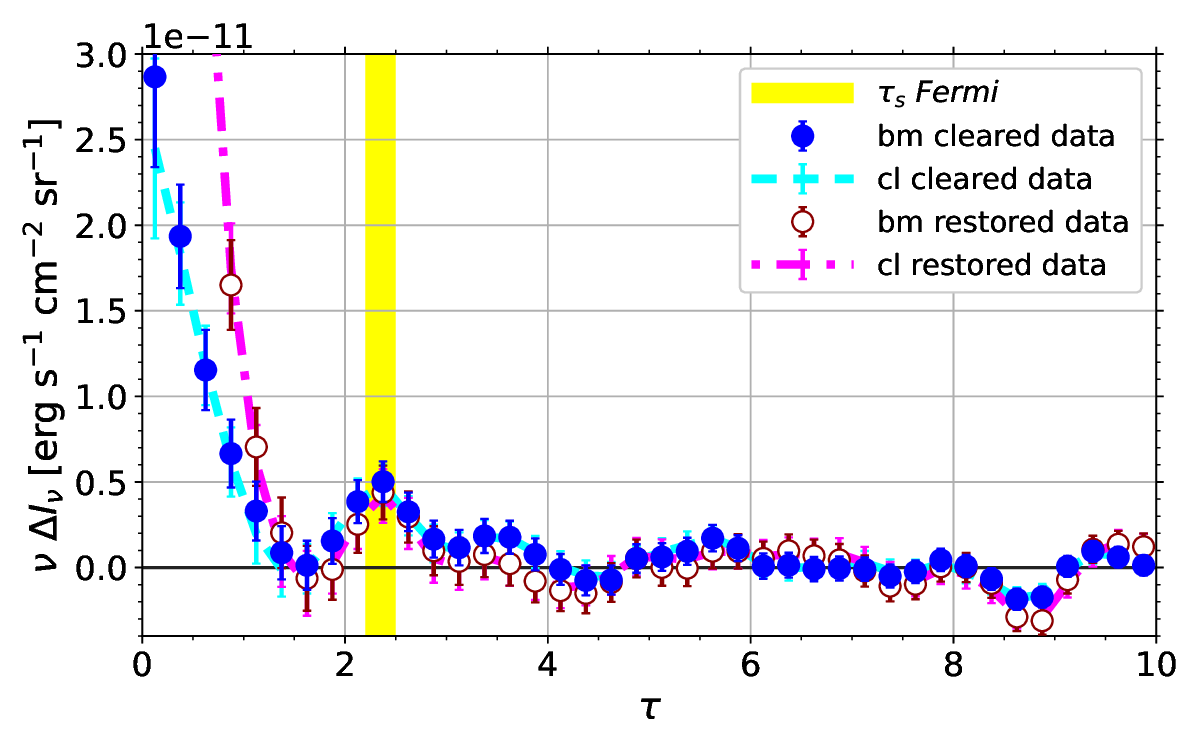}
    \caption{
    Rescaled and radially-binned excess in brightness, averaged over the nominal 44 cluster sample in the high-frequency channel, shown as a function of normalised radius $\tau$, for beam (symbols) and cluster (curves)
    co-addition, using both \clean\ (blue filled circles and dashed cyan curve) and \orig\ (red circles and dot-dashed magenta curve) data; see legend.
    The range of virial shock radii based on \emph{Fermi}
    is shown as in Fig.~\ref{fig:mock}.
    }
    \label{fig:band7_flux}
\end{figure}

Figure \ref{fig:band7_flux} shows, for both cleared and restored data sets, the radial, $\tau$-profile of the mean excess brightness,
averaged over the 44 clusters in our sample, as defined in \append{\S\ref{append:intensity}} [namely, using Eqs.~(\ref{eq:excess_Intensity_bm}) and (\ref{eq:excess_Intensity_cl}), with $1\sigma$ uncertainty intervals (error bars) computed using Eqs.~(\ref{eq:err_Intensity_bm}) and (\ref{eq:err_Intensity_cl})].
Nominal parameters are used.

Both beam and cluster co-addition methods are shown in the figure, and are
found to be in good agreement with each other, here and quite generally, except where point sources make a strong contribution.
This agreement indicates that the diffuse features are fairly evenly stacked among the clusters, rather than dominated by a handful of outlier clusters.
We verify this behaviour by manually checking the excess in 
individual clusters, 
estimating the mean and standard deviation of the excess, and removing the strongest contributing clusters, as demonstrated for the virial signal in \S\ref{subsub:sig_prof}. 
For simplicity, we focus on beam co-addition for the remainder of \S\ref{subsubsec:HighFreqIntensity}.

As expected, a strong signal is identified in the centres of clusters, and is partly removed by the modified CLEAN algorithm.
The original signal, as inferred from the \oridata, which more faithfully represents the signal including point sources, is peaked approximately at the centre of each cluster, with an excess intensity of $\Delta I \simeq 1.2 \times 10^{-10}$ erg s$^{-1}$ cm$^{-2}$ sr$^{-1}$ when averaged over clusters.
This component, likely attributed to a combination of point sources, such as radio galaxies, and diffuse emission, such as from radio halos and minihalos, is further discussed in \S\ref{sec:centralemission}.

Outside the clusters, at large, $\tau\gtrsim 3$ radii, well beyond the virial radius, the signal is in general consistent with noise, with only expectedly small, $\lesssim 1\sigma$ fluctuations.
The cleared and the restored data are not in good agreement here, indicating that deviations of the restored data from the background
are largely associated with point sources.
The low-significance cleared-data excess emission around $\tau\sim3.5$ ($\sim2\sigma$) and around
$\tau\sim5.5$ ($\sim2.5\sigma$)
are discussed below in \S\ref{subsub:sig_prof}.

In the peripheries of clusters, around $2\lesssim\tau\lesssim3$, both data sets show a localised, significant excess peaked
in the $2.25\leq\tau<2.5$ bin,
coincident with the virial shock signal found in {\RK} by stacking \emph{Fermi} data (shown as a vertical yellow band in the figures), as well as with other observations and previous theoretical predictions outlined in \S\ref{sec:Intro}.
This excess, when averaged over clusters, is $\Delta I\sim 5.0 \times 10^{-12}$ erg s$^{-1}$ cm$^{-2}$ sr$^{-1}$ in the cleared data, and
$\sim10\%$ smaller in the restored data.
The similar
radius, magnitude, and shape of the excess in both data sets indicate that this signal is not associated with, nor strongly contaminated by, point sources.
We henceforth refer to this signal as the virial excess.

\subsubsection{Significance profile}
\label{subsub:sig_prof}

\begin{figure}
    \centering
    \includegraphics[width=0.45\textwidth,trim={0 0.5cm 0 0},clip]{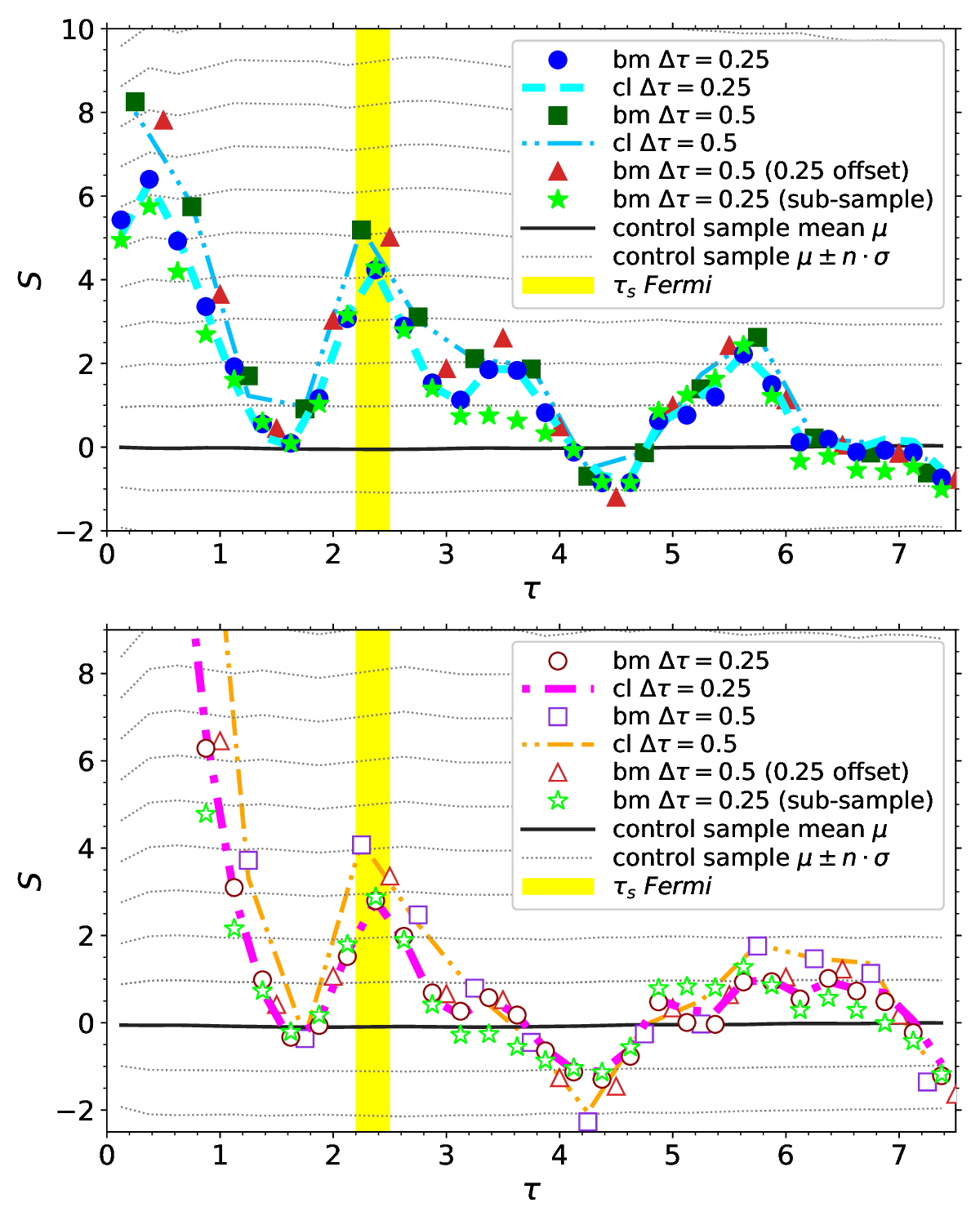}
    \caption{
    Significance profile $S(\tau)$ of the high-frequency excess in \clean\ (top panel) and \orig\ (bottom panel) data.
    Results shown (see legend) for both beam
    (symbols) and cluster (curves) co-addition, with both $\Delta \tau = 0.25$ (same notations as Fig.~\ref{fig:band7_flux}) and $\Delta \tau = 0.5$ (squares and double-dot-dashed curves with no offset; triangles for a 0.25 offset in $\tau$),
    along with control sample ($\mu\pm n\sigma$ in solid and dotted black) curves and the
    shock radius range of \emph{Fermi} (shaded yellow); notations as in Fig.~\ref{fig:mock}.
    Also shown (light-green stars) are the nominal significance profiles after removing three outlier clusters which show a $>2.5\sigma$ excess at $\tau \simeq 3.5$, apparently associated with very extended NVSS sources (see \S\ref{subsub:sig_prof}).
	Nominal parameters are used.
    }
    \label{fig:band7_sig}
\end{figure}

We quantify the significance of the local excess above the background using both the nominal $S$ derived in
\S\ref{sec:StackedQuantities}
for a normal distribution, and the control samples described in \S\ref{subsec:controlsamples}.
Figure \ref{fig:band7_sig} shows the radial profile of the excess significance $S$, computed according to Eqs.~(\ref{eq:photon_add}) and (\ref{eq:cluster_add}),
as a function of $\tau$, for both the \clean\ (top panel) and the \orig\ (bottom) data, using both beam (abbreviated ``bm''; symbols) and cluster (abbreviated ``cl''; curves) co-addition.
The sensitivity of the results to alternative parameter choices is explored in \append{\S\ref{append:sensi}}.
Our analysis is based on the $0\leq\tau<10$ region, sufficiently large to capture three times the radius of the \gama-ray excess, or equivalently, more than four times the FWHM even in the lowest-frequency channel; the significance figures below focus on the $0\leq\tau<7.5$ range only for visibility.

As discussed in \S\ref{subsec:controlsamples},
we overplot the confidence-level estimates based on the corresponding control samples,
extrapolated to high orders in the  normal-distribution approximation.
Namely,
the figure shows (as dotted lines) the mean of the beam co-added $S$, offset by integer multiples of its standard deviation, based on the respective control samples.
The nominal confidence levels
are
consistent with these extrapolated confidence levels
(\eg compare dashed vs. dot-dashed curves in Fig.~\ref{fig:mock}),
within the accuracy discussed in
\S\ref{subsec:controlsamples}.
Namely, outside the central $\tau<0.5$, our nominal significance estimates are
consistent with the
extrapolated levels
within $\sim4\%$ ($\sim10\%$) for the cleared (restored) data.

Consider first the restored data, shown in the bottom panel of Fig.~\ref{fig:band7_sig}.
Here, the central signal within $\tau\lesssim 1$ is highly significant, reaching the $\sim18\sigma$ level.
Beyond a local minimum (the 'dip', henceforth) seen around $\tau\sim 1.5$, the nominal stacked signal
(circles and thick dot-dashed curve)
reaches the $\sim3\sigma$ confidence level in the peak, $2.25 \leq \tau < 2.5$ bin, with only $\lesssim 1\sigma$ fluctuations at large radii beyond $\tau\simeq 3$.
Both the signal and the fluctuations appear to be somewhat extended in $\tau$.
Using thicker, $\Delta \tau=0.5$ bins (squares and thin double-dot dashed curve)
raises the significance of the virial excess to the $\sim4\sigma$ level in the $2 \leq \tau < 2.5$ bin, while the fluctuations reach $\sim 2\sigma$.
The significance is also  increased if these $\Delta \tau=0.5$ bins are first offset by $0.25$ in $\tau$ (triangles), but only to the $\sim3.4\sigma$ level, in the $2.25 \leq \tau < 2.75$ bin.
The location of the peak is consistent with the stacked \emph{Fermi} data analysis in either
choice of angular binning, and in both stacking methods.

Next, consider the cleared data (top panel), after point sources were removed by CLEAN and not restored.
Here, the central peak is diminished to the $\sim6\sigma$ level, as expected due to its point source contribution, whereas the virial excess becomes more significant, with a  $\sim4.2\sigma$
peak in the same, $2.25 \leq \tau < 2.5$ radial bin.
Thus, while point source removal does not appreciably change the magnitude or position of this virial signal (see Fig.~\ref{fig:band7_flux}), it raises its significance by lowering the background noise level.
Such behaviour is consistent with this excess arising from diffuse, rather than point-source, emission.
The peak remains broad; using thicker, $\Delta \tau=0.5$ bins again raises the significance, here to the $\sim5\sigma$ confidence level both in the $2.0 \leq \tau < 2.5$ bin and in the offset, $2.25 \leq \tau < 2.75$ bin.
The nominal excess is comprised of the cumulative contribution of $\sim70\%$ of the clusters in our sample, with a mean $S \simeq 1.1 \pm 0.7$ in the $2<\tau<3$ range. No sub-sample strongly governs the signal, so removing the few most significant,  $\{3.3,2.4,2.2\ldots\}\sigma$ contributions does not significantly alter the stacked results.

Unlike the virial excess, discussed in some detail below in \S\ref{subsec:virial_excess}, the central, $\tau\lesssim1.5$ signal is highly sensitive to point source removal, which
substantially lowers its brightness (Fig.~\ref{fig:band7_flux}) and corresponding significance (Fig.~\ref{fig:band7_sig}).
While most of the central flux is thus attributed to point sources, the residual stacked signal in the cleared data suggests a $\sim 6\sigma$ diffuse central component, extending in
out to $\sim 1 R_{500}$.
This diffuse signal shows some $|\Delta S/S|\lesssim7\%$, or equivalently $|\Delta S|\lesssim0.4$, difference between the two stacking methods (compare symbols and corresponding curves in Fig.~\ref{fig:band7_sig}).
In contrast, the virial peak shows only a $\sim0.4\%$ difference between the two stacking methods, suggesting that the virial excess is more evenly distributed among clusters than the central cleared signal.

Although the noise level is reduced in the cleared data,
point source removal appears to amplify some of the $\lesssim1\sigma$ fluctuations, found in restored data outside the clusters ($\tau>3$), resulting in apparently extended, $\sim 2\sigma$ excess signals in cleared data near $\tau\simeq 3.5$ and near $\tau \simeq 5.5$.
Consider first the $\tau\sim3.5$ excess.
This signal is seen only in the high-frequency channel, and not in other frequencies (see \S\ref{subsub:lowf_sigprof}), so we cannot substantiate its viability.
Furthermore, this signal is dominated by only
three clusters, each showing a $>2.5\sigma$ excess at $\tau\simeq 3.5$. Two of these clusters
harbour a (partly masked) extended, $>1$ Jy NVSS point sources at the same radius.
Excluding these three clusters lowers this excess to $<1\sigma$, with only a minute effect on the virial signal (see stars in Fig.~\ref{fig:band7_sig}).
Hence, we disregard this $\tau\simeq 3.5$ signal as a fluctuation.
In contrast, the $\tau\sim5.5$ signal --- like the virial excess --- reflects multiple small contributions from 
a large fraction ($\sim70\%$)
of our clusters, and is persistent, seen in all channels.
However, this signal is less significant than the virial excess and more sensitive to CLEAN, and it becomes far less significant or even vanishes entirely when the declination selection criterion is varied  (see \append{\S\ref{append:sensi}}). We further discuss this $\tau\simeq 5.5$ signal in \S\ref{subsec:summary_discussion}.

\subsubsection{Virial excess}
\label{subsec:virial_excess}

Figure \ref{fig:band7_sig} and the preceding discussion indicate that the virial excess in the high-frequency channel is robust and significant at the 4--$5\sigma$ confidence level.
The signal peaks around $\tau\simeq 2.4$, coincident with its \gama-ray counterpart, identified earlier by stacking \emph{Fermi} data.
The excess is somewhat extended in $\tau$, contributing to bins over the $2.0\lesssim\tau\lesssim2.8$ range.
The location, magnitude, and shape of the signal inferred from beam vs. cluster co-addition are in good agreement with each other.
This agreement supports the robustness of these statistics, and indicates that the stacked excess arises from many of the clusters in our sample, rather than being dominated by a handful of outlier clusters.
The agreement between the excess parameters in cleared vs. restored data indicates that the excess is diffuse, and not dominated or distorted by point sources.
These conclusions are confirmed by inspecting the individual significance profiles of all clusters in our sample.

\begin{bfigure}
    \centering
    \DrawFig{\includegraphics[width=0.5\textwidth,trim=10 40 0 80, clip]{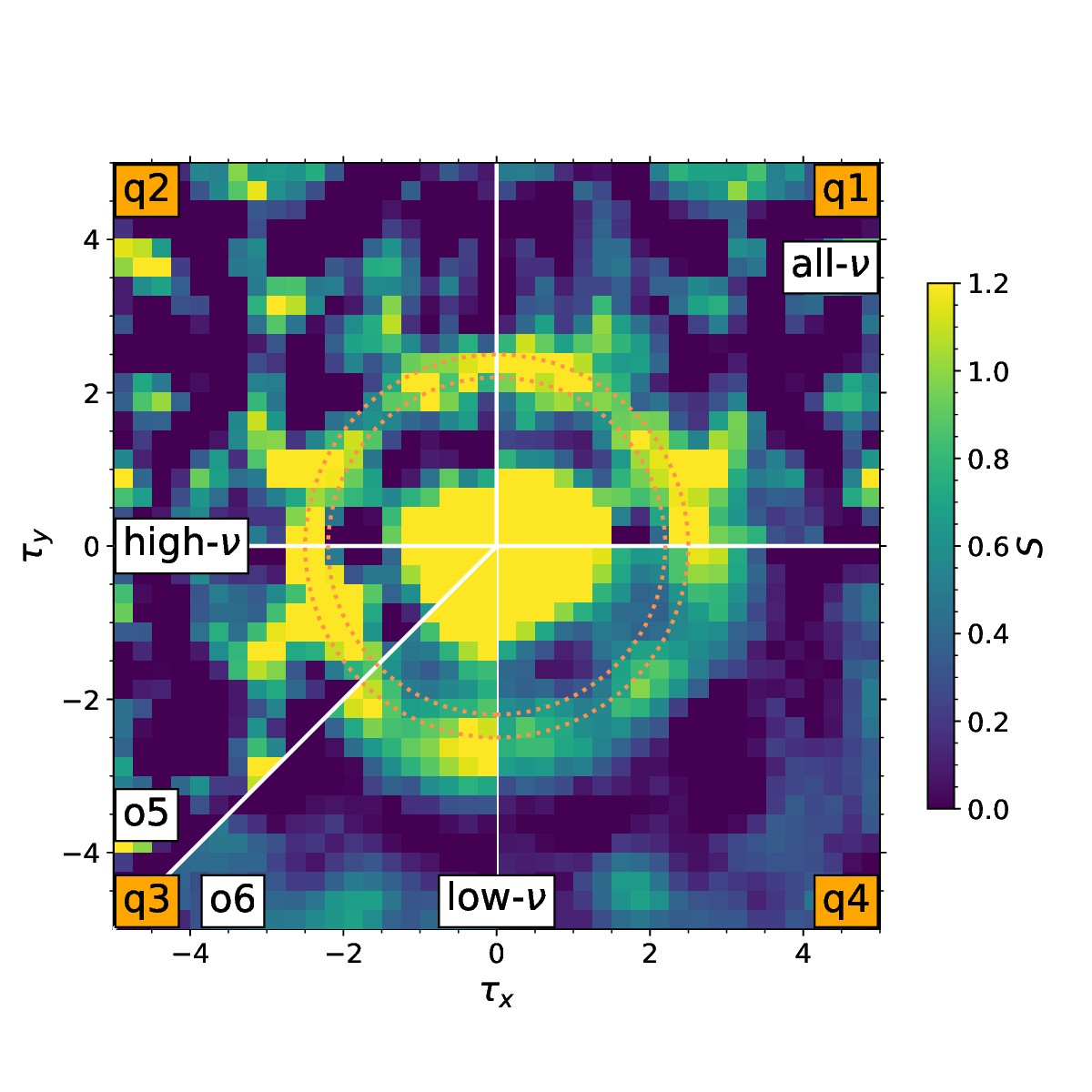}}
    \caption{
    Folded significance map of the \cleandata\ excess, after clusters were rescaled, randomly rotated, and beam co-added (see labels and colourbar).
    The data are folded onto either quadrants or (in quadrant $q3$) octants, to better highlight the signal.
    Shown are the high-frequency channel quadrant $q2$ and octant $o5$), the co-added low-frequency channels (quadrant $q4$ and octant $o6$), and the two results combined (quadrant $q1$); see labels.
    The $2.2\lesssim\tau\lesssim 2.5$ best-fit range of shock radii based on the {\RK} stacking of \emph{Fermi} data is also shown (between orange dotted circles).
	The same figure with a larger $\tau$ range is shown in \S\ref{append:2D_sig_maps}.
    }
    \label{fig:2dsig}
\end{bfigure}

The rescaled, stacked virial signal is sufficiently strong to be seen even without full radial binning.
Figure \ref{fig:2dsig} shows a partly-folded map of the significance $\MySig(\tau_x,\tau_y)$
of the stacked excess, prepared as
in the radial plots but without radial binning.
Namely, an excess flux map is first generated for each of the 44 clusters in our sample, using the \clean\ data.
Each such map is rescaled by the respective $R_{500}$, and rotated randomly about the centre of the cluster to wash out any co-added large-scale gradients.
The resulting maps are then stacked using beam co-addition, according to Eq.~(\ref{eq:photon_add}).
To better show the signal, the resulting image is then folded onto a quadrant or an octant (see labels).
The nominal, high-frequency signal is folded onto quadrant 2 (upper left sector).
To further emphasize the signal, we fold the same image again, onto octant 5 (just below quadrant 2).
The bottom quadrant and octant show the low-frequency signal, obtained by co-adding the seven low-frequency channels as discussed in \S\ref{subsec:7lowerbands}.
Quadrant 1 (upper right) combines the high-frequency channel with the co-added, low-frequency channels.
Both the central, $\tau\lesssim1$ signal, and a virial ring peaked around $\tau\sim 2.4$ are evident.
The $2.2 \lesssim \tau \lesssim 2.5$ range, identified (\RK) as a virial ring in stacked \gama-rays, is also shown (dotted orange circles).
Figures with a larger $\tau$ extent and modified selection criteria are shown in \S\ref{append:2D_sig_maps}.

In order to further test the robustness of the virial signal, we show (see \S\ref{subsec:LowFrequencyVirial}) that it presents in all channels, and carry out a series of sensitivity and convergence tests in \append{\S\ref{append:backgroundremove}} and \S\ref{append:sensi}.
In particular, we examine the sensitivity of the signal to the cluster selection criteria, the background modelling order, the background fitting area, and different radial bin sizes.
Overall, the results are found to be insensitive to reasonable variations in these parameters.

\subsubsection{Ruling out a sidelobe origin}
\label{subsec:sidelobe}

Given the strong central signal, especially before point sources are removed, sidelobes become a major concern: one must carefully test if the virial ring may arise simply from stacked interferometric ripples around the central signals.
In principle, such a spurious effect should be ruled out by the CLEAN algorithm, as the identified point sources near the centre were convolved with the PSF before they were removed, along with their sidelobes, and
the remaining central signal in the cleared data is not sufficiently strong to generate the virial excess as a sidelobe ripple.
Nevertheless, putative artefacts due to inaccuracies in the algorithm should be ruled out before the virial excess can be securely established.

In subsequent parts of the analysis, we rule out a putative virial excess stemming from the residuals of sidelobe ripples of the central signal that were not removed by CLEAN.
This conclusion is derived in several independent methods, including:
(i) a significant virial excess signal is found also in those clusters that do \emph{not} show a central signal (as discussed below and shown in Fig.~\ref{fig:band7_weak_strong});
(ii) Monte-Carlo simulations of
PSF-convolved central emission models do not show the observed virial ring
(\append{\S\ref{append:centre_restored}});
(iii) multiple channels show the signal at similar ring radii, despite variations among their PSF patterns (\S\ref{subsec:7lowerbands});
(iv) removing one or a few outlier clusters with bright NVSS sources in their centre strongly diminishes the raw (restored, unmasked) stacked central signal, with no significant effect on the virial excess
(\append{\S\ref{append:psf}});
(v) intentional attempts to generate an artificial virial excess by choosing control, centrally-bright clusters, fail to produce a significant signal (\append{\S\ref{append:psf}});
(vi) the spectrum of the virial excess is found to be a robust power-law (\S\ref{subsec:virial_spec}), whereas the central signal shows a non-trivial spectrum varying spatially and among clusters (\S\ref{append:centre_restored});
and
(vii) the virial excess is well-fit by the expected model: spatially coincident with the previously stacked $\gamma$-ray signal
and of a normalisation comparable to that inferred in Coma \citep{2017ApJ...845...24K}, not far from equipartition (\S\ref{sec:summary}).

Although most clusters show a bright central signal, at least before point sources are removed, some of them do not.
The first method we invoke to test for sidelobe artefacts is to focus on those clusters that do not show a central excess.
In such clusters, an excess signal near $\tau\simeq 2.4$ clearly cannot be dismissed as sidelobes of emission from the centre.
We define such centre-faint clusters as those showing an intensity deficit,
$\Delta I<0$, in their central, $\tau < 0.5$ region, after the large-scale background was removed.
For the selection of these clusters, we use the raw data, \ie the restored data without additional NVSS point-source masking, in order to include the full imprint of all
point sources that may have left some sidelobes contamination in the cleared data even after CLEAN.
We adopt nominal parameters, focusing on the high-frequency channel.

\begin{bfigure}
    \centering
    \includegraphics[width=0.45\textwidth,trim={0 0.5cm 0 0},clip]{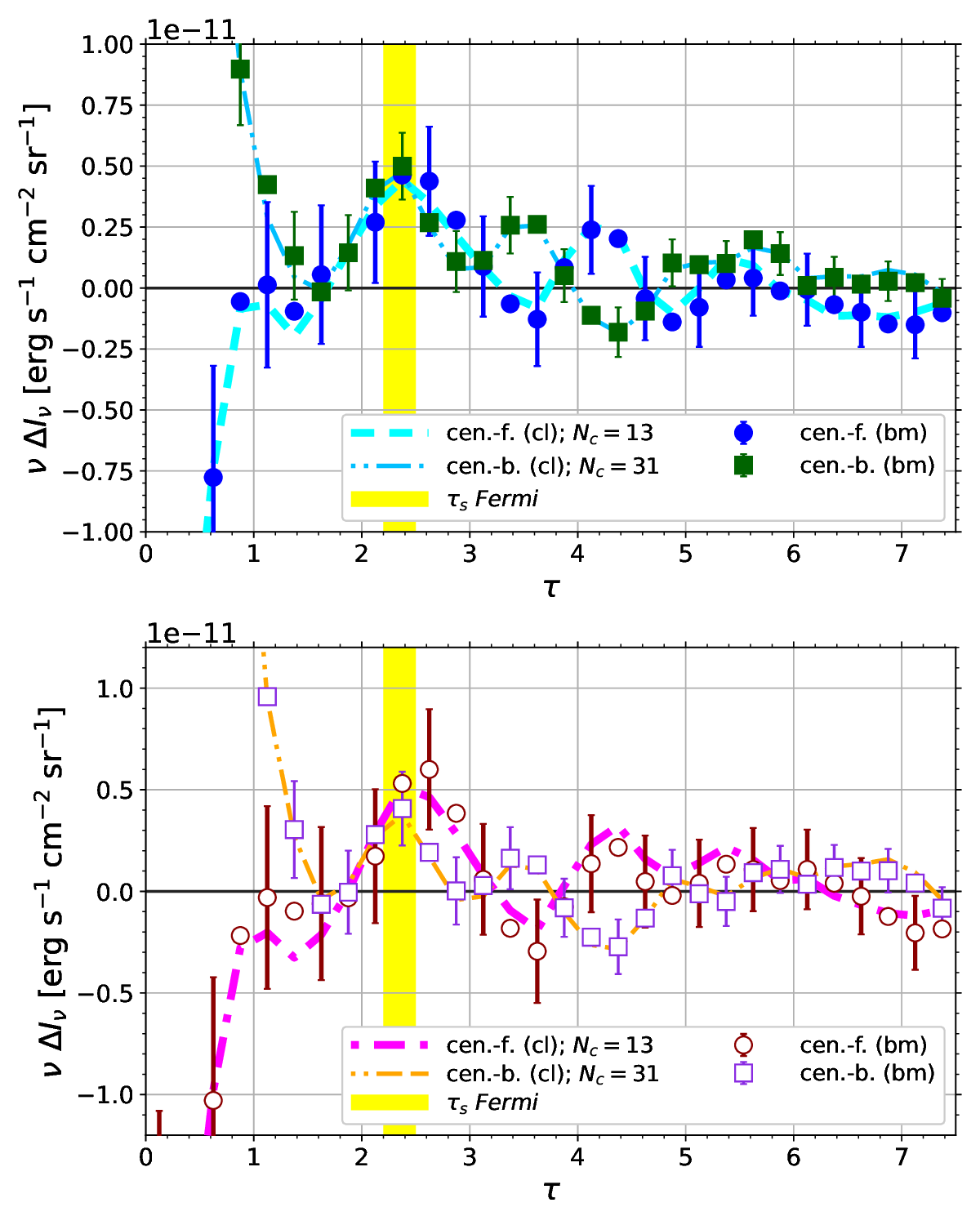}
    \caption{
    Same as Fig.~\ref{fig:band7_flux}, but showing the excess brightness separately for centre-faint (abbreviated ``cen.-f.''; circles and dashed curves) vs. centre-bright (``cen.-b.''; squares and double dot-dashed) cluster sub-samples, as defined in \S\ref{subsec:73MHz}.
    Results shown using both cleared (top panel) and restored (bottom) data, for both beam (symbols) and cluster (curves; error bars are omitted for visibility) co-additions, for the nominal analysis in the high-frequency channel (see legend, which also specifies the number $N_c$ of clusters in each sub-sample).
    }
    \label{fig:band7_weak_strong}
\end{bfigure}

After removing the background, 13 out of the 44 clusters in the sample are classified as centre-faint, while 31 clusters are classified as centre-bright.
Figure~\ref{fig:band7_weak_strong} shows the stacked excess intensity profiles of these centre-faint vs. centre-bright clusters,
for both \clean\ (top panel, with masking) and \orig\ (bottom, without masking) data.
As the cleared data show, the centre-faint clusters make a substantial contribution to the virial ring signal, comparable to and even exceeding that of the centre-bright clusters, when accounting for the different sizes of each sub-sample.
Namely, the virial excess in the cluster-faint clusters is significant at the $\sim 2.5\sigma$ confidence level,
slightly higher than anticipated from Poisson statistics when taking into account the smaller size of this sub-sample.

The mean intensity of the centre-faint clusters
agrees with that of the centre-bright clusters in the virial radius bin.
As expected, the two sub-samples show different, $\lesssim2\sigma$ fluctuations outside the clusters, beyond $\tau\simeq 3$.
The centre-faint signal is somewhat broader than found for the full sample, extending out to $\tau\simeq 3$, in resemblance of
the low-frequency signal discussed in \S\ref{subsec:7lowerbands}.
We conclude that the agreement between the virial signal as inferred from centre-faint vs. centre-bright clusters supports the virial signal as a genuine effect, rather than an artefact due to a putative failure of CLEAN to remove PSF sidelobes of the emission from the centres of clusters.

\subsection{Low-frequency channels}
\label{subsec:7lowerbands}

Next, consider the seven low-frequency channels, in order to test the results of the nominal, high-frequency analysis in \S\ref{subsec:73MHz} and to extract some spectral information.
In these frequencies, the resolution is substantially lower than in the high-frequency channel, PSF sidelobes are more pronounced, and the control samples indicate noise characteristics that require somewhat larger correction factors $\bar{\eta}$, as provided in Table \ref{tab:maps_sum}.
In order to obtain robust results, we therefore co-add the seven channels, taking into account the correlation factor $\psi$, unless otherwise stated.

\subsubsection{Brightness profile}
\label{subsubsec:LowFreqIntensity}

\begin{figure}
    \centering
    \includegraphics[width=0.45\textwidth,trim={0 0.5cm 0 0},clip]{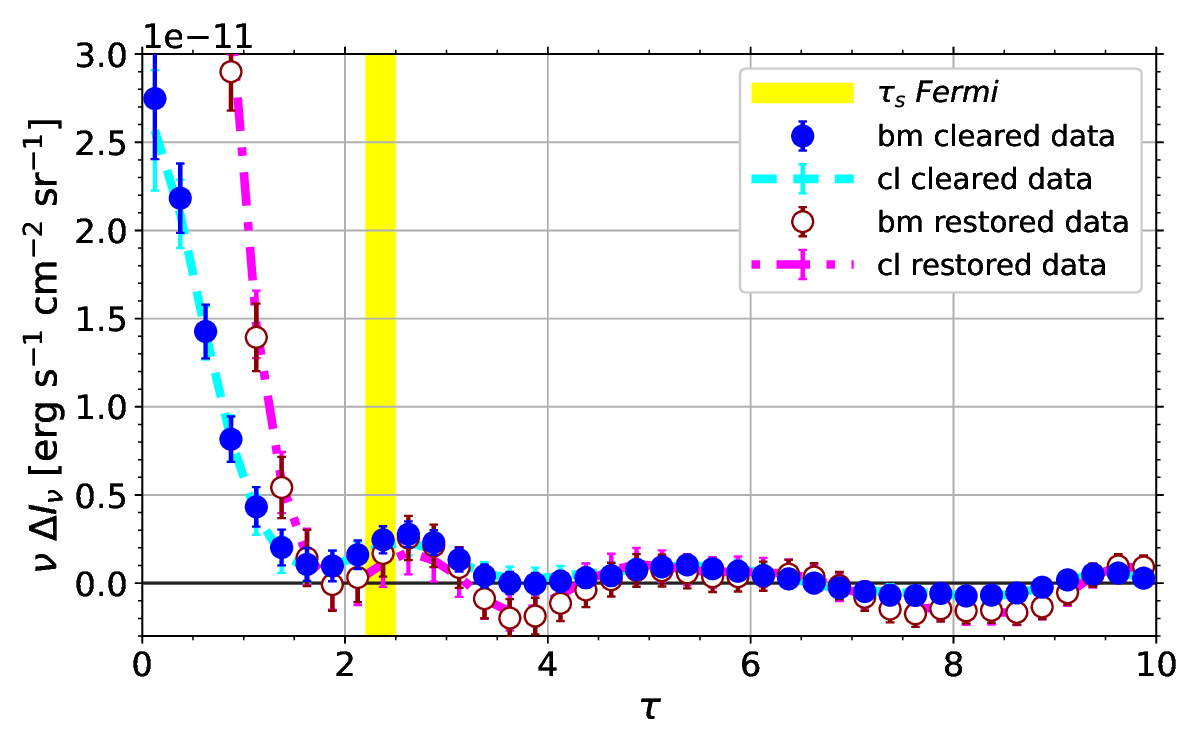}
    \caption{
    Brightness excess profile of the seven co-added
    low-frequency channels, averaged over clusters.
    Notations are the same as in Fig.~\ref{fig:band7_flux}.
    }
    \label{fig:band0-6_flux}
\end{figure}

Figure \ref{fig:band0-6_flux} presents the mean excess brightness profiles of the co-added low-frequency channels, for both the \clean\ and \orig\ data.
The figure shows $\Delta I(\tau)=\nu \Delta I_\nu$, which is optimised for an $I_\nu\propto \nu^{-1}$ spectrum
, computed as described in \append{\S\ref{append:intensity}}
[using Eq.~(\ref{eq:Intensity_sum}), with uncertainties computed according to Eq.~(\ref{eq:err_Intensity_sum})].
Again, we use nominal parameters as in \S\ref{subsec:73MHz}, and find that the two stacking methods yield consistent results.
For simplicity, we focus here (throughout \S\ref{subsubsec:LowFreqIntensity}) on beam co-addition.

The brightness profile in the low-frequency channels is similar to the nominal, high-frequency behaviour discussed in \S\ref{subsec:73MHz}, with expected differences due to the poorer resolution available here.
Strong central emission in the restored data is found to peak, when averaged over clusters and channels, at $\Delta I\simeq 1.1 \times 10^{-10}$ erg s$^{-1}$ cm$^{-2}$ sr$^{-1}$ (using beam co-addition),
quite similar in shape and in magnitude to the high-frequency channel.
Outside the clusters, at $\tau\gtrsim 3$, a very broad, $\sim2\sigma$
excess is seen throughout the $4.5\lesssim\tau\lesssim6$ range,
coincident with the $\tau \sim 5.5$ high-frequency excess discussed in \S\ref{subsub:sig_prof} and in \S\ref{subsec:summary_discussion}.

In the periphery of the clusters, a significant excess is found around $\tau\sim2.6$, broadly consistent with the virial signal seen in the high-frequency channel in terms of location, width, and amplitude.
The excess is again quite similar in cleared and restored data, suggesting that it is mostly diffuse in nature.
However, some differences between the two data sets suggest that contamination by point sources is stronger and more difficult to correct at low frequencies, due to the poor resolution and a possible abundance of soft-spectrum sources.

This virial excess in the cleared data peaks, when averaged over clusters and channels, at $\Delta I \sim 2.8 \times 10^{-12}$ erg s$^{-1}$ cm$^{-2}$ sr$^{-1}$ (for beam co-addition).
This value is similar to its high-frequency counterpart, supporting the detection of the high-frequency signal and affirming that it is not a PSF artefact, as discussed above (in \S\ref{subsec:sidelobe}) and
further below (in \S\ref{sec:summary} and \append{\S\ref{append:psf}}).
The virial excess at low frequencies is slightly ($\sim8\%$) less significant than in the high-frequency channel, possibly because the signal is convolved with more extended PSFs at low frequencies, rendering the signal broader (see \append{\S\ref{append:backgroundremove}});
a more careful analysis of the signal in different channels, along with a derivation of the implied spectrum, is deferred to \S\ref{subsec:spec_lwa} (backward modelling) and \S\ref{subsec:virial_spec} (forward modelling).

The virial signal is slightly broader in the low-frequency channels than it is in the high-frequency channel, as one might expect from the lower resolution.
Here, the signal peaks
in the $2.5\leq\tau<2.75$ bin, adjacent to the
$2.25\leq\tau<2.5$ bin of the high-frequency peak.
The broader signal at low frequencies shows a significant excess coincident with the $2.2\lesssim\tau\lesssim 2.5$ range of the stacked \emph{Fermi} signal and with the high-frequency LWA channel, but it extends
slightly beyond $\tau\simeq 3$.
In addition to the lower resolution, this broadening could be in part attributed to
the contribution of faint, soft point sources.
This possibility is supported by the substantial differences between the profiles obtained from cleared vs. restored data, for example the pronounced minimum around $\tau\simeq 3.7$
in the latter.
We revisit the issue in more detail in \S\ref{subsec:LowFrequencyVirial}.

\subsubsection{Significance profile}
\label{subsub:lowf_sigprof}

\begin{figure}
    \centering
    \includegraphics[width=0.45\textwidth,trim={0 0.5cm 0 0},clip]{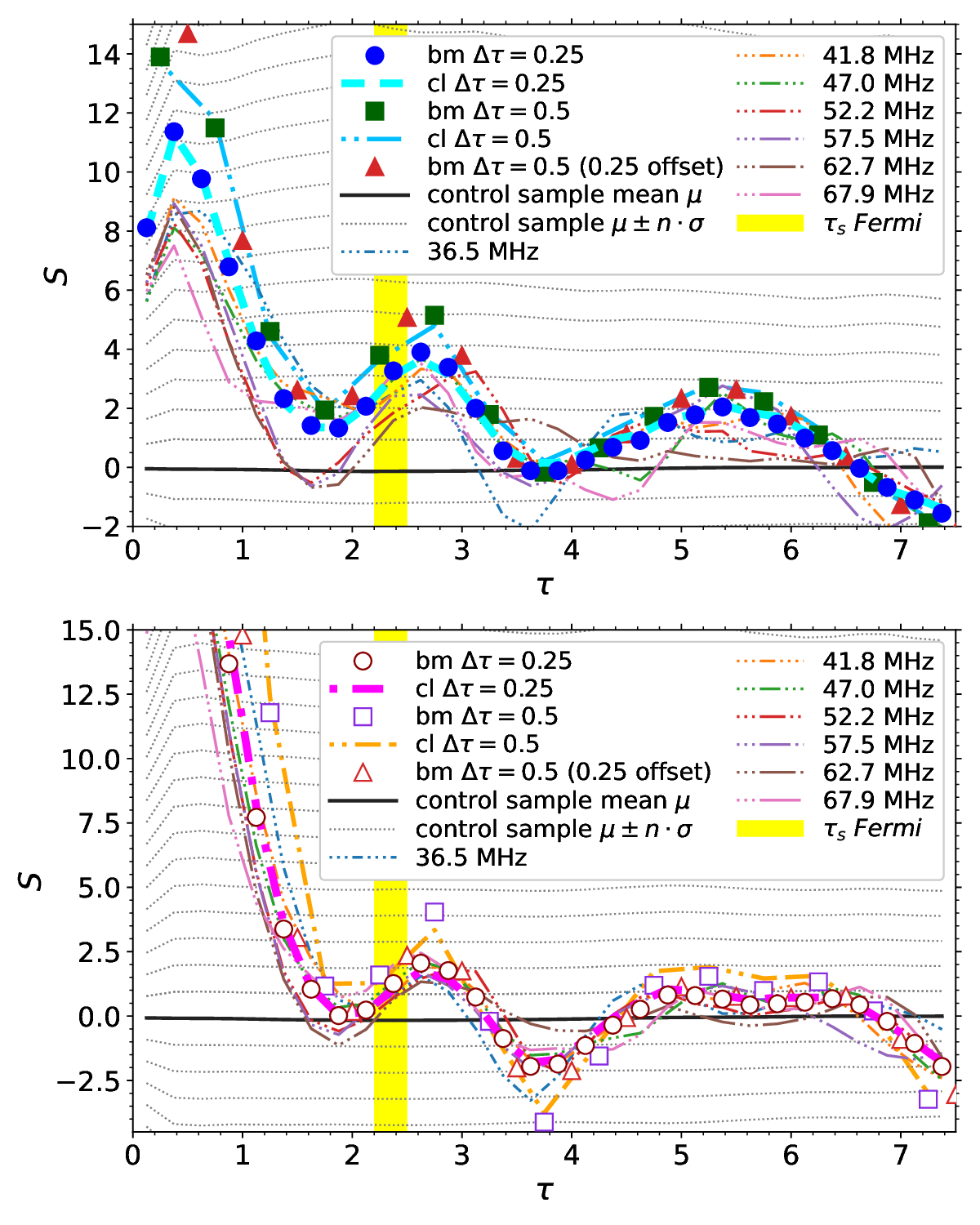}
    \caption{
    Same as Fig.~\ref{fig:band7_sig}, but for the co-added low-frequency channels, according to Eq.~(\ref{eq:band_stacking}).
    The significance of the excess brightness in individual channels is also shown (triple dot-dashed curves;
    longer dashing for higher frequencies; see legend), for beam co-addition only (for visibility).
    }
    \label{fig:band0-6_sig}
\end{figure}

Following the same procedure applied to the nominal, high-frequency channel in \S\ref{subsub:sig_prof}, next we compute the significance of the excess in the low-frequency channels.
Figure~\ref{fig:band0-6_sig} shows the resulting significance profiles $S$ for the \clean\ (top panel) and \orig\ (bottom) data, both for individual channels (thin curves without symbols)
and for the excess co-added over the seven low-frequency channels (symbols and thick curves).
Confidence levels extrapolated
from the corresponding (channel co-added) control samples are also shown (dotted curves), using the same procedure as for Fig.~\ref{fig:band7_sig}.

Like in the high-frequency case, the central signal is very significant at low frequencies, with a $\sim 25\sigma$ excess in the restored data. The dip here is seen at $\tau \sim 1.75$, just outside its high-frequency counterpart, presumably because of the lower spatial resolutions.
When using the cleared data, the central signal is reduced to the $\sim 12\sigma$ level, consistent with a dominant point-source contribution as found in the high-frequency case.

For $\Delta\tau = 0.25$ resolution,
both cleared and restored data sets show a virial signal peaked around $\tau=2.6$, seen separately in almost every individual channel, as well as in the co-added result.
The exception is the $52.2\MHz$ channel, where the peak is found farther out, around $\tau\simeq 3$.
The virial signal
is not strongly affected by the removal of point sources, although it does strengthen from $\sim 2\sigma$ in the restored data to $\sim 3.8\sigma$ in the cleared data, as expected from a diffuse signal contaminated by point sources, and as seen in the nominal analysis.
As in the high-frequency case, using thicker, $\Delta\tau= 0.5$ bins raises the significance, as expected for an extended signal.
In the cleared data, we again find a $\sim5\sigma$ excess in the $2.25\leq\tau<2.75$ bin, just like in the high-frequency channel.
Without an offset, the $\sim 5\sigma$ peak shifts from the $2\leq \tau<2.5$ in the high-frequency channel, to the adjacent, $2.5\leq\tau<3$ bin at low frequencies.
The nominal excess is comprised of the cumulative contribution of $\sim60\%$ of the clusters in our sample.

Beyond $\tau\simeq 3.5$, the significance profile is consistent with noise, with $\lesssim 1\sigma$ fluctuations in the restored data, and the aforementioned $\sim 2\sigma$ excess at $4.0 \lesssim \tau \lesssim 6.5$ in the cleared data.
This excess, peaked around $\tau\sim 5.5$, is robust, like its high-frequency counterpart, and is comprised of small contribution from most of the clusters in our sample.

\subsubsection{Virial excess}
\label{subsec:LowFrequencyVirial}

Focusing in more detail on the virial excess, we examine its properties based on the low-frequency channels, alone and in relation to the high-frequency results.
All low-frequency channels show some virial excess, in significance levels ranging between $\sim2.0\sigma$ and $\sim3.7\sigma$ in the cleared data.
Co-adding the cleared, low-frequency channels yields a significant virial signal,
reaching the $\sim(3.8\mbox{--}5)\sigma$ level.
This is comparable
to the $(4$--$5)\sigma$ level obtained using the cleared high-frequency channel alone, thanks to its superior resolution.
The low-frequency virial signal is, like in the high-frequency channel, robust and not sensitive to selection criteria or analysis details  (see \append{\S\ref{append:sensi}}).

As in the high-frequency case, the co-added low-frequency signal is sufficiently strong to be visible even without full radial binning, as illustrated in the
folded
maps in Fig.~\ref{fig:2dsig}. The virial, $\tau\sim 2.5$ arc is visible in the low-frequency quadrant 4 (bottom right), although it is somewhat more diffuse and less pronounced than in the high-frequency quadrant 2 (top left), due to the lower resolution. As expected, further folding the low-frequency result onto octant 6 (bottom triangle) or co-adding it to the high-frequency channel (top right quadrant 1) gives a more pronounced signal.

The low-frequency virial signal is somewhat less robust than it is in the high-frequency channel, presenting larger differences between beam and cluster co-addition methods.
In particular, we find $\sim7\%$ differences between the two stacking methods in the cleared data.
However, these discrepancies are not significant, and they appear to be at least in part associated with residual point sources. As expected, the \cleandata\ signals show better agreement between the two stacking methods than found in the restored data.

In low frequencies, the virial signal peaks consistently at radii somewhat larger than the $\tau\sim2.4$
peak found in the high-frequency channel, an effect attributed in part to the better resolution of the latter.
Six of the seven low-frequency channels peak in the $2.5 \leq \tau < 2.75$ bin when using the cleared data,
just outside the adjacent, $2.25\leq\tau<2.5$ bin of the high-frequency peak.
With a finer, $\Delta\tau\simeq0.167$
(6 bins per $\tau$; see \append{\S\ref{append:sensi}}) resolution, we again find the low-frequency peak adjacent to the high-frequency peak
with the same high resolution, indicating that the effect is small.
We conclude that
the peak shifts from $\tau\simeq 2.6$ in the co-added low-frequencies, to $\tau\simeq 2.4$ in the better resolved, high-frequency.

Interestingly, when using the restored data, some of the peaks shift to slightly larger radii.
This suggests that the discrepancy with respect to the high-frequency channel may in part be associated with low-frequency contamination from soft point sources, not fully removed by CLEAN.
Another possibility is that the virial emission from individual clusters has a broad distribution, extending out to $\tau\simeq 3$ or even beyond that, with some preference around $\tau\simeq 2.4$ that is picked up when sufficient high-resolution data are stacked. Such an interpretation could explain why the virial signals both for the low-frequency results and for the centre-faint, high-frequency results (see \S\ref{subsec:sidelobe}) have a similarly broad $\tau$ distribution, with a peak similarly shifted to $\tau\simeq 2.6$.

The comparable peak radii found in the different low-frequency channels
provide additional evidence that the virial excess is unlikely to be an artefact due to uncleaned sidelobe contamination from the central signal. Indeed, such putative ripples would show in different radii, corresponding to the frequency-dependent PSF, as we verify directly in \append{\S\ref{append:centre_restored}}.

\subsubsection{Spectra}
\label{subsec:spec_lwa}

\begin{figure}
    \centering
    \includegraphics[width=0.45\textwidth]{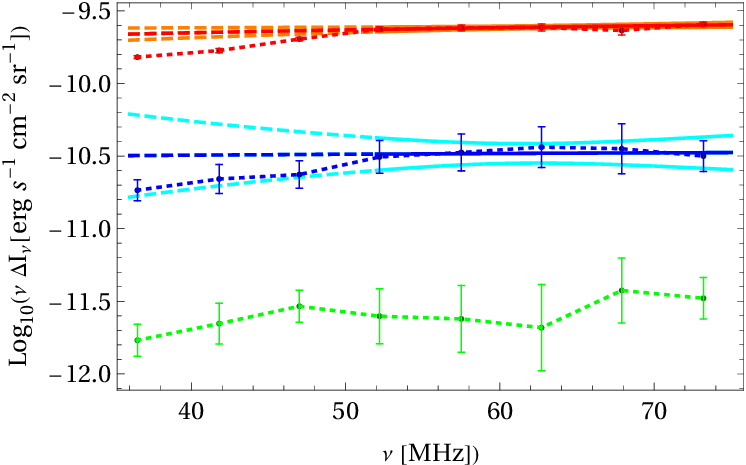}
    \caption{
    Spectrum (error bars with dotted lines to guide the eye) of the raw (restored data, non-masked; top red) and the cleaned (cleared data, NVSS-masked; middle blue), central ($0\leq\tau<\taumax=0.17$) signal, as well as the virial signal (in a broad, $2.25\leq\tau<3.0$ region; \cleandata; bottom green).
    At $\nu>50\MHz$ frequencies, where the PSF depends weakly on frequency, the central signal can be reasonably well fit as a power-law spectrum (solid curves with $1\sigma$ confidence levels), and extrapolated (dashed curves) to lower frequencies.
    For the virial signal, systematic (PSF and background) and statistical uncertainties are prohibitively large for such a backward determination of the spectrum 
    }
    \label{fig:Spectra}
\end{figure}

Next, consider a crude estimate of the
spectra of the central and virial signals, before deprojecting and modelling them in \S\ref{sec:centralemission} and \S\ref{sec:virialemission}.
The specific brightness of the two signals in all eight channels is shown in Fig.~\ref{fig:Spectra}, as a function of frequency.
For this purpose, we define the central signal as the $\tau<\taumax$ excess, exploring different choices of $\taumax$; the figure focuses on the most central, $\taumax=0.17$ region accessible with present resolution.
As the virial signal is extended radially at low frequencies, we define it for spectral purposes as the mean excess in the broad, $2.25\leq\tau<3.0$ range.
When considering the central signal, we examine both the raw excess, in the restored data without NVSS masking, and the cleaned excess, in the cleared data with masking.
The former includes the contribution of point sources, and is useful as it more accurately reflects the spectrum of any putative sidelobe effects due to the central signal.
The latter is a closer representation of the underlying, diffuse central signal, although some point source contamination is likely to persist.

In principle, the spectrum should be estimated by forward modelling, where different model signals are injected into control samples, which are then analysed with the real pipeline in order to determine the best-fitting model.
We use this method to analyse the spectrum of the central signal in \S\ref{sec:centralemission}, and of the virial signal in \S\ref{sec:virialemission}.
Here, we use a
crude backward modelling
to directly fit the binned spectrum in Fig.~\ref{fig:Spectra}.
The results of such modelling may be useful in regimes where projection effects are favourable
and PSF corrections can be approximately accounted for.
As in \S\ref{subsec:controlsamples}, let us define the spectral index $\alpha$ through $I_\nu\propto \nu^{-\alpha}$.

The raw central excess is the strongest among the three signals in the figure, and its statistical uncertainties (red error bars on the top curve) are negligible.
At frequencies above $50\MHz$, where the PSF no longer varies strongly with frequency (see Table \ref{tab:maps_sum}) and the resolution is sufficient for $\taumax=0.17$, the data are found to be
well-fit by a pure power-law, $\alpha=0.8\pm0.1$, with $\chi^2_\nu\simeq 0.4$ chi-squared per degree of freedom. At lower frequencies, the spectrum appears to be harder, but this is likely due to the strong frequency dependence of the PSF.
For larger $\taumax$, the spectrum remains
well-fit by a pure power law above $50\MHz$, but the integrated spectrum gradually softens.
Such softening suggests a considerable contribution from underlying diffuse emission, possibly in the form of radio halos and minihalos which often exhibit such a behaviour.
At larger scales, the steepening appears to saturate at some $1.5\lesssim\alpha\lesssim2$; for $\taumax=1$, we find $\alpha=1.5\pm0.1$ ($\chi^2_\nu\simeq 1.5$).
At small scales, extrapolation to $\taumax=0$ gives $\alpha=0.7\pm0.1$, consistent with a population of point sources.

It is interesting to compare the spectra of possible sidelobe ripples from the central signal, to the $\alpha\sim 1$ spectrum of the virial shock, as expected based on theory and on previous high-energy studies, and favoured in \S\ref{sec:virialemission} based on forward modelling.
The raw, steep, central signal can be approximated
as multiple point sources stacked close to the centre of the cluster,
so a strong, sharp ripple is expected to show $\alpha\simeq0.7$, substantially harder than the virial signal.
In contrast, the cumulative emission from an extended central region can give rise to a
more diffuse ripple with a much softer, $\alpha\sim 1.5$ spectrum.
Therefore, mistaking a ripple for an $\alpha\simeq 1$ virial signal is unlikely.
Such contamination cannot, however, be ruled out entirely based on spectral arguments alone, as the inferred spectrum is sensitive to the precise handling of point sources. For instance, removing one cluster with a bright NVSS source in its centre changes the raw spectrum in Fig.~\ref{fig:Spectra} from $0.8\pm0.1$ to $1.0\pm0.2$, which is similar to the expected virial signal. Naively, the ripples should remain softer than $\alpha=1$, but making such a determination requires a more careful analysis.

The \clean\ central signal is much weaker than the raw signal, its statistical errors are larger, and some spectral curvature emerges in backward modelling.
Consequently, while the signal can still be fit by a power-law spectrum, the uncertainty in the resulting spectral index, ranging from $\alpha=0.9\pm1.0$ at $\taumax=0.17$ to $\alpha=1.4\pm0.7$ at $\taumax=1$, is substantial.
Note that this mostly diffuse signal is quite similar to the expectation from stacked radio halos, in terms of its spatial and spectral distributions.
The virial signal is even weaker, strongly dominated by statistical uncertainties, and
the location of
its peak radius is
frequency-dependent,
so a backward determination of the spectrum is not useful.
The spectra of the two signals in the cleared data are
addressed more carefully below, using forward modelling
(see \S\ref{sec:centralemission}, \S\ref{subsec:virial_spec} and \S\ref{append:CentralResults}).

\section{Central excess}
\label{sec:centralemission}

Before studying the virial excess, it is important to separate out the central, $\tau\lesssim 1.5$ excess, to minimize its risk of biasing the measurement of the
virial signal.
We examine the central signal in both restored and cleared data.
As in \S\ref{subsec:spec_lwa},
the raw excess is studied using the restored data without masking NVSS sources, while the cleaned excess after point source removal is studied using the cleared data after masking NVSS
sources.

The virial signal could in principle contribute, in projection and
especially where the PSF is extended,
to the central excess.
As we show in
\append{\S\ref{append:mutual_effects}},
this contribution is only at the $\lesssim 5\%$ level for the raw data, and so can be neglected.
For the cleared signal, this contribution can reach $\sim20\%$, so the central and virial signals are simultaneously modelled in \S\ref{subsec:virialmodel}.
For simplicity, and to $\sim20\%$ accuracy,
in this section,
we neglect the virial contribution even for the cleared data.

We utilise the different appearances of the central signal in restored and cleared data to better model the emission and its effect on the virial signal.
Importantly, we examine if sidelobes of the central signal may be able to mimic the virial excess.
In \S\ref{sec:CentralModels}, we outline models for the central component in each data set. The fitting procedure, through $\chi^2$ minimisation, is described in \S\ref{subsec:Fitting_procedure}.
We outline the results in \S\ref{subsec:central_excess_results} and provide the technical details in \append{\S\ref{append:CentralResults}}.

\subsection{Central excess models}
\label{sec:CentralModels}

We model the central emission as a combination of a point source (subscript '{\pnt}') and an extended (subscript '{\ext}') component, with respective power-law spectral indices $\alpha_{\pnt}$ and $\alpha_{\ext}$.
The range of redshifts spanned by the clusters in our sample is small, so the specific flux projected along the line of sight
in normalised coordinates $\bm{\tau}=\{\tau_x,\tau_y\}$ can be approximately described by a simple one-dimensional model,
\begin{equation}
\label{eq:central_emision_point}
F_{cen} \equiv F_{\pnt} \tilde{\nu}^{-\alpha_p} \delta(\tau) + A_{\ext} \tilde{\nu}^{-\alpha_e} \frac{\Theta(\tau_{\rm cut}-\tau)}{(\tau^2 + \tau_c^2)^{\zeta}}  \coma \end{equation}
where $\tilde{\nu}\equiv \nu/\nu_h$ is the normalised (to $\nu_h=73.152\MHz$) frequency, $\tau_c$ is the core radius, and $\Theta$ is the Heaviside step function.
The normalizations $F_{\pnt}$ and $A_{\ext}$, and the normalised core radius $\tau_c$ and cutoff radius $\tau_{\rm cut}$, are for simplicity assumed to be identical for all clusters.
As the typical core radius, $0.1R_{500}$, is smaller than the FWHM of the LWA PSF even in the high-frequency channel, model results are not sensitive to the exact choice of $\tau_c$ around this value; hence, we henceforth use $\tau_c=0.1$.
For illustrative purposes, we define the total flux in the high-frequency channel, $F_{73}\equiv F_{\pnt}+F_{\ext}$, where
$F_{\pnt}$ and
$F_{\ext}\equiv 2\pi \tau_{\rm cut}^{2-\zeta}A_{\ext}/(2-\zeta)$
are the specific fluxes of the point-like and extended components, respectively, integrated over $\tau$ at $\nu=\nu_h$.

Recall that bright sources in the LWA maps are deconvolved
with the PSF, and the associated emission is either removed \citep[for cleared data;][]{2018AJ....156...32E} or restored with a Gaussian restoring beam (for restored data; see typical FWHM in Table \ref{tab:maps_sum}).
When modelling the cleared data, Gaussian restoration is irrelevant, and we simply convolve the model with the PSF.

In contrast, in the restored data, which shows a much brighter central excess than in the cleared data, most ($\sim 75\%$; see \S\ref{append:CentralResults}) of this excess has been identified by CLEAN and restored.
Indeed, the central signal is stronger and more sharply peaked in the restored data than it is in the cleared data, as seen from the stacked intensity (by a factor of $\sim4$; Figs.~\ref{fig:band7_flux} and \ref{fig:band0-6_flux}) and significance (Figs.~\ref{fig:band7_sig} and \ref{fig:band0-6_sig}) profiles.
Therefore, our model for the restored central excess  (\ref{eq:central_emision_point}) is convolved with
the Gaussian restoring beam and then fit to the data.
This model
can then be convolved with the PSF, to constrain the maximal putative sidelobe artefacts.

\subsection{Fitting procedure}
\label{subsec:Fitting_procedure}

For both restored and cleared data sets, we run Monte-Carlo simulations and minimize $\chi^2$ to determine the best-fitting parameters.
The procedure involves sampling a grid of possible model parameter values, and testing each parameter set against $N_{\rm mock}$ mock samples.
Each mock sample contains 44 mock clusters, like the real sample, distributed randomly across the allowed regions of the sky as described in \S\ref{subsec:controlsamples}.
For each mock cluster,
we inject the model emission Eq.~\eqref{eq:central_emision_point},
convolve the mock sky image with either the PSF or a restoring Gaussian beam, as explained above,
and pass the resulting mock image through our nominal
stacking and analysis pipelines.
We use $N_{\rm mock} = 30$ such mock samples for each choice of parameter set, sufficient
for convergence within $\lesssim0.1\%$ in $\chi^2$.

Each control sample $j$ yields a significance
distribution $\mathcal{M}_{j}(\nu, \tau,\Myc)=S(\nu,\tau, \Myc)$, as described in \S\ref{sec:StackedQuantities}.
We average among the $N_{\rm mock}$ samples of a given parameter set in order
to estimate its model outcome, $\mathcal{M}(\nu,\tau,\Myc) \equiv N_{\rm mock}^{-1} \Sigma_{j=1}^{N_{\rm mock}} \mathcal{M}_{j}(\nu,\tau,\Myc)$.
The $\chi^2$ value can then be computed, as usual, by summing over all relevant channels, radial bins, and clusters,
\begin{equation}
    \chi^2 =
    \sum_{\Myfr}
    \sum_{\tau} \sum_{\Myc=1}^{N_c} \left[\MySig(\nu,\tau,\Myc) - \mathcal{M}(\nu,\tau,\Myc)\right]^2
    \coma
    \label{eq:chi2_0}
\end{equation}
as $\MySig(\nu,\tau,\Myc)$ is
approximately a unit normal
random variable.
As in \S\ref{sec:result}, we separately analyse the high-frequency channel and the combined low-frequency channels.
The seven low-frequency channels are jointly
fitted with the same parameters of the model.
For the central excess modelling, we compute $\chi^2$ in the radial range $0\le \tau \le 1.5$ (and verify that the results are not sensitive to exact choice of outer radius).

A simple alternative to Eq.~\eqref{eq:chi2_0} is obtained by choosing the normalised stacked signal $\MySig(\nu,\tau)$ as the unit normal random variable, instead of its individual cluster counterpart $\MySig(\nu,\tau,\Myc)$.
This choice leads to
\begin{equation}
    \chi^2 =
    \sum_{\Myfr}
    \sum_{\tau} \left[\MySig(\nu,\tau) - \mathcal{M}(\nu,\tau)\right]^2
    \coma
    \label{eq:chi2}
\end{equation}
reducing the number $\DF$ of degrees of freedom (DOF) by a factor $\sim N_c$.
Here, we again averaged over mock samples, $\mathcal{M}(\nu,\tau) \equiv N_{\rm mock}^{-1} \Sigma_{j=1}^{N_{\rm mock}} \mathcal{M}_{j}(\nu,\tau)$, where  $\mathcal{M}_{j}(\nu, \tau)=S(\nu,\tau)$ is obtained by stacking over clusters, as defined in \S\ref{sec:StackedQuantities}, in either beam-weighted or cluster-weighted methods.
This prescription aims to better fit the stacked signals (shown in Figs.~{\ref{fig:band7_sig}} and \ref{fig:band0-6_sig}, typically giving a lower $\chi^2$ per DOF), at the expense of the sensitivity to the mass-dependence of individual cluster models.
The method is more conservative than Eq.~\eqref{eq:chi2_0}, typically providing similar parameter estimates at lower confidence levels, so we primarily use Eq.~\eqref{eq:chi2}.
Quantitatively, Eqs.~\eqref{eq:chi2_0} and \eqref{eq:chi2} typically yield $\lesssim 20\%$ differences between their best fit parameter estimates.
The values of $\zeta$ and $\tau_{\rm cut}$ change slightly more, by $\sim 30\%$, but
these variations typically lie within
the $1\sigma$ uncertainty of these parameters.
A larger and more significant difference is obtained for the normalisation $\dot{m}\xi_e\xi_B^{(2+p)/4}$ of the virial signal.
Here, using Eq.~\eqref{eq:chi2} lowers the best fit by $\sim (20 \text{--} 55)\%$; see discussion in \S\ref{subsec:TS}.

The frequency summations in Eqs.~(\ref{eq:chi2_0}) and (\ref{eq:chi2}) do not take into account correlations between channels, and so tend to overestimate the $\chi^2$ expected for co-adding independent low-frequency channels.
Indeed, these equations
reasonably identify the best-fitting parameters, but the correlations
can lead to
unreasonably small parameter uncertainties.
We may crudely correct for correlations by choosing the corrected $\MySig(\tau)$, defined in  Eq.~(\ref{eq:band_stacking}), as the random variable. Along with the corresponding, corrected model
$\mathcal{M}(\tau) = N_\nu^{-(1+\psi)/2} \Sigma^{N_\nu}_{\nu=1} \mathcal{M}(\nu, \tau)$,
we then obtain
\begin{equation}
    \chi^2 =
    \sum_{\tau} \left[\MySig(\tau) - \mathcal{M}(\tau)\right]^2
    \coma
    \label{eq:chi2_2}
\end{equation}
further reducing $\DF$
by a factor of $\sim N_\nu$.
Equation (\ref{eq:chi2_2}) cannot be used to model the spectrum, and so is mainly invoked below only to gauge the correlations.

For model parameters other than the spectrum, we can compare the best fits obtained from Eqs.~(\ref{eq:chi2}) and (\ref{eq:chi2_2}).
While the best-fit values agree, the former equation leads to consistently and predictably underestimated uncertainties, due to channel correlations.
The results become fully consistent with each other if one
multiplies the $\chi^2$ of Eq.~(\ref{eq:chi2}) by a constant correction factor $\MyChiCorr$. When co-adding cleared data in all low-frequency channels, this correction yields consistent results for all model parameters, provided that $\MyChiCorr\simeq 0.25$ (see \append{\S\ref{append:controlsample}}).

An alternative way to correct for correlations is by generalising Eq.~(\ref{eq:band_stacking}) to all model parameters. Namely, assuming that the uncertainty in any model parameter $a$ should satisfy $\sigma_{a,N} = N^{(\psi - 1)/2} \sigma_a$. Here, $\sigma_a$ ($\sigma_{a,N}$) denotes the uncertainty obtained from one channel ($N$ channels), and $\psi$ was calibrated in \S\ref{sec:StackedQuantities}.
We find that Eq.~(\ref{eq:chi2}) yields a similar relation, $\sigma_{a,N}^* = N^{(\Psi - 1)/2} \sigma_a^*$, but with a different correlation parameter $\Psi$ affected by the numerical procedure, found to be $\Psi \simeq 0.45$ for the restored data and $\Psi \simeq0.01$ for the cleared data.
Hence, when deriving the uncertainty $\sigma_{a,N}^{*}$ in $a$ from Eq.~(\ref{eq:chi2}) with summation over $N$ channels, we correct the result as
\begin{equation}
\label{eq:uncertainty_corr}
\sigma_{a,N} = N^{(\psi - \Psi)/2} \sigma_{a,N}^{*} \coma
\end{equation}
and again recover a good agreement with the outcome of Eq.~(\ref{eq:chi2_2})
(see demonstrations in \append{\S\ref{append:controlsample}}).
Anticipating
parabolic $\chi^2$ minima
near
the best-fit values, we find that $\MyChiCorr\simeq N^{(\Psi - \psi)}\simeq 0.26$
for the cleared data,
consistent with the independent
finding above,
and
$\MyChiCorr\simeq 0.46$ for the restored data.

Unless otherwise stated, we henceforth compute $\chi^2$ using Eq.~\eqref{eq:chi2} and correct the co-added results for the low-frequency channels using Eq.~\eqref{eq:uncertainty_corr}.
Other methods are occasionally demonstrated to yield similar results.
We verify that the results are not sensitive to the choice of $N_{\rm mock}$.
For brevity, in the following text, we sometimes quote $\chi^2$ (or, when relevant, $\MyChiCorr\chi^2$) values only for beam co-addition, but provide both beam (denoted 'bm') and cluster ('cl') values in the tables and most figures.

\subsection{Central modelling results}
\label{subsec:central_excess_results}

For the raw data, the Gaussian-convolved
central model, including both point-like and extended components, provides a good fit
in both high-frequency ($\chi^2$ per DOF value of $\chi_n^2\simeq 0.8$)
and co-added low-frequency
($\MyChiCorr\chi_n^2\simeq 0.6$)
channels, but the large parameter uncertainties suggest that the model is oversimplified.
Here, the spectral indices of the two components typically fall in the range $0.7\lesssim \alpha\lesssim 1.3$, but the point-like component softens considerably in the higher-frequency channels, and we are unable to  determine the spectrum of each component more precisely, as the results are sensitive to the co-addition method and channel selection.
The detailed modelling is outlined in \append{\S\ref{append:centre_restored}}.

In the cleared data, the extended emission component alone, convolved with the PSF, reproduces the central signal nicely in both high-frequency ($\chi^2_n\simeq 0.2$) and co-added low frequency ($\MyChiCorr\chi^2_n \simeq 0.1$) channels, with a spectral index $\alpha_e = 1.08^{+0.24}_{-0.24}$ (beam co-addition), consistent with the backward estimate in \S\ref{subsec:spec_lwa}.
This extended excess likely arises from a combination of faints sources, diffuse emission from the cluster cores, and bright point-source residuals that were not fully removed by CLEAN.
The detailed modelling is provided in \append{\S\ref{append:centre_cleared}}.

We find that in both raw and cleared data, residuals due to the central signal cannot account for the virial excess, thus supporting the validity of the latter.
The worst-case putative post-CLEAN sidelobe artefacts can be estimated from the PSF-convolved model inferred (with restoring beam convolution) from the raw data.
The spurious ripples thus introduced in the stacked quantities are frequency-dependent and inconsistent with the data, failing to mimic the virial excess; see \append{\S\ref{append:centre_restored}} for details.
In general, the first ripple is found at radii too small to account for the virial excess, whereas the second ripple is too extended and too weak for confusion with the virial excess.
Similarly, the spectrum of the raw central excess appears to vary with $\tau$ and with $\nu$, so its putative sidelobe residuals are very unlikely to reproduce the robust flat spectrum (see \S\ref{subsec:virial_spec}) of the virial excess.

\section{Virial excess signal}
\label{sec:virialemission}

With the central signal quantified in \S\ref{sec:centralemission}, we now turn to the virial excess, focusing on the cleared data, where much of the point-source contamination has been removed.
We test if the observed signal agrees with simple models for the synchrotron emission from the virial shock, which are briefly provided in \S\ref{subsec:virialmodel} based on a derivation in \append{\S\ref{append:model}}.
To this end, we combine linear regression
with TS tests, as described in \S\ref{subsec:TS}.
The results are then presented in \S\ref{subsec:virial_p2} for the special case of
a flat injection spectrum, $p=2$,
giving rise to a cooled photon spectrum $\alpha_{vir} \simeq 1$ from the virial shock.
The results are then generalised for an arbitrary spectrum in \S\ref{subsec:virial_spec}.
We examine the dependence of the virial excess signal upon the cluster mass in \S\ref{subsec:mass_depen}.
Finally, the magnetic field downstream of the shock is estimated in \S\ref{subsec:magnetic}.

\subsection{Virial shock synchrotron model}
\label{subsec:virialmodel}

Our virial shock modelling assumes that the shock deposits a fraction $\xi_e$ of the downstream thermal energy in CREs, and a fraction $\xi_B$
in magnetic fields. While CRE cooling is dominated by Compton losses off the CMB for the relevant, $B<1\muG$ fields, a fraction $\sim B^2/(8\pi u_{cmb})$ of the energy is radiated as synchrotron emission, where $u_{cmb}$ is the CMB energy density. The CREs cool rapidly, so their radiative signature is expected to be well-localised near the virial shock, reflecting the temporal and spatial changes in energy injection rate across the shock.
Therefore, while we model the emission from a virial shock, for simplicity, by invoking a stationary flow in spherical symmetry, the actual shock morphology can be highly irregular, and the energy injection rate through the shock can show strong variations with position and time.

The co-addition procedures we apply to the LWA data
effectively
average out these spatial and temporal variations, preferentially picking up the component of the virial shock signal that is most conducive to our stacking and circular binning.
The stacking and analysis procedures can also pick up foreground and background contaminations,
such as the changes in background due to the quenching of star-forming galaxies that crossed inside the virial shock.
We consider two different spatial models for the virial-shock component of the stacked radio signal: emission from a thin spherical shell at a normalised radius $\tau_s$ (henceforth the shell model), and emission from a thin ring of the same radius $\tau_s$ on the plane of the sky (henceforth the {\planar} model).

For concreteness, and following {\RK}, we adopt an isothermal $\beta$-model to describe the gas distribution in a cluster, and assume hydrostatic equilibrium.
The anticipated signal, derived for both shell and {\planar} models in \append{\S\ref{append:model}}, then follows the scaling
\begin{equation}
    \nu I_\nu \propto \dot{m} \xi_e \xi_B^\frac{2+p}{4} \tau_{\sh}^{-\frac{p+4}{2}} \times \frac{M_{14}^\frac{8 + p}{6}}{(1+z)^4} \times
    \nu^{-\frac{p-2}{2}} \tilde{f}(\tau/\tau_{\sh})
    \coma
\label{eq:shock_prop}
\end{equation}
where we defined $M_{14} \equiv M_{500}/10^{14} \Msun$
and the normalised angular dependence $\tilde{f}$ of the relevant (shell or {\planar})
model.
Given the cluster mass and redshift, the free parameters of the model are the normalised shock radius $\tau_{\sh}$, the CRE injection spectral index $p$, and the product
$\dot{m}\xi_e\xi_B^{(2+p)/4}$,
which weighs the gas accretion rate, the CRE acceleration efficiency, and the magnetisation efficiency.
The proportionality constant in Eq.~\eqref{eq:shock_prop} depends on $p$, the cosmology, and the range $\{\gmin,\gmax\}$ of CRE Lorentz factors
[see Eq.~\eqref{eq:shock_nuInu}].
In the anticipated limit of a strong shock, $p=2$, so $\nu I_\nu$ becomes frequency independent, the free parameters reduce to $\tau_{\sh}$ and $\dot{m}\xi_e\xi_B$, and the proportionality constant depends on the CRE energy range only through the combination $\ln (\gmax/\gmin)$
[see Eq.~\eqref{eq:StrongShock_nuInu}].

\subsection{Fitting procedure}
\label{subsec:TS}

To determine the best-fitting parameters and their uncertainties, we apply the same procedures here as in \S\ref{subsec:Fitting_procedure}.
Namely, for each model and set of parameters, we average over multiple Monte-Carlo simulations of control samples incorporating the injected model, compare the outcome to the data, and minimize the resulting $\chi^2$.
Again, the analysis is performed
separately for the high-frequency channel and for the co-added seven low-frequency channels.

The $\chi^2$ minimisation is equivalent to maximising the likelihood $\mathcal{L}$, given
by
\begin{equation}
    \ln \mathcal{L} = - \frac{1}{2} \chi^2.
    \label{eq:likelyL}
\end{equation}
For the co-added low-frequency channels, we use the corrected value $\MyChiCorr\chi^2$ instead of $\chi^2$, as defined and tested in \S\ref{subsec:Fitting_procedure}, to crudely correct for inter-channel correlations. Again, this has no effect on the estimated parameters, and only corrects their estimated uncertainty intervals.

To quantify the significance of the virial excess,
we use the test statistic \citep[\eg][]{MattoxEtAl96_TS}
\begin{equation}
    {\rm TS} \equiv -2 \ln \frac{\mathcal{L}_{-}}{\mathcal{L}_{+}} = \chi^2_{-} - \chi^2_{+} \coma
\end{equation}
where subscript '$-$' ('$+$') refers to the model without (with) the virial shock.
After maximising the likelihood of each model over its free parameters,
the resulting TS is assumed to approximately have a $\chi^2$ distribution with $\DF = \DF_{+}-\DF_{-}$ degrees of freedom, corresponding to the extra parameters added by the virial shock
\citep{Wilks1938}.
The implied significance of the excess is denoted $\sigma_{\rm TS}$.

Beam and cluster co-additions give similar results for the virial excess; for brevity,
the following text focuses mainly on beam co-addition.
In particular, quoted values of TS, $\chi^2_n$, and, when relevant, also $\MyChiCorr\chi^2_n$, refer to beam co-addition unless otherwise stated.
Both beam-weighted and cluster-weighted results are provided
in most figures
and in the supplementary material.

Figures~\ref{fig:band7_sig} and \ref{fig:band0-6_sig} show that the virial excess is localised in the $2 \lesssim \tau \lesssim 3$ bins in the high-frequency data, and in the $2 \lesssim \tau \lesssim 3.5$ bins in the low-frequency data.
Therefore, our nominal $\chi^2$ analysis focuses on the radial range $0 \leq \tau \leq \tau_{\rm fit}$
with
$\tau_{\rm fit} = 5$, including both central and virial signals. We verify that the results are not sensitive to
reasonable variations in
$\tau_{\rm fit}$
(ranging from $4$ to at least $7.5$).

The parameters inferred for the central and virial models are not entirely independent of each other.
In \S\ref{sec:centralemission}, we modelled
the central source parameters while focusing on small radii and neglecting the virial contribution.
In \S\ref{subsec:virial_p2}
below, we simultaneously model the virial excess and the main parameters of the central signal.
Comparing the results indicates that the virial excess has a small but noticeable, $\lesssim15\%$ effect on the inferred parameters of the central model.
The opposite effect --- a putative bias in the virial model parameter estimates due to contamination by the central component --- leads to a somewhat larger effect, with
$\lesssim25\%$ changes in best-fitting parameters.
Such effects are discussed in \S\ref{append:mutual_effects}.

In addition to the statistical errors associated with the real and control samples, propagated into the model-parameter estimates, there are also systematic effects associated with the model assumptions, the properties of the cluster catalogue, etc.
Based on the study of different model and analysis variants, we find that the systematic uncertainties are typically no larger than the statistical errors.
An exception is the radio signal normalisation, $\dot{m}\xi_e\xi_B^{(2+p)/4}$, where the systematic error may be as large as a factor of $\sim 4$.

While replacing our nominal $\chi^2$ estimate in Eq.~\eqref{eq:chi2} by the more standard, \ie non-stacked Eq.~\eqref{eq:chi2_0}, does not significantly change the best-fitting values of most parameters, it does tend to lower the best-fitting normalisation $\dot{m}\xi_e\xi_B^{(2+p)/4}$
of the virial signal by $\sim 20 \text{--} 55 \%$.
Hence, we present results based on both equations in the following sections.
The difference in the best-fitting $\dot{m}\xi_e\xi_B^{(2+p)/4}$ can be attributed at least in part to the dispersion of excess significance among individual clusters, and the partial correlation between this excess and cluster mass.

\begin{bfigure*}
    \centering
    \includegraphics[width=\textwidth,trim={0 0.5cm 0 0},clip]{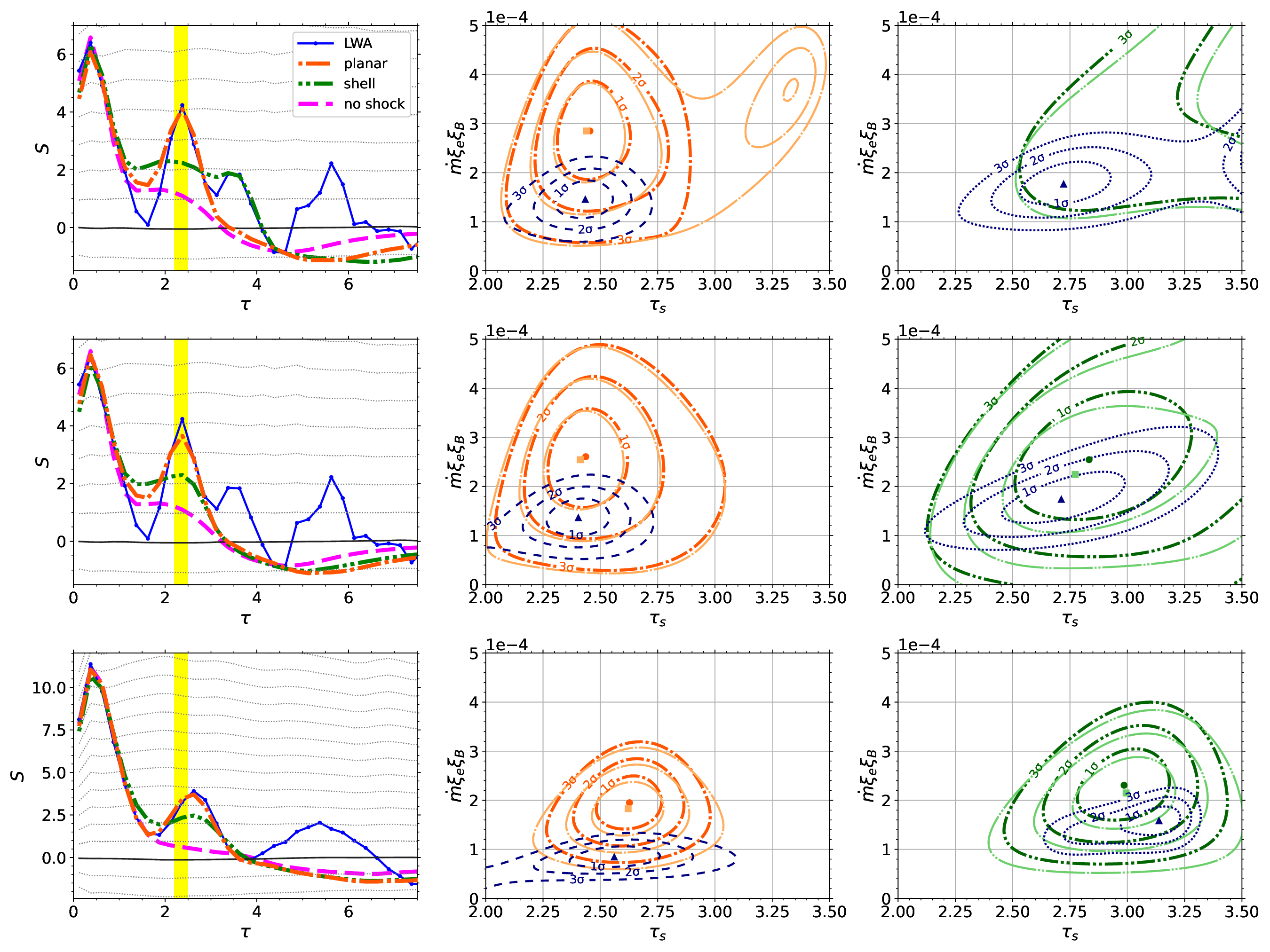}
    \caption{
    Modelling both central and virial signals in the high-frequency (top row for $\tau_{\rm fit}=5$ and middle row for $\tau_{\rm fit}=3$) and co-added low-frequency (bottom, $\tau_{\rm fit}=5$) \cleandata\ channels.
    The left panels show the significance profiles of LWA data (solid blue curve) and the best fit {\planar} (dot-dashed orange), shell (double dot-dashed green), and no shock (dashed magenta) models.
    The control sample curves and vertical shock range are the same as in Fig.~\ref{fig:band7_sig}.
    The middle and right panels show the best-fit parameters (symbols) and the $1\sigma$--3$\sigma$ confidence level contours (same curves styles as in the left panels) for nominal analysis with beam (disks and thicker, short dashed contours) and cluster (squares and thinner, long dashed contours) co-addition.
    Switching to the standard $\chi^2$ of Eq.~\eqref{eq:chi2_0} (blue triangles with dashed or dotted confidence levels) tends to lower $\dot{m}\xi_e\xi_B^{(2+p)/4}$ and increase the confidence level; see \S\ref{subsec:TS}.
	The contours in the top row are biased by the $\sim 2\sigma$ contamination around $\tau \simeq 3.5$, but not when using the standard $\chi^2$ (see text for details).
    }
    \label{fig:model}
\end{bfigure*}

\subsection{Results for a strong virial shock ($p=2$)}
\label{subsec:virial_p2}

\begin{table*}
	\caption{Best fit for
	joint central and virial shock nominal (cleared data; beam co-addition) modelling.
	} 
    \centering
\hspace{-0.3cm}
\begin{tabular}{ccccccccccc}
Model & Channel & $F_{73}$ & $\alpha_e$ & $\tau_s$ & $\dot{m}\xi_e\xi_B^{\frac{2+p}{4}}$ $(10^{-4})$ & $p$ & $\chi^2$ & $\DF$ & TS ($\sigma$) \\
(1) & (2) & (3) & (4) & (5) &
(6) & (7) & (8) & (9) & (10) \\ \hline
\planar & high & $1.63^{+0.23}_{-0.20}$ & — & $2.45^{+0.11}_{-0.09}$  & $2.85^{+0.66}_{-0.65}$ & \textbf{2} & 17.2 & 15 & 18.5 (3.9$\sigma$)  \\
& low & $2.01^{+0.25}_{-0.20}$ & $1.20^{+0.27}_{-0.26}$  & $2.63^{+0.09}_{-0.10}$  & $1.95^{+0.36}_{-0.35}$ & \textbf{2} & 137.1 & 134 & 30.5 (5.2$\sigma$)  \\
& low & \textbf{2.01} & \textbf{1.20} & $2.62^{+0.10}_{-0.10}$  & $2.27^{+0.71}_{-0.50}$ & $2.02^{+0.17}_{-0.16}$  & 137.1 & 133 & 29.7 (4.8$\sigma$) \\
shell & high$^{\dagger}$
& $1.65^{+0.23}_{-0.23}$ & — & $2.78^{+0.31}_{-0.21}$ & $2.29^{+1.17}_{-0.95}$ & \textbf{2} & 13.2 & 7 & 7.6 (2.3$\sigma$) \\
& low & $1.95^{+0.25}_{-0.23}$ & $1.14^{+0.28}_{-0.26}$  & $2.99^{+0.13}_{-0.13}$ & $2.31^{+0.49}_{-0.47}$ & \textbf{2} & 162.0 & 134 & 24.0 (4.5$\sigma$)  \\
& low & \textbf{1.95} & \textbf{1.14} & $3.00^{+0.13}_{-0.13}$ & $2.73^{+0.63}_{-0.59}$ & $2.04^{+0.12}_{-0.11}$ & 160.0 & 133 & 23.8 (4.2$\sigma$) \\
\end{tabular}
    \label{tab:fitting_virial}
\begin{tablenotes}
\item
    {\bf Columns:}
    (1) Shock model;
    (2) The high-frequency (high) or the seven co-added low-frequency (low) channels;
    (3) Flux of the central emission normalised to the 73MHz frequency, in units of $10^{-23}$ erg s$^{-1}$ cm$^{-2}$ Hz$^{-1}$;
    (4) Spectral index of the extended central emission;
    (5) Shock radius normalised to $R_{500}$;
    (6) Virial shock normalisation, in $10^{-4}$ units;
    (7) Injected spectral index of CREs;
    (8) $\chi^2$ values of the fit (without the $\mathcal{C}$ correction; see \S\ref{subsec:Fitting_procedure});
    (9) Number of degrees of freedom;
    (10) TS value (equivalent significance in parenthesis).
    See text (\S\ref{subsec:virial_p2}) for technical details.
    Values in \textbf{boldface} are fixed rather than fitted.\\
    $^{\dagger}$ --- Using $\tau_{\rm fit}=3$ to avoid a $\tau\sim3.5$ fluctuation; see \S\ref{subsec:model_high_nu}.
\end{tablenotes}
\end{table*}

We begin with the simple case where the shock is strong.
Here, CREs are injected with a flat, $p=2$ spectrum, so the resulting $\nu I_\nu$ of synchrotron emission is frequency independent.
We fit the cleared data to the joint model, comprised of both a central extended source and virial shock emission.
For the virial shock component, we fit the parameters $\tau_s$ and $\dot{m}\xi_e\xi_B$.
For the central component of the model, we fit the parameters $F_{73}$ and $\alpha_e$, but fix the parameters $\zeta=0.9$ (1.2) and $\tau_{\rm cut}=0.9$ (1.6) for high (low) frequency data to their best-fit values, for simplicity and because
these values are poorly constrained.
Our results --- including the virial shock parameters --- are not sensitive to the exact choices of $\zeta$ and $\tau_{\rm cut}$
(see \append{\S\ref{append:centre_cleared}} and Table \ref{tab:fitting_virial_detail}). We find that biases introduced by the central signal to the virial modelling and vice versa are small (see \append{\S\ref{append:mutual_effects}}).

Figure \ref{fig:model} shows the best-fitting significance profiles (left column) and parameter estimates (middle and right columns) for the high-frequency (top and middle rows) and co-added low-frequency (bottom row) channels.
Confidence contours in such figures are (unless otherwise stated) based on fitting $\chi^2$ as a fourth-order polynomial in the two free parameters, for better visibility.
Models
are shown both in the absence (dashed significance profiles) and in the presence (other dashings) of the virial shock, in both planar (dot-dashed)
and shell (double dot-dashed)
model variants.
Table \ref{tab:fitting_virial} demonstrates the
best-fitting parameters and the implied TS significance levels of the virial shock, for different model variants (more details and models are provided in  Table \ref{tab:fitting_virial_detail}).
In the low-frequency co-addition results shown in the table, parameter uncertainties are corrected using Eq.~(\ref{eq:uncertainty_corr}) and the TS values are based on $\mathcal{C}\chi^2$ (where $\mathcal{C} \simeq 0.25$; see \S\ref{subsec:Fitting_procedure}).

We discuss the results for the high-frequency channel in \S\ref{subsec:model_high_nu} and for the combined low-frequency channels in \S\ref{subsec:model_low_nu}, obtained in both cases by simultaneously fitting the central and virial signals with $p=2$.

\subsubsection{High-frequency channel}
\label{subsec:model_high_nu}

The high-frequency models are shown in the top and middle rows of Fig.~\ref{fig:model}.
The top row uses the nominal data, featuring in particular a $\sim2 \sigma$ excess around $\tau\simeq 3.5$, shown above (see Fig.~\ref{fig:band7_sig}) to arise from two clusters with extended NVSS point sources.
To reduce the sensitivity of the model to these and other structures outside the virial radius, the middle row of Fig.~\ref{fig:model} shows high-frequency channel results when limiting the fitting radial range to $\tau_{\rm fit} = 3$.

The nominal {\planar} model (orange dot-dashed curves) fits the virial shock signal well, with $\chi_n^2(\tau \le 5) \simeq 1.0$, and is not sensitive to the inclusion (top row) or exclusion (middle row) of $\tau>3$ data.
The best-fitting parameters are (for $\tau_{\rm fit}=5$, unless otherwise stated)
\begin{equation}
    \tau_s=2.45^{+0.11}_{-0.09} ~ (2.44^{+0.11}_{-0.09})
    \label{eq:best_tau_s}
\end{equation}
and
\begin{equation}
    \dot{m}\xi_e\xi_B=2.85^{+0.66}_{-0.65} ~ (2.79^{+0.70}_{-0.64}) \times 10^{-4}
    \label{eq:best_xi}
\end{equation}
for beam (cluster; henceforth omitted) co-addition.
The best-fitting shock position coincides with that inferred by {\RK} based on the \emph{Fermi} \gama-ray signal.
We find that $\mbox{TS}\simeq 18.5$, corresponding to a $3.9\sigma$ detection of the virial shock for $\DF=2$ degrees of freedom.
If we were to use the $\tau_s = 2.4$ of {\RK} as a prior, the significance would increase to $\mbox{TS}\simeq 18.2$ ($4.3\sigma$ for $\DF=1$).

The above results are based on the nominal $\chi^2$ of Eq.~(\ref{eq:chi2}).
If we use, instead, the more standard, non-stacked least-squares of Eq.~(\ref{eq:chi2_0}), the best-fitting shock position $\tau_s = 2.40^{+0.08}_{-0.08}$ is practically unchanged, but $\dot{m}\xi_e\xi_B = 1.48^{+0.27}_{-0.27} \times 10^{-4}$ is nearly halved (blue dashed contours), and the significance of the virial shock increases to $\TS\simeq33.2$ ($5.4\sigma$ for $\DF=2$, or
$5.8\sigma$ for $\DF=1$ with the {\RK} prior).
Figure \ref{fig:band7_model_TS} presents the radial significance profiles based on Eq.~(\ref{eq:chi2_0}).
Also shown is the TS-based significance profile (thick red solid curve)
corresponding to the nominal {\planar} model, as a function of $\tau_s$.

\begin{bfigure}
    \centering
    \includegraphics[width=0.45\textwidth,trim={0 0.4cm 0 0},clip]{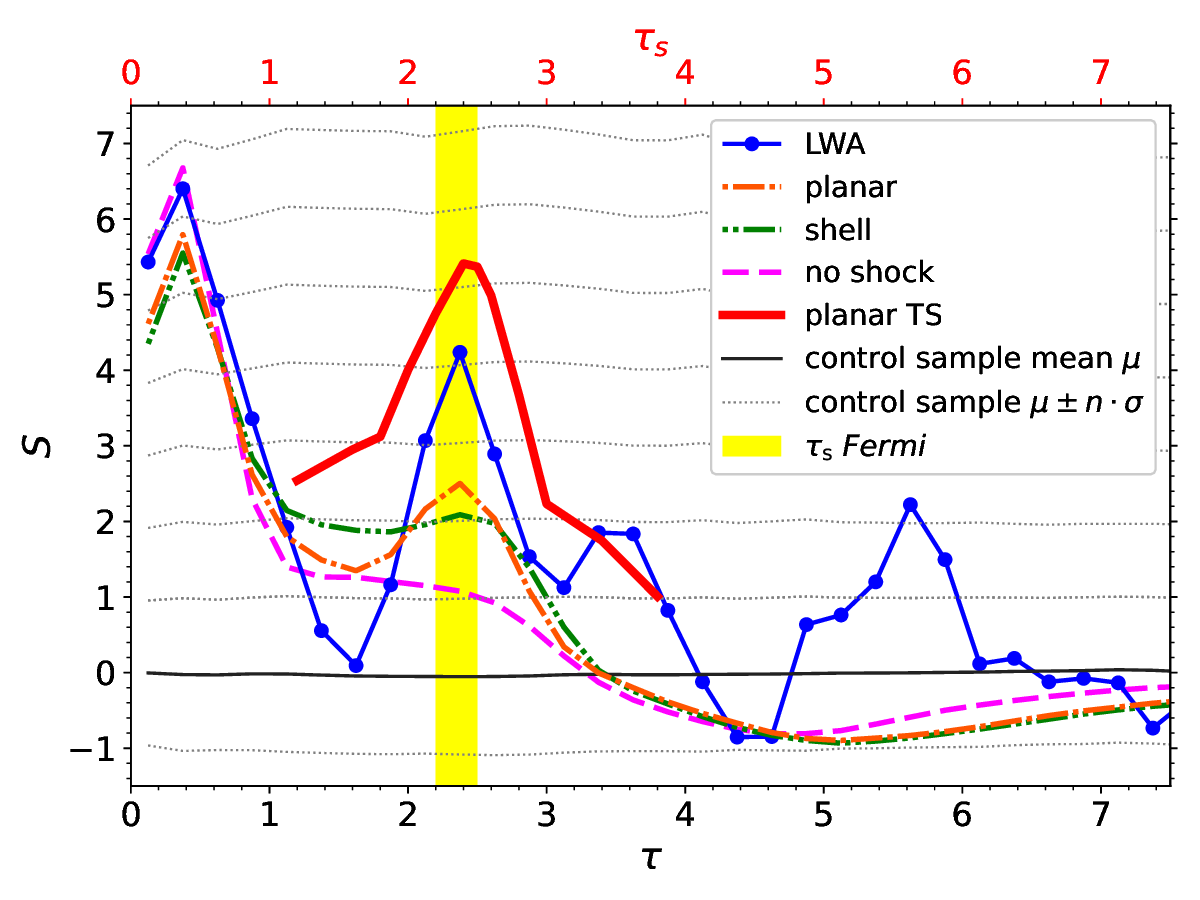}
    \caption{
    Modelling the high frequency
    data as in Fig.~\ref{fig:model} (with the same notations), but using the least-squares estimate (\ref{eq:chi2_0}). The TS-based significance profile (thick solid red curve) of the nominal planar model is shown as a function of $\tau_s$.
    }
    \label{fig:band7_model_TS}
\end{bfigure}

Like the planar model, the shell model is not sensitive to the choice of $\tau_{\rm fit}$, but only when using the standard least-squares Eq.~(\ref{eq:chi2_0}).
In this case, we find both $\tau_s = 2.70^{+0.17}_{-0.14}$
and  $\dot{m}\xi_e\xi_B = 1.85^{+0.31}_{-0.31} \times 10^{-4}$ somewhat larger
than their planar case counterparts in this method, with a high-significance, $\chi^2_n (\tau \leq 5) \simeq 1.2$ and $\TS\simeq41.1$ ($6.1\sigma$ for $\DF=2$; no priors used henceforth) detection of the virial shock.
The nominal Eq.~(\ref{eq:chi2}) is more sensitive to confusion with the $\tau\sim3.5$ artefact, so using $\tau_{\rm fit}\gtrsim5$ leads to an exaggerated $\tau_s\simeq 4$.
For $\tau_{\rm fit}=3$, the best fit is $\tau_s = 2.83^{+0.26}_{-0.22}$ and $\dot{m}\xi_e\xi_B = 2.55^{+0.88}_{-0.81}$, consistent with the above but with a weaker,
$\chi^2_n(\tau\leq3) \simeq 1.9$ and  $\TS\simeq 7.6$ ($2.3\sigma$ for $\DF=2$) detection due to the smaller radial range used.

We conclude that the virial shock signal is identified at a high confidence level in the high-frequency channel.
Both planar and shell models give good fits to the data, and with similar normalisations $\dot{m}\xi_e\xi_B$.
While the two models reproduce the same peak locations in flux, in the shell model this peak radius is smaller than the shock radius, due to the combined effect of projection and PSF convolution.
Note that in the high-frequency channel, neither model reproduces the deep local minimum at $\tau\simeq 1.5$.

\subsubsection{Low-frequency channels}
\label{subsec:model_low_nu}

The low-frequency model is demonstrated in the bottom row of Fig.~\ref{fig:model} and in Table \ref{tab:fitting_virial} (and in more detail in Table \ref{tab:fitting_virial_detail}).
The fitted models show similar behaviour to that found in the high-frequency channel, with comparable parameters and a similarly high-confidence detection of the virial shock.

The nominal {\planar} model gives
\begin{equation}
    \tau_s = 2.63^{+0.09}_{-0.10}~(2.62^{+0.10}_{-0.10})
    \label{eq:best_tau_s_lowf}
\end{equation}
and
\begin{equation}
\dot{m}\xi_e\xi_B=1.95^{+0.36}_{-0.35}~(1.83^{+0.36}_{-0.36}) \times 10^{-4}
    \label{eq:best_xi_lowf}
\end{equation}
for beam (cluster, henceforth omitted) co-addition.
The $\MyChiCorr\chi_n^2(\tau \le 5) \simeq 0.3$ fit yields $\TS\simeq30.5$ ($5.1\sigma$ for $\DF=2$).
Again, replacing Eq.~(\ref{eq:chi2}) by the standard Eq.~(\ref{eq:chi2_0}) leaves $\tau_s$ unchanged, halves $\dot{m}\xi_e\xi_B = 0.84^{+0.15}_{-0.13} \times 10^{-4}$, and increases the significance to $\TS\simeq37.5$ ($5.8\sigma$ for $\DF=2$).

The shell model yield $\tau_s = 2.99^{+0.13}_{-0.13}$ and $\dot{m}\xi_e\xi_B = 2.31^{+0.49}_{-0.47} \times 10^{-4}$
with $\MyChiCorr\chi^2_n \simeq 0.3$ and $\TS\simeq24.0$ ($4.5\sigma$ for $\DF=2$)
when using the nominal Eq.~(\ref{eq:chi2}).
Like the high-frequency case, the shell model gives a somewhat larger $\tau_s$ compared to its {\planar} counterpart.
Using the standard Eq.~(\ref{eq:chi2_0}) instead of Eq.~(\ref{eq:chi2}) gives a slightly larger $\tau_s = 3.14^{+0.06}_{-0.09}$, a $\sim30\%$ smaller $\dot{m}\xi_e\xi_B = 1.58^{+0.19}_{-0.20} \times 10^{-4}$, and substantially raises the significance to $\TS\simeq70.0$ ($8.1\sigma$ for $\DF=2$).

We conclude that the low-frequency channels support the high-frequency behaviour, again indicating the presence of the virial shock at a very high confidence level, and with parameter estimates and confidence levels consistent with their high-frequency counterparts.
Co-adding the two virial signals, namely in the high-frequency channel and in the co-added low-frequency channels, with equal weights, the significance estimated by the standard Eq.~(\ref{eq:chi2_0}) indicates a detection of the virial shock at confidence levels reaching $\TS\simeq69.2$ ($8.0\sigma$ for $\DF=2$) for the {\planar} model and $\TS\simeq106.3$ ($10.0\sigma$ for $\DF=2$) for the shell model.

\subsection{Results for virial shock with arbitrary $p$}
\label{subsec:virial_spec}

\begin{figure}
    \centering
    \includegraphics[width=0.4\textwidth,trim={0 0.4cm 0 0},clip]{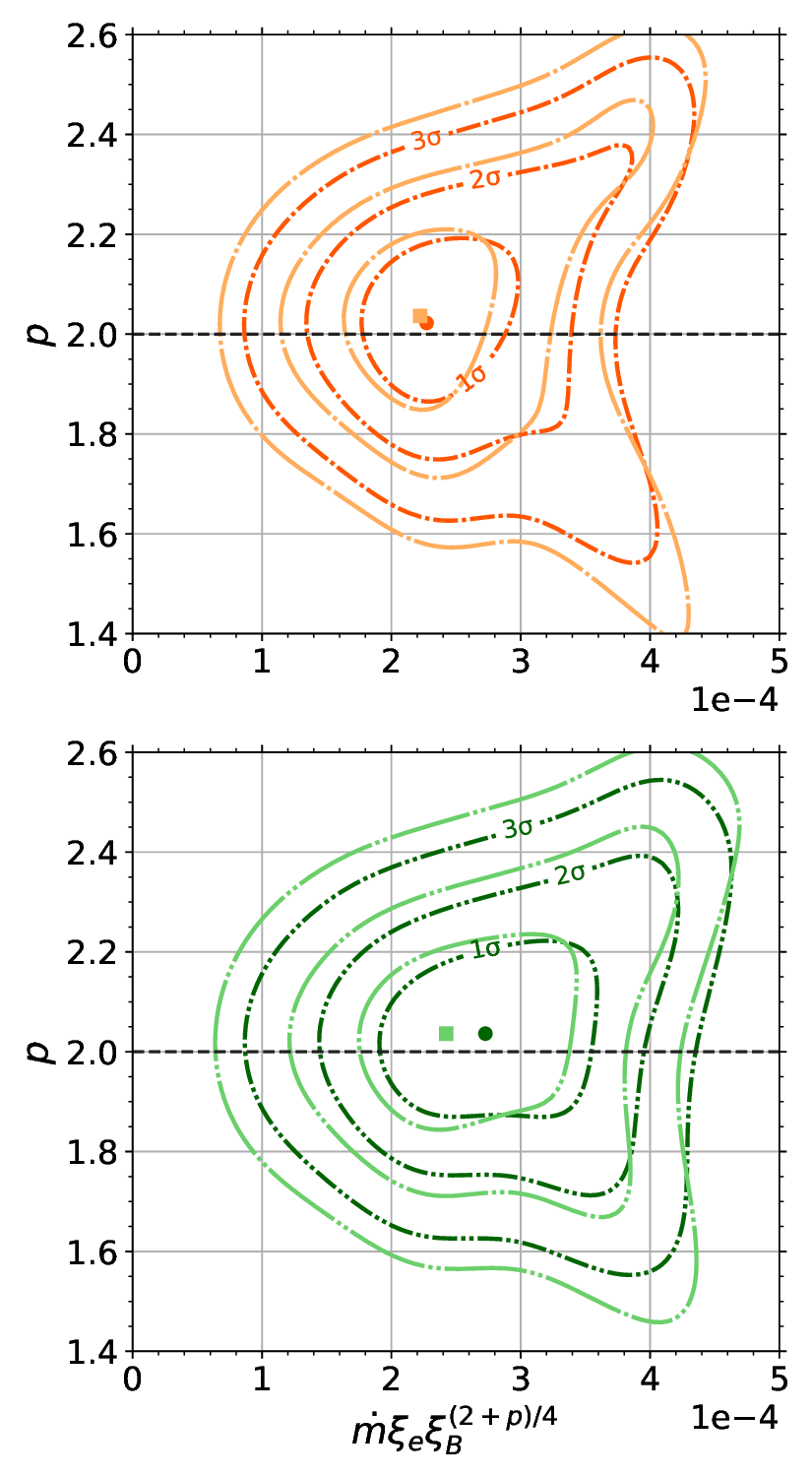}
    \caption{
    Best fit (symbols and $1\sigma$ to $3\sigma$ confidence contours) for virial shock parameters  $\dot{m}\xi_e\xi_B^{(2+p)/4}$ and $p$, for the {\planar} (top panel) and shell (bottom) models, using low-frequency \clean\ data.
    Each panel shows the contours and best-fitting parameter for nominal analysis with beam and cluster co-addition (same contour and symbol notations as in Fig.~\ref{fig:model}).
    The strong shock limit $p=2$ is also shown (dashed line).
    }
    \label{fig:spec_p}
\end{figure}

We now relax the strong shock assumption, modelling the virial signal with $p$ as a free parameter rather than fixed to its strong-shock, $p=2$ limiting value.
For this purpose, we use the cleared data in all low-frequency channels, as in \S\ref{subsec:model_low_nu}; the results are again summarised in Table \ref{tab:fitting_virial} and in Table \ref{tab:fitting_virial_detail}.
Due to the significant uncertainties surrounding the virial shock emission, the spectrum estimated through forward modelling is more accurate than the direct estimate in \S\ref{subsec:spec_lwa}.

To simplify the analysis, the central emission parameters are fixed to their best-fitting values inferred for the respective $p=2$ case, leaving only the virial shock signal parameters $p$, $\tau_s$, and the generalised normalisation $\dot{m}\xi_e\xi_B^{(2+p)/4}$.
The best-fitting shock radius again yields $\tau_s = 2.62^{+0.10}_{-0.10}$ ($2.62^{+0.11}_{-0.11}$) for the {\planar} (shell) model with nominal $\chi^2$, consistent with the results in \S\ref{subsec:model_low_nu}, so we focus on the relation between the remaining two parameters, $p$ and $\dot{m}\xi_e\xi_B^{(2+p)/4}$.

Figure \ref{fig:spec_p} shows the confidence level contours of these parameters, for the {\planar} (top panel) and shell (bottom) model.
Here, contours are produced by interpolating and smoothing the $\chi^2$ data, instead of fitting it to a polynomial, as the latter would distort the contours in this particular case.
We find the spectral index
\begin{equation}
    p = 2.02^{+0.17}_{-0.16}~(2.04^{+0.12}_{-0.11}) \coma
    \label{eq:spec_p}
\end{equation}
for the {\planar} (shell) model, with $\TS\simeq29.7$ ($4.8\sigma$ for $\DF=3$) for the {\planar} model and $\TS\simeq23.8$ ($4.2\sigma$ for $\DF=3$) for the shell model.
The best-fit spectral index is consistent with the strong shock limit $p = 2$, justifying the assumption of a flat,  $dN/dE \propto E^{-2}$ injected CRE spectrum.
Consequently, the associated normalisation $\dot{m}\xi_e\xi_B^{(2+p)/4} =
2.27^{+0.71}_{-0.50} \times 10^{-4}~(2.73^{+0.63}_{-0.59} \times 10^{-4})$ for the {\planar} (shell) model agrees with the findings in \S\ref{subsec:model_low_nu}.
By adopting the standard $\chi^2$ of  Eq.~\eqref{eq:chi2_0} instead of the nominal Eq.~\eqref{eq:chi2}, the best-fitting result for $p$ is again consistent with $2.0$, and the other parameters and confidence levels are unchanged with respect to the $p=2$ results.

The flat spectrum of the virial excess is robust, showing no sensitivity to the co-addition method or to cluster selection, nor any evidence for spectral curvature.
For example, for the planar (shell) model, the four lowest frequency, 37--52 MHz channels yield $p = 2.04^{+0.15}_{-0.15}~(2.04^{+0.10}_{-0.10})$, while the three higher frequency, 57--68 MHz channels give a similar $p = 2.04^{+0.24}_{-0.29}~(2.02^{+0.21}_{-0.22})$.
This robustness contrasts with the central-signal spectrum in our modelling, which strongly depends on assumptions, co-addition method, choice of channels, etc.
While this sensitivity may reflect our inaccurate modelling of the cluster centre, it may well arise from a spatially-dependent spectral curvature of the combined point-like and extended components of the central excess.
In such a case, the robustness of the virial excess spectrum provides further evidence that it does not arise from putative sidelobe artefacts.

\subsection{Mass dependence}
\label{subsec:mass_depen}

The synchrotron brightness of the virial shock is expected to be a strong, $I_\nu\propto M^{5/3}$ function of the cluster mass; see Eq.~\eqref{eq:shock_prop}.
We test whether massive clusters indeed show a stronger virial excess signal by splitting our sample into two sub-samples according to cluster mass.
Choosing $M_{500} = 5 \times 10^{13} M_{\sun}$ as the approximate median gives a sub-sample of 23 more massive clusters with a median $M_{500} \simeq 1.0 \times 10^{14} \Msun$, and a sub-sample of 21 less massive clusters with a median $M_{500} \simeq 3.3 \times 10^{13} \Msun$.

\begin{figure}
    \centering
    \includegraphics[width=0.45\textwidth,trim={0 0.5cm 0 0},clip]{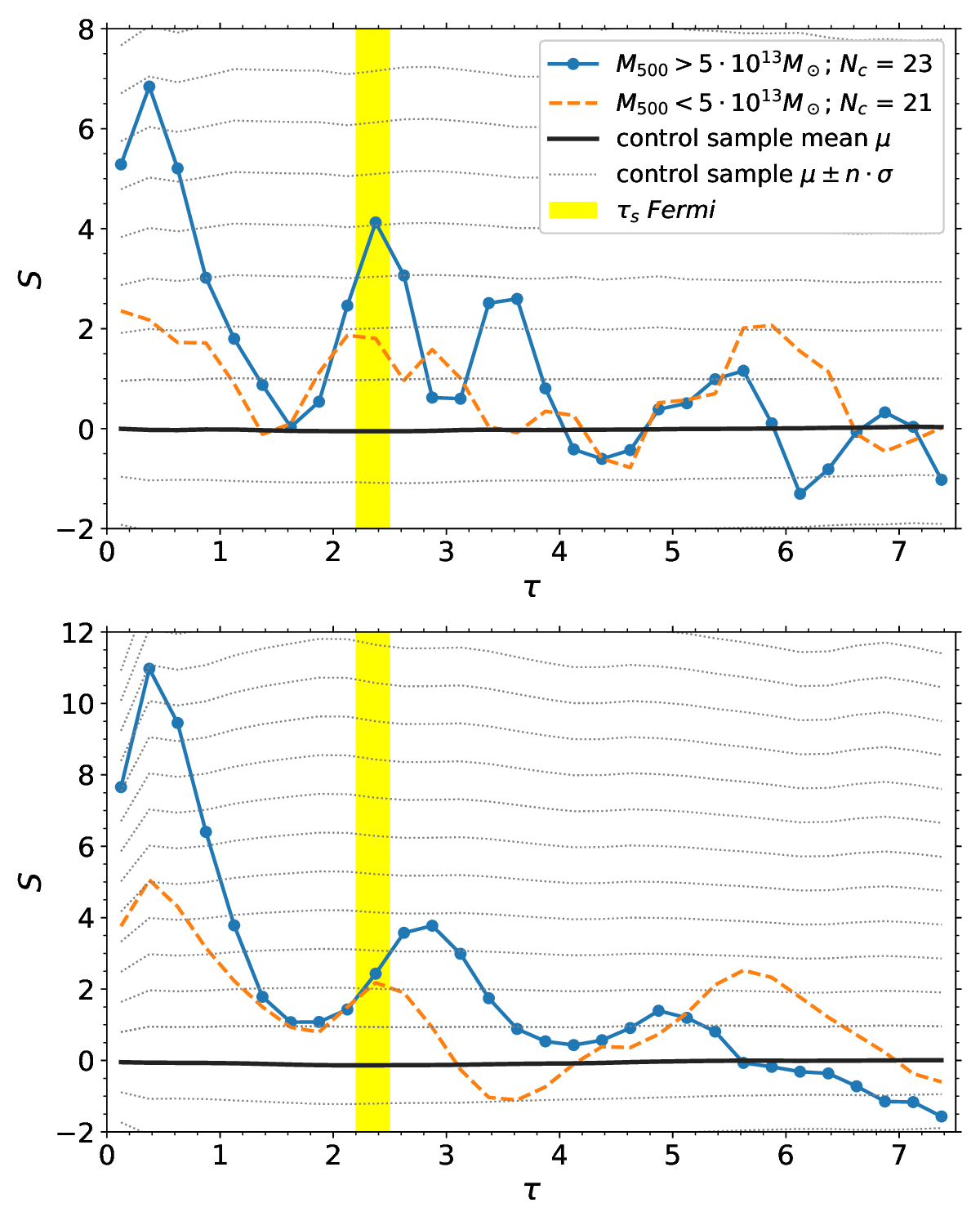}
    \caption{
    Significance of excess brightness profiles in  high-frequency (top panel) and low-frequency (bottom) cleared data, for the 23 high mass ($M_{500} > 5 \times 10^{13} \Msun$; solid blue curves) and 21 low mass ($M_{500} < 5 \times 10^{13} \Msun$; dashed orange) clusters.
    Other notations are as in Fig.~\ref{fig:band7_sig}.
    }
    \label{fig:mass_test}
\end{figure}

We stack the two sub-samples in the same method as in \S\ref{sec:StackedQuantities}, using nominal parameters. Figure \ref{fig:mass_test} shows the significance of the excess brightness of the two sub-samples, using both high (top panel) and low  (bottom) frequency cleared data with beam co-addition.

In both high and low frequencies, we find a significant, $\sim 4\sigma$ excess in the stacked massive clusters, but only a $\sim2\sigma$ excess in the stacked low mass clusters, in
qualitative agreement with the expected mass dependence. The high-mass excess is of comparable significance to that of the full sample, despite having only half the clusters.
Due to the smaller size of the sub-samples, some spurious signals (such as a  $\sim2.5\sigma$ excess around $\tau \simeq 3.5$ in the high-frequency channel) become stronger than in the full sample.

The estimated virial excess brightness per cluster in the high-frequency channel (stacked low-frequency channels) is $\Delta I \simeq 6.7 \times 10^{-12}  \erg\se^{-1}\cm^{-2}\sr^{-1}$ ($3.6 \times 10^{-12} \erg\se^{-1}\cm^{-2}\sr^{-1}$) in the massive-cluster sub-sample, which is higher by a factor of $1.3$ with respect to the full sample, and by a factor of $\sim 2.0 ~(1.5)$ with respect to the low-mass sub-sample. 
These results are again consistent with a strong mass-dependence of the signal. 
For the low-mass sub-sample, the significance of the signal is too low for a forward modelling that could reliably test the $M^{5/3}$ dependence.

Fitting the central and virial models to the massive sub-sample alone yields results consistent with those derived for the full sample. The inferred spectral indices and shock position do not change appreciably, while the central flux $F_{73}$ increases
by $\lesssim75\%$ ($\lesssim45\%$) for the restored (clear) data of the massive sub-sample.
The best-fitting virial shock normalisation $\dot{m}\xi_e\xi_B^{(2+p)/4}$ in the massive sub-sample is the same as in the full sample when using the standard $\chi^2$ of Eq.~\eqref{eq:chi2_0}, but smaller by $\sim30\%$ ($\sim20\%$) for the {\planar} (shell) model when adopting the nominal $\chi^2$ of Eq.~\eqref{eq:chi2}, as expected from the dispersion among clusters and the partial correlation with mass.
In conclusion, the results are consistent with massive clusters dominating the signal, but a larger sample is needed to quantify the mass dependence of the model parameters.

\begin{table*}
    \centering
    \caption{Magnetic field estimates for select galaxy clusters.
    }
\begin{adjustbox}{width=1\textwidth}
\begin{tabular}{lcccccc|ccc|ccc}
Name & $\beta$ & $n_{0}$ & $r_c$ & $k_{\rm B}T$ & $z$ & $R_{500}$ & $\dot{m}\xi_e$ & $\xi_B$ & $B$ & $\dot{m}$ & $\xi_B'$ & $B'$ \\
 &  & ($10^{-3}$ cm$^{-3}$) & (kpc) & (keV) &  & (Mpc) & (\%) & (\%) & (${\rm \mu G}$) &  & (\%) & (${\rm \mu G}$) \\
(1) & (2) & (3) & (4) & (5) & (6) & (7) & (8) & (9) & (10) & (11) & (12) & (13) \\ \hline
A 376    & 0.57 &  5.85 & 119.1 & 3.69 & 0.05 & 0.81 & 0.6 & 4.8 $\pm$ 1.1 & 0.24 $\pm$ 0.03 & 0.9 $\pm$ 0.1 & 1.0 $\pm$ 0.1 & 0.10 $\pm$ 0.00 \\
A 576    & 0.57 &  7.61 & 105.6 & 3.67 & 0.04 & 0.83 & & & 0.24 $\pm$ 0.03 & 0.9 $\pm$ 0.1 & 1.0 $\pm$ 0.1 & 0.10 $\pm$ 0.00 \\
 A 779    & 0.34 & 10.35 &  11.5 & 1.76 & 0.02 & 0.45 & & & 0.24 $\pm$ 0.03 & 2.5 $\pm$ 0.2 & 0.6 $\pm$ 0.1 & 0.08 $\pm$ 0.00 \\
 A 1795   & 0.63 & 29.55 &  86.4 & 5.76 & 0.06 & 1.22 & & & 0.26 $\pm$ 0.03 & 0.7 $\pm$ 0.0 & 1.1 $\pm$ 0.1 & 0.12 $\pm$ 0.01 \\
 A 2065   & 0.57 & 12.11 & 113.6 & 5.15 & 0.07 & 1.05 & & & 0.31 $\pm$ 0.04 & 1.1 $\pm$ 0.1 & 0.9 $\pm$ 0.1 & 0.13 $\pm$ 0.01 \\
 A 2256   & 0.93 &  5.05 & 417.6 & 6.45 & 0.06 & 1.12 & & & 0.23 $\pm$ 0.03 & 0.3 $\pm$ 0.0 & 1.7 $\pm$ 0.2 & 0.13 $\pm$ 0.01 \\
 A 2634   & 0.38 &  4.11 &  78.1 & 3.45 & 0.03 & 0.75 & & & 0.35 $\pm$ 0.04 & 3.0 $\pm$ 0.2 & 0.5 $\pm$ 0.0 & 0.11 $\pm$ 0.01 \\
 HCG 51   & 0.34 &  4.71 &  32.8 & 1.41 & 0.03 & 0.44 & & & 0.24 $\pm$ 0.03 & 3.7 $\pm$ 0.2 & 0.5 $\pm$ 0.0 & 0.08 $\pm$ 0.00 \\
 NGC 6329 & 0.38 & 10.61 &  13.9 & 1.64 & 0.03 & 0.48 & & & 0.19 $\pm$ 0.02 & 1.7 $\pm$ 0.1 & 0.7 $\pm$ 0.1 & 0.07 $\pm$ 0.00 \\
\end{tabular}
\end{adjustbox}
    \label{tab:mag}
\begin{tablenotes}
\item
    Results shown for the seven cluster with $\beta$-models in \citet{FukazawaEtAl04} based on results (\ref{eq:best_tau_s}) and (\ref{eq:best_xi}), using either $\dot{m}\xi_e = 0.6 \%$ inferred from the $\gamma$-ray stacking (columns 8--10) or by assuming $\xi_e=\xi_B$ equipartition (columns 11-13).
\item
    {\bf Columns:} (1) Cluster name; (2) $\beta$-model index; (3) Central particle number density converted from the central electron number density in \citet{FukazawaEtAl04}.; (4) Core radius; (5) Temperature;
    (6) Redshift; (7) $R_{500}$;
    (8) $\dot{m}\xi_e = 0.6 \%$ from \RK;
    (9) Magnetisation efficiency $\xi_B$ based on $\dot{m}\xi_e = 0.6 \%$;
    (10) Magnetic field $B$ based on $\xi_B$.
    (11) Accretion parameter [using Eq.~(\ref{eq:accretion_rate1})] with $f_\beta = 1/3$, used to extrapolate the $\beta$-model to the shock position (see \append{\S\ref{append:decouple_xib}});
    (12) Magnetisation efficiency $\xi_B'$ based on $\dot{m}$ and assuming $\xi_e = \xi_B$;
    (13) Magnetic field $B'$ based on $\xi_B'$.
\end{tablenotes}
\end{table*}

\subsection{Magnetic field estimate}
\label{subsec:magnetic}

Synchrotron emission from the virial shock provides a new probe of the magnetic field in galaxy cluster peripheries.
To estimate these magnetic fields, we need to first separate $\xi_B$ from the measured quantity $\dot{m}\xi_e\xi_B^{(2+p)/4}$. One method is to adopt the $\dot{m}\xi_e \simeq 0.6 \%$ value inferred ({\RK})
by modelling the $\gamma$-ray emission from the virial shock.
A second method is to estimate the accretion parameter $\dot{m}$ from a model for the gas distribution, and then break the degeneracy between $\xi_e$ and $\xi_B$ by assuming equipartition, \ie $\xi_e$ = $\xi_B$ (see \append{\S\ref{append:decouple_xib}}).

In either method, a $\xi_B$ estimate gauges the magnetic field $B$ given a model for the gas distribution. We invoke a rudimentary $\beta=2/3$ model for all 44 clusters in our sample, but focus on nine clusters for which an isothermal $\beta$-model was derived by \citet{FukazawaEtAl04}.
Table \ref{tab:mag} provides the $\dot{m}$, $\xi_B$ and $B$ estimates for these seven clusters, obtained in both of the above methods.
The $\beta$-model and virial shock parameters are also provided in the table.
The table focuses on the nominal analysis --- the {\planar} model for the high-frequency, cleared data, with the $\tau_s$ of  Eq.~\eqref{eq:best_tau_s} and $\dot{m}\xi_e\xi_B$ of Eq.~\eqref{eq:best_xi} --- but the text provides a range of possible values covering all model variants.

In the first method, we adopt $\dot{m}\xi_e \simeq 0.6 \%$, so our estimates of $\dot{m}\xi_e\xi_B^{(2+p)/4}$ translate into magnetisation efficiencies ranging from
\begin{equation}
    \xi_B \simeq (1.6 \pm 0.2) \%
    \label{eq:xi_b}
\end{equation}
to
\begin{equation}
    \xi_B \simeq (7.7 \pm 1.0) \%
    \label{eq:xi_bb}
\end{equation}
for the entire 44 cluster sample.
Using these $\xi_B$ values to estimate the peripheral magnetic fields of the seven modelled clusters then gives, on average,
\begin{equation}
    B \simeq 0.1\text{--}0.3 ~{\rm \mu G}
    \label{eq:mean_B}
\end{equation}
(with smaller statistical uncertainties), where we assumed that density and temperature are both suppressed at $\tau_s$ by a factor $f_\beta = 1/3$ with respect to $\beta$-model estimates; see \append{\S\ref{append:decouple_xib}}.
For the rest of the clusters in the sample, we compute $B$ by assuming a $\beta=2/3$ model and hydrostatic equilibrium (see \append{\S\ref{append:decouple_xib}}).
In this method, Eqs.~\eqref{eq:xi_b} and \eqref{eq:xi_bb} correspond to a mean $B \simeq 0.2\text{--}0.6 ~{\rm \mu G}$ for all 44 clusters.

In the second method, by estimating $\dot{m}$ for the seven clusters and assuming equipartition between  $\xi_e$ and $\xi_B$, we derive an average magnetisation efficiency  $\xi_B$ in the range $0.8\%$ to $1.8\%$, with statistical uncertainties $\lesssim 0.1\%$. Such efficiencies translate into $B \simeq 0.1\text{--}0.2 ~{\rm \mu G}$, somewhat lower than in the first method, where $\xi_B$ was effectively found to be larger than $\xi_e$.
Here, invoking $\beta=2/3$ and hydrostatic equilibrium for all clusters yields $B\simeq 0.2\mbox{--}0.3\muG$.

\section{Summary and discussion}
\label{sec:summary}

We stack and radially bin the low-frequency, (36--73) MHz LWA data around 44 massive, high-latitude, extended MCXC clusters, after rescaling each cluster to its $R_{500}$ radius.
Overall, the results indicate a high-significance excess emission peaked around $\tau\equiv r/R_{500}\sim 2.5$, \ie near the virial radius and coincident with a similarly-stacked $\gamma$-ray signal previously found in \emph{Fermi}-LAT data and identified as inverse-Compton emission from virial shock-accelerated CREs (\RK).
The radius,
spectrum, and brightness of the LWA signal are consistent with the anticipated \citep{WaxmanLoeb00, KeshetEtAl03} synchrotron emission from the same virial shock CREs, as they gyrate in the post-shock magnetic fields.
The signal is also consistent with radio to \gama-ray \citep{2017ApJ...845...24K, 2018ApJ...869...53K} and SZ \citep{HurierEtAl19, 2020ApJ...895...72K} signals detected in individual clusters.
We show that the results are robust to analysis details, in particular point source removal, foreground and background modelling, and sidelobe residuals from central point sources and diffuse emission.

\subsection{Stacking}
\label{subsec:summary_stacking}

The clusters, with mass $M_{500} > 10^{13} M_{\sun}$ and
angular radii $0\dgrdot2 < \theta_{500} < 0\dgrdot4$ (constrained by resolution from below and foreground structures from above), are selected from low-foreground parts of the sky (excluding the Galactic plane, Loop-I, and the Fermi bubbles) with high, $\delta>20\dgr$ declinations
to ensure a well-behaved LWA PSF (see \S\ref{sec:Data}).
The high, 73 MHz frequency data have a substantially better angular resolution than in the seven lower-frequency, $36\text{--}68\text{ MHz}$ channels, so we analyse the high-frequency channel and the combined low-frequency channels separately.
To better understand the data and the impact of point source cleaning, we study both the cleared data (after CLEAN) and the restored data (after point sources identified by CLEAN were restored with an effective, Gaussian beam).

The data are stacked among rescaled clusters, and binned radially as a function of $\tau$.
The stacking is carried out in two different methods (\S\ref{sec:analysis}), giving equal weights to either flux (beam co-addition) or clusters (cluster co-addition), in order to better resolve the origin of the excess.
Monte-Carlo simulations of control cluster samples are used to calibrate the noise level in each channel, estimate the correlations between the co-added low-frequency channels, and quantify the origins of the signals (see \S\ref{subsec:controlsamples} and \append{\S\ref{append:controlsample}}).
These control samples consist of random points, referred to as control clusters, in the relevant LWA sky, which undergo the same pipeline applied to the real MCXC sample.

The stacked data show a strong signal from the centres of clusters, and a high-significance signal from their peripheries, near the expected position of the virial shock.
This virial excess is strong enough to be seen in the stacked image without radial binning, at least when four-folded (Fig.~\ref{fig:2dsig}).
At low, $\Delta\tau=0.5$ radial resolution,
both high- and low-frequency
results show a strong central emission at $\tau\lesssim 1$, and a peripheral, $\sim 5\sigma$ excess peaked in the $2.25 \leq \tau < 2.75$ bin.
In our nominal, higher, $\Delta\tau=0.25$ resolution, the excess peak shifts somewhat between channels: a $\gtrsim 4\sigma$ signal in the $2.25 \leq \tau < 2.5$ bin for the high-frequency data (referred to as the nominal virial excess; see Fig.~\ref{fig:band7_sig} and \S\ref{subsec:73MHz}), and a $\simeq 3.8\sigma$ signal in the adjacent, $2.5 \leq \tau < 2.75$ bin for the co-added low-frequency channels (Fig.~\ref{fig:band0-6_sig} and \S\ref{subsec:7lowerbands}).
This small shift in peak position is likely caused by the lower spatial resolution of the low-frequency channels, broadening the excess in each cluster and thus shifting the weighted average outwards.
Testing a finer, $\Delta\tau\simeq0.167$ resolution indicates that the shift is indeed small,
with peaks found at the $2.33\leq\tau<2.5$ bin in the high-frequency channel and at $2.5\leq\tau<2.67$ in the low-frequency data (see
\S\ref{append:sensi}).

We use several independent methods to verify that the virial excess is not an artefact originating from beam-sidelobe residuals of the strong central emission missed by CLEAN (as outlined in \S\ref{subsec:sidelobe}). These tests rule out such an artefact, because:
(i) a similar virial excess is found both in clusters that show a central signal and in clusters that do not (Fig.~\ref{fig:band7_weak_strong});
(ii) convolving the best-fitting models for the raw central emission with the PSF yields sidelobe residuals at cluster peripheries which are insignificant or inconsistent with the observed signal (Fig.~\ref{fig:fake_center_restored});
(iii) similar virial-ring radii are inferred in multiple channels, whereas the respective PSF patterns and resulting sidelobe positions vary substantially  (\S\ref{subsec:7lowerbands});
(iv) clusters
with  bright central NVSS sources can be included in, or excluded from, the sample, with a major change in stacked central signal but no significant effect on the virial excess (Fig.~\ref{fig:43c_test});
(v) multiple attempts (see \append{\S\ref{append:psf}}) to artificially mimic the virial ring by stacking non-cluster sources or bright regions in the sky, all fail to produce a significant signal near the virial radius (Fig.~\ref{fig:local_max});
(vi) The robust power-law spectrum of the virial excess is unlikely to emerge from sidelobe residuals of the raw central excess, which varies among clusters and with radius and is not a single power-law (compare \S\ref{subsec:virial_spec} and \S\ref{append:centre_restored});
and
(vii) the virial excess agrees with predictions for the synchrotron emission from virial shock CREs, showing a radius consistent with the similarly-stacked $\gamma$-ray signal (\eg Figs.~\ref{fig:band7_sig} and \ref{fig:bins}) and a normalisation consistent with a previous Coma signal \citep{2017ApJ...845...24K}, not far from equipartition (see discussion below).

The synchrotron emission from the virial shock should be a strong, $I_\nu \propto M^{5/3}$ function of the cluster mass (see \append{\S\ref{append:model}}).
By splitting the cluster sample into two sub-samples of comparable sizes according to mass, with $M_{500} = 5 \times 10^{13} \Msun$ as the approximate median, we find a $\sim4\sigma$ virial excess in the massive sub-sample and a weaker, $\sim2\sigma$ in the low-mass sub-sample.
The massive sub-sample dominates the signal, with model parameters and a confidence level comparable to those of the full sample, in spite of having only half the number of clusters.
These results qualitatively agree with the expected mass-dependent virial signal and support our virial shock model.
Note that inverse-Compton emission is expected to show a weaker, linear dependence on $M$; indeed, no significant correlation was found by {\RK} between \gama-ray emission and cluster mass.
The small size of our sample precludes a more accurate determination of the signal dependence upon the mass (\S\ref{subsec:mass_depen}) or dynamical state (\append{\S\ref{append:sensi}}) of the cluster.

\subsection{Modeling}

We model the LWA data as a superposition of a smooth background, central emission, and a virial signal.
Different models are inspected for each of these components: different polynomials for the background (see \append{\S\ref{append:backgroundremove}}),
different combinations of point-like and extended emission for the central source (\S\ref{sec:CentralModels}), and either a projected sphere (shell model) or a ring/cylinder aligned parallel to the line of sight (planar model) for the virial excess (\S\ref{subsec:virialmodel} and \append{\S\ref{append:model}}).
Monte-Carlo simulations are used to inject different models into random parts of the LWA sky, and analyse the outcome with the same pipeline used for the real sample.
Least-squares minimisation, applied to the individual or stacked signals, is used to determine the best-fitting parameters and their uncertainties, and TS-tests are used to estimate the confidence level of incorporating each component in the model.
In our nominal analysis, all three components are modelled simultaneously.
However, the central and virial components are sufficiently separated spatially from each other to be studied independently, with only a modest bias to the model normalisation of each component.

We model the central signal as a combination of point-like and extended components
(see \S\ref{sec:centralemission}).
In the {\oridata} without point-source masking, the central excess is dominated by point-like emission with a spectral index
$\alpha_p = 1.04^{+0.05}_{-0.04}$.
In the \cleandata, the residual, diffuse and fainter signal is dominated by an extended component with index $\alpha_e = 1.08^{+0.24}_{-0.24}$,
presumably arising from stacked diffuse emission and point sources below the CLEAN threshold.
When modelling the cleared-data central emission and the virial excess simultaneously,
the normalisation of the central model decreases by $\lesssim 15\%$,
while the spectral index does not change appreciably (see \append{\S\ref{append:mutual_effects}}).

We model the virial excess as synchrotron emission, assuming that the shock deposits a fraction $\xi_e$ of the downstream thermal energy in CREs, and a fraction $\xi_B$ in magnetic fields. The model has three free parameters: the normalised shock radius $\tau_s$, the CRE energy spectral index $p$, and the brightness normalisation $\dot{m}\xi_e\xi_B^{(2+p)/4}$, the latter combining the accretion rate with the acceleration and magnetisation efficiencies.
Least-squares minimisation and TS-tests then indicate a $\sim4\sigma\mbox{--}8\sigma$ (for different channels and spatial models) virial shock contribution; see Fig.~\ref{fig:band7_model_TS}, Table \ref{tab:fitting_virial}, and Table \ref{tab:fitting_virial_detail}.
The previous detection of $\gamma$-rays ({\RK}) from similarly stacked galaxy clusters suggests that $\tau_s\sim 2.3\pm0.1$ ($2.2\lesssim \tau_s\lesssim 2.5$ accounting for systematics)
and $p\simeq 2.1\pm0.2$, whereas analyses of Coma indicate that $2.0\lesssim p\lesssim2.2$  \citep{2018ApJ...869...53K} and $\dot{m}\xi_e\xi_B\simeq 10^{-4}$ \citep{2017ApJ...845...24K}.
One could combine all LWA channels and use these previous results as priors, further boosting the confidence of the detection, as demonstrated in \S\ref{sec:virialemission}.
However, we adopt a conservative approach, generally analysing high and low-frequency channels separately, and avoiding such priors.

Combining the
seven low-frequency channels, we find that $p = 2.02^{+0.17}_{-0.16}$ (Fig.~\ref{fig:spec_p}), consistent with the inverse-Compton measurements and with the flat, $p \simeq 2$ spectrum expected in the strong shock limit.
Invoking the strong shock limit $p = 2$, as expected in virial shocks, simplifies the analysis, leaving only two free parameters in the virial shock model.
Applying the standard $\chi^2$ minimisation and TS-test to the full, \ie not co-added, high (low) frequency data, the {\planar} model shows a $\sim5.4\sigma$ ($\sim5.8\sigma$) shock contribution, with $\tau_s = 2.43^{+0.08}_{-0.09}$ ($2.56^{+0.13}_{-0.13}$) and $\dot{m}\xi_e\xi_B = 1.45^{+0.25}_{-0.25} \times 10^{-4}$ ($0.84^{+0.15}_{-0.13} \times 10^{-4}$).
The shell model yields an even more significant, $\sim6.1\sigma$ ($\sim8.1\sigma$) detection, for a somewhat larger and brighter virial ring with $\tau_s = 2.72^{+0.13}_{-0.13}$ ($3.14^{+0.06}_{-0.09}$) and $\dot{m}\xi_e\xi_B = 1.77^{+0.30}_{-0.27} \times 10^{-4}$ ($1.58^{+0.19}_{-0.20} \times 10^{-4}$).

There is a considerable overlap between the parameter estimates in the high and low-frequency data; combining the two regimes (with equal weight, say) indicates a joint significance reaching $\sim8.0\sigma$ for the {\planar} model and $10.0\sigma$ for the shell model.
As our nominal method, we adopt a more conservative approach, in which the $\chi^2$ analysis is applied after co-adding clusters, thus replacing the standard $\chi^2$ metric \eqref{eq:chi2_0} by its stacked counterpart \eqref{eq:chi2}.
This method yields somewhat lower confidence levels (for example, a $\sim3.9\sigma$ instead of $\sim 5.4\sigma$ {\planar} model contribution in the high-frequency channel), but gives a better visual agreement with the stacked profiles (\S\ref{subsec:virial_p2}).
The results of both
methods are shown in Fig.~\ref{fig:model}, and provided in Table \ref{tab:fitting_virial} and Table \ref{tab:fitting_virial_detail}.

The best-fitting parameters of the central and virial models show in general modest variations as a function of the frequency range, the model variant, and the analysis details, within their expected uncertainties.
For example, estimates of the central and virial normalisation parameters affect each other somewhat, due to the partial spatial overlap between the two signals, but this effect is limited to $\lesssim15\%$ variations in $F_{73}$ and $\lesssim25\%$ variations in $\dot{m}\xi_e\xi_B^{(2+p)/4}$ (see \append{\S\ref{append:mutual_effects}}).
However, the parameter $\dot{m}\xi_e\xi_B^{(2+p)/4}$ incurs a substantial, factor $\sim 4$ systematic uncertainty, as it depends on model assumptions and on analysis details.  This systematic uncertainty exceeds the factor $\sim2$ uncertainty previously attributed to the \gama-ray normalisation parameter $\dot{m}\xi_e$, due to the additional dependence on the magnetic field.
In particular, the shell model
gives $\dot{m}\xi_e\xi_B^{(2+p)/4}$ estimates $\sim20\%$--$90\%$ higher than its planar counterpart at low frequencies, as it distributes the virial signal over a wider $\tau$ range; in the high frequency channel, the shell and planar estimates agree within
$\sim20\%$.
Applying the $\chi^2$ analysis after, rather than before, co-adding clusters, raises the
$\dot{m}\xi_e\xi_B^{(2+p)/4}$ estimates by a factor 0.5--1.3, probably due to the dispersion of the virial excess strength among individual clusters, combined with a limited correlation between virial emission and cluster mass.

\subsection{Discussion}
\label{subsec:summary_discussion}

Sensitivity and consistency tests are applied to each stage of the analysis, as demonstrated throughout the text and outlined in
\append{\S\ref{append:backgroundremove}}
and \S\ref{append:sensi}.
These tests include varying the parameters controlling cluster selection, data analysis, and stacking procedures.
Cluster selection criteria tests include varying
the range of angular sizes $\theta_{500}$,
the range of masses $M_{500}$ (see \S\ref{subsec:mass_depen}),
the galactic latitude cut $b$, and the declination $\delta$.
Data analysis tests include Monte-Carlo simulations with control samples (\S\ref{subsec:controlsamples}), variations of the flux cutoff and masking radius of NVSS point sources, and tests of the background model, in particular variations of the order $N_b$ of the polynomial fit and the fitted angular radius $\theta_b$.
Stacking procedure tests include testing beam vs. cluster co-addition (\S\ref{sec:result}), least-squares computation both before and after stacking (\S\ref{sec:centralemission} and \S\ref{sec:virialemission}), folding data 4--8 times (Fig.~\ref{fig:2dsig}) instead of radial binning, and varying the size $\Delta \tau$ and offset of the radial binning.
Overall, the results are found to be quite robust.

Nevertheless, while the general properties of the virial excess are robust, its precise stacked properties, inferred significance, and surrounding features are affected somewhat by the small size of our nominal, 44 cluster sample.
In particular, as the projected signal of a virial shock is not expected to be exactly circular, some information and flux are lost when randomly rotated clusters are stacked or when the data are radially binned.
Consequently, the peak of the virial excess shifts slightly as a function of frequency and selection criteria, and the normalisation $\dot{m}\xi_e\xi_B^{(2+p)/4}$ may be somewhat underestimated.
As an example of a surrounding feature, consider the $\sim 2.5 \sigma$ extended, $\tau \simeq 5.5$ excess that persists in all channels and effectively raises the background.
This signal is robust to most of our sensitivity tests, but vanishes entirely when radially binned if the $20\dgr$ lower limit on the declination $\delta$ is either raised or lowered (see \append{\S\ref{append:sensi}}); had we not chosen $\delta>20\dgr$ as our nominal cut, this peripheral signal would vanish and the nominal significance of the virial excess would increase.
As another example, the $\tau\sim1.5$ dip, which may in the future distinguish between the shell (shallow dip) and planar (deep dip) models, is sensitive to our selection criteria.

The nominal data are consistent with both planar and shell models;
fitting the model after (before) cluster co-addition, \ie adopting Eq.~\eqref{eq:chi2} [adopting Eq.~\eqref{eq:chi2_0}],
slightly favours the planar (shell) model.
Interestingly, a similar situation was found in \gama-rays by {\RK} when stacking \emph{Fermi} data around clusters.
In principle, projection effects would naturally lead to the shell model, assuming that no preferential orientation survives the co-addition of multiple clusters.
However, the signal could still appear planar if there is a substantial quenching of the radio background inside the virial radius \citep[see for example][and references therein]{Ando2023}. Such quenching could thus deepen the $\tau\sim 1.5$ dip to the observed $\sim1.5\sigma$, if the former is comparable in magnitude to the virial shock signal.
In contrast, if magnetic fields lie preferentially in the plane of the shock, then the $\sim \sin^2\phi$ dependence upon pitch angle $\phi$ would render the shock signal somewhat less limb-bright, \ie less planar-like.
Note that in the Coma cluster, which is excluded from these stacking analyses due to its large $\theta_{500}$, the planar model provides a better fit, but the distribution of LSS around this particular cluster happens to lie preferably in the plane of the sky \citep{2018ApJ...869...53K}.

A possibly related issue is the $\sim 2.5\sigma$, extended excess peaked at $\tau\simeq 5.5$ (see Figs.~\ref{fig:band7_sig} and \ref{fig:band0-6_sig}). 
This signal passes most of our sensitivity tests (see \S\ref{subsub:sig_prof}, \S\ref{subsub:lowf_sigprof}, and \S\ref{append:sensi}), and although the radially-binned excess is not robust to changes in the declination cut, this may be due to the poor statistics, as the feature appears to survive in 2D significance maps (see Fig.~\ref{fig:2dsig} and \S\ref{append:2D_sig_maps}). 
Curiously, this excess coincides with a similarly stacked $\sim2\sigma$ excess in $\gamma$-rays (\RK) and with the $\sim2\sigma$ outer drop in SZ $y$-parameter \citep{Anbajagane2022}.
These putative signals, if confirmed, would then suggest that virial shock are non-spherical, with minimal radii around $2.4R_{500}$ and maximal radii that are a factor $2$--$3$ larger, as found in the Coma cluster \citep{2017ApJ...845...24K, 2020ApJ...895...72K}. 
In such a case, the excess brightness projected at $2.4\lesssim\tau\lesssim 6$ would effectively raise the subtracted foreground estimate, leading to a deeper $\tau\sim 1.5$ dip and to a more planar appearance of the virial excess; indeed, such an effect would be difficult to distinguish from background quenching.
Finally, note that a sidelobe origin can be ruled out for the $\tau\sim5.5$ excess, as for the virial excess, and for similar reasons. In particular, 
the $\tau\sim5.5$ excess presents in multiple channels at similar radii (Fig.~\ref{fig:band0-6_sig}), 
convolving the raw central emission with the PSF never shows similar ripples at such large $\tau$ (\S\ref{append:CentralResults}), and all our attempts fail to artificially produce such a $\tau\simeq 5.5$ excess by stacking non-cluster sources or bright sky regions (\S\ref{append:psf}).

Synchrotron emission from the virial shock provides a new probe of the magnetic field in the periphery of galaxy clusters.
With $\dot{m}\xi_e \simeq 0.6\%$ inferred form \gama-ray stacking (\RK), we derive a magnetisation efficiency ranging from
$\xi_B \simeq (1.6 \pm 0.2)\%$ to $\sim(7.7\pm1.0)\%$; see Eqs.~(\ref{eq:xi_b}) and (\ref{eq:xi_bb}).
These estimates are broadly consistent with $\xi_B \sim 1\%$, inferred in the Coma cluster by correlating VERITAS and \emph{WMAP} signals \citep{2017ApJ...845...24K}.
If we also adopt $\dot{m}=1.1$, inferred by in A2319 by combining galaxy counts and \gama-ray signals \citep{2020ApJ...895...72K}, then we find $\xi_B/\xi_e\simeq 10$, within an uncertainty factor of 3 or so.
This result (which incurs a substantial systematic uncertainty due to the combination of estimates from different clusters), would suggest more energy in magnetic fields than in CREs (although not necessarily in CRIs).
Extrapolating available isothermal $\beta$-models
for seven clusters in the sample, and correcting for the lower peripheral pressure, we estimate mean magnetic fields $\left<B\right> \simeq (0.1\text{--}0.3)~{\rm \mu G}$ inside the virial shocks; see \S\ref{subsec:magnetic} and Table \ref{tab:mag}.
Note that for such magnetic fields, the radio-emitting CREs cool over $\lesssim 0.4~{\rm Gyr}$ timescales, thus radiating over a radial range no larger than the extent of the detected virial signal, in agreement with our model.

Our results validate previous reports (see \S\ref{sec:Intro}) of inverse-Compton and SZ signals from virial shocks, as well as a preliminary correlation between WMAP and VERITAS data in Coma, all inferring similar shock parameters.
The detected level of synchrotron emission suggests an all-sky  $\nu I_\nu \simeq 10^{-11} (\xi_e \dot{m}/0.01) (\xi_B/0.01) \erg\se^{-1}\cm^{-2}\sr^{-1}$, dominating the extragalactic low-frequency radio background \citep{2017ApJ...845...24K}.
Synchrotron emission from virial shocks should surface in other present-day and future low-frequency interferometers, and --- albeit with a relatively stronger foreground --- at higher-frequency radio telescopes.
Future work, incorporating additional tracers to support the signal and break the degeneracy between $\dot{m}$ and $\xi_e$ \citep[\eg using galaxy counts, see][]{2020ApJ...895...72K}, and free of our spherically symmetric binning, could map out the cosmic-web evolution.
The $\xi_B\gtrsim1\%$ magnetic fields injected downstream of virial shocks should play an important role in the evolution of the ICM.
As CREs are consistently found to carry nearly a percent of the energy downstream of virial shocks, their CRI counterparts should hold at least $10\%$ of the energy, leading to a range of nonthermal phenomena as they permeate the ICM 
\citep{KushnirEtAl09,Keshet10}.

\vspace{0.5cm}

\section*{Acknowledgements}
We are most grateful to Michael W. Eastwood for his key contributions.
We thank I. Reiss, G. Ilani, A. Ghosh, E. Waxman, and the anonymous referee for helpful suggestions. 
This research was supported by the Israel Science Foundation (grants No. 1769/15 and 2126/22), by the IAEC-UPBC joint research foundation (grant 300/18), and by the Ministry of Science, Technology \& Space, Israel, and has received funding from the GIF (grant I-1362-303.7/2016).
The OVRO-LWA is supported by the National Science Foundation under grant number AST-1828784.

\section*{Data Availability} The data generated from computations are reported in the paper, and any additional data will be made available upon reasonable request to the corresponding author.

\bibliographystyle{mnras}

\bibliography{Virial}

\newpage
\clearpage

\renewcommand{\appendixname}{Supp.Mat.}

\appendix
\setcounter{page}{1}

\onecolumn

\begin{center}
{\Huge Supplementary Material}
\end{center}

\vspace{-1cm}

\section{Sky maps}
\label{append:sky_maps}

\begin{bfigure*}
    \begin{center}
    \DrawFig{\includegraphics[width=0.95\textwidth,trim={0 0.7cm 0 0},clip]{figure/allsky73MHz_clear.eps}}
    \DrawFig{\includegraphics[width=0.95\textwidth,trim={0 0.7cm 0 0},clip]{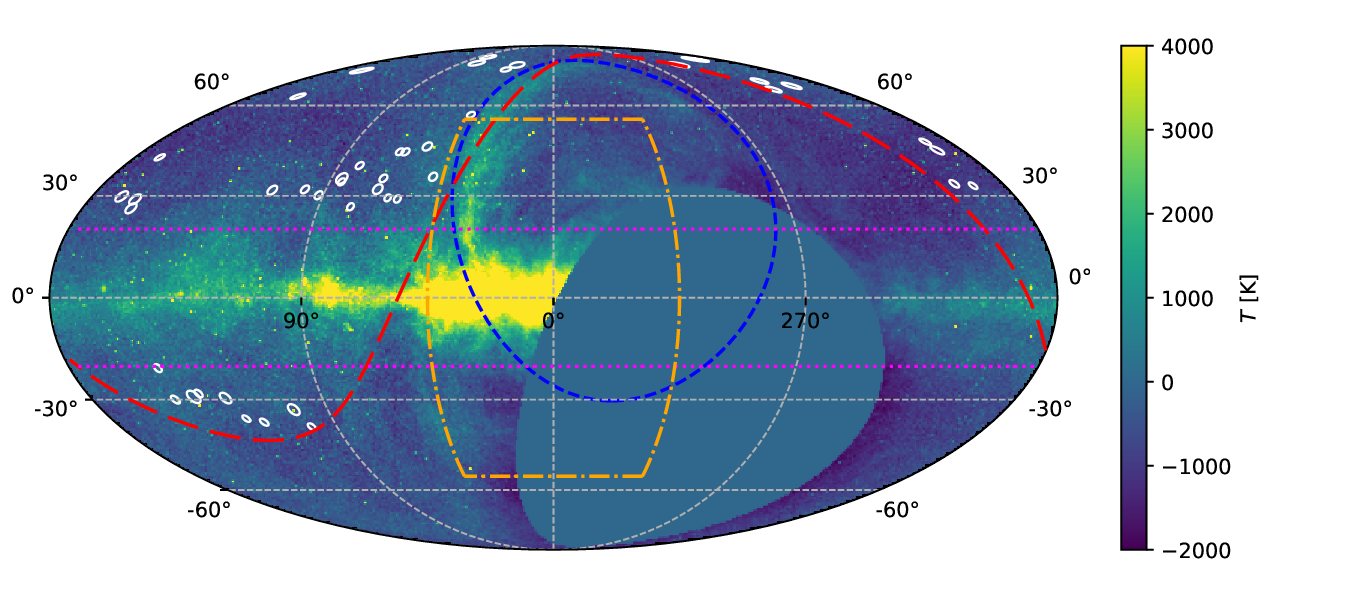}}
    \end{center}
    \caption{
	Larger version of the sky map in  Fig.~\ref{fig:allsky}, here showing both cleared (top panel) and restored (bottom) data.
    }
    \label{fig:allsky_restored}
\end{bfigure*}

Figure \ref{fig:allsky_restored} shows a larger version of the cleared-data sky map of Fig.~\ref{fig:allsky}, in comparison to restored-data map.

\section{Stacked intensity}
\label{append:intensity}

While flux co-addition involves a simple summation of the observed
$\Delta T \delta \Omega$,
stacking the intensity involves a normalisation by solid angle that can be carried out in more than one method. In analogy to the $S$ co-addition, here too we consider weighing the sum either by
beams,
\begin{equation}
\label{eq:excess_Intensity_bm}
\Delta I(\Myfr, \tau)^{(bm)} 
=
\frac{\sum_{\Myc=1}^{N_c} \sum_{\MyH=1}^{N_p(\tau,\Myc)} \Delta F(\Myfr,\Myc,\MyH)}{\delta\Omega\sum_{\Myc=1}^{N_c} N_p(\tau,\Myc) } 
=
\frac{2\nu^3k_B}{c^2}\frac{\sum_{\Myc=1}^{N_c} \sum_{\MyH=1}^{N_p(\tau,\Myc)} \Delta T(\Myfr,\Myc,\MyH)}{\sum_{\Myc=1}^{N_c} N_p(\tau,\Myc) } \coma 
\end{equation}

\noindent
or by clusters,
\begin{equation}
\label{eq:excess_Intensity_cl}
\Delta I(\Myfr, \tau)^{(cl)} 
=
\frac{1}{N_c} \sum_{\Myc=1}^{N_c} \left[
\frac{\sum_{\MyH=1}^{N_p(\tau,\Myc)} \Delta F(\Myfr,\Myc,\MyH)}{\delta\Omega\,N_p(\tau,\Myc)} \right] 
=
\frac{2\nu^3k_B}{c^2 N_c} \sum_{\Myc=1}^{N_c} \left[
\frac{\sum_{\MyH=1}^{N_p(\tau,\Myc)} \Delta T(\Myfr,\Myc,\MyH)}{N_p(\tau,\Myc)} \right] \fin 
\end{equation}

\twocolumn

The corresponding uncertainty estimates, accounting for antenna temperature noise, are then estimated as
\begin{align}
& \sigma_I(\nu,\tau)^{(bm)}
 \simeq \frac{2 \Myfr^3 k_B \eta(\nu, \tau) T_0(\nu)\pxN}{c^2 \sum_{\Myc=1}^{N_c} N_p(\tau,\Myc)}
\left[\sum_{\Myc=1}^{N_c} N_{\rm beam}(\tau,\Myc)\right]^{1/2} \nonumber \\
& \quad \quad \quad = \frac{2 \Myfr^3 k_B \eta(\nu, \tau) T_0(\nu) }{c^2}
\left[\frac{\pxN}{\sum_{\Myc=1}^{N_c} N_p(\tau,\Myc)}\right]^{1/2}
\label{eq:err_Intensity_bm}
\end{align}
and
\begin{align}
& \sigma_I(\nu,\tau)^{(cl)} =
\frac{2 \Myfr^3 k_B \eta(\nu, \tau) T_0(\nu)}{c^2 N_c}
\sqrt{\sum_{\Myc=1}^{N_c} \frac{\pxN^2 N_{\rm beam}(\tau,\Myc)}{N_p^2(\tau,\Myc)}} \nonumber \\
& \quad \simeq \frac{2 \Myfr^3 k_B \eta(\nu, \tau) T_0(\nu) \pxN^{1/2}}{c^2 N_c} \left[\sum^{N_c}_{\Myc=1}\frac{1}{N_p(\tau,\Myc)}\right]^{1/2} \fin
\label{eq:err_Intensity_cl}
\end{align}

Note that due to our choice of random variables, $\MySig^{(bm)}=\Delta I^{(bm)}/\sigma_I^{(bm)}$ applies identically, whereas $\MySig^{(cl)}=\Delta I^{(cl)}/\sigma_I^{(cl)}$ hold only under the condition
$\sum_\Myc (N_p^{-1}\sum_\MyH \Delta T)=(N_c^{-1}\sum_\Myc N_p^{-1})^{1/2}\sum_\Myc (N_p^{-1/2}\sum_\MyH \Delta T)$.

We also estimate the mean intensity, co-added over the $N_\nu$ low-frequency channels, as
\begin{equation}
\label{eq:Intensity_sum}
    \Delta I(\tau) = \frac{1}{N_\nu} \sum_{\nu=1}^{N_\nu} \Delta I(\nu, \tau)  \coma
\end{equation}
for both beam and cluster co-addition.
The associated uncertainty is estimated, for simplicity, as
\begin{equation}
    \sigma_I(\tau) = \frac{1}{N_\nu} \left[N_\nu^{\psi}\sum_{\nu=1}^{N_\nu} \sigma_I(\nu,\tau)^{2}
    \right]^{1/2} \coma
\label{eq:err_Intensity_sum}
\end{equation}
taking into account the same effective
correction for correlations between channels as used in Eq.~\eqref{eq:band_stacking}, seen to provide the correct results within the limits of full or zero correlations.
Again, due to our choice of random variables, the channel co-addition satisfies  $\MySig(\tau)=\Delta I(\tau) /\sigma_I(\tau)$ only under some conditions, in particular $\sum_\nu \Delta I=(N_\nu^{-1}\sum_\nu \sigma_I^{2})^{1/2}\sum_\nu (\Delta I/\sigma_I)$ for beam co-addition.
We confirm that the estimate Eq.~\eqref{eq:err_Intensity_sum} is consistent with our control samples.

\section{Background removal}
\label{append:backgroundremove}

\begin{figure}
    \centering
    \includegraphics[width=0.45\textwidth,trim={0 0.5cm 0 0},clip]{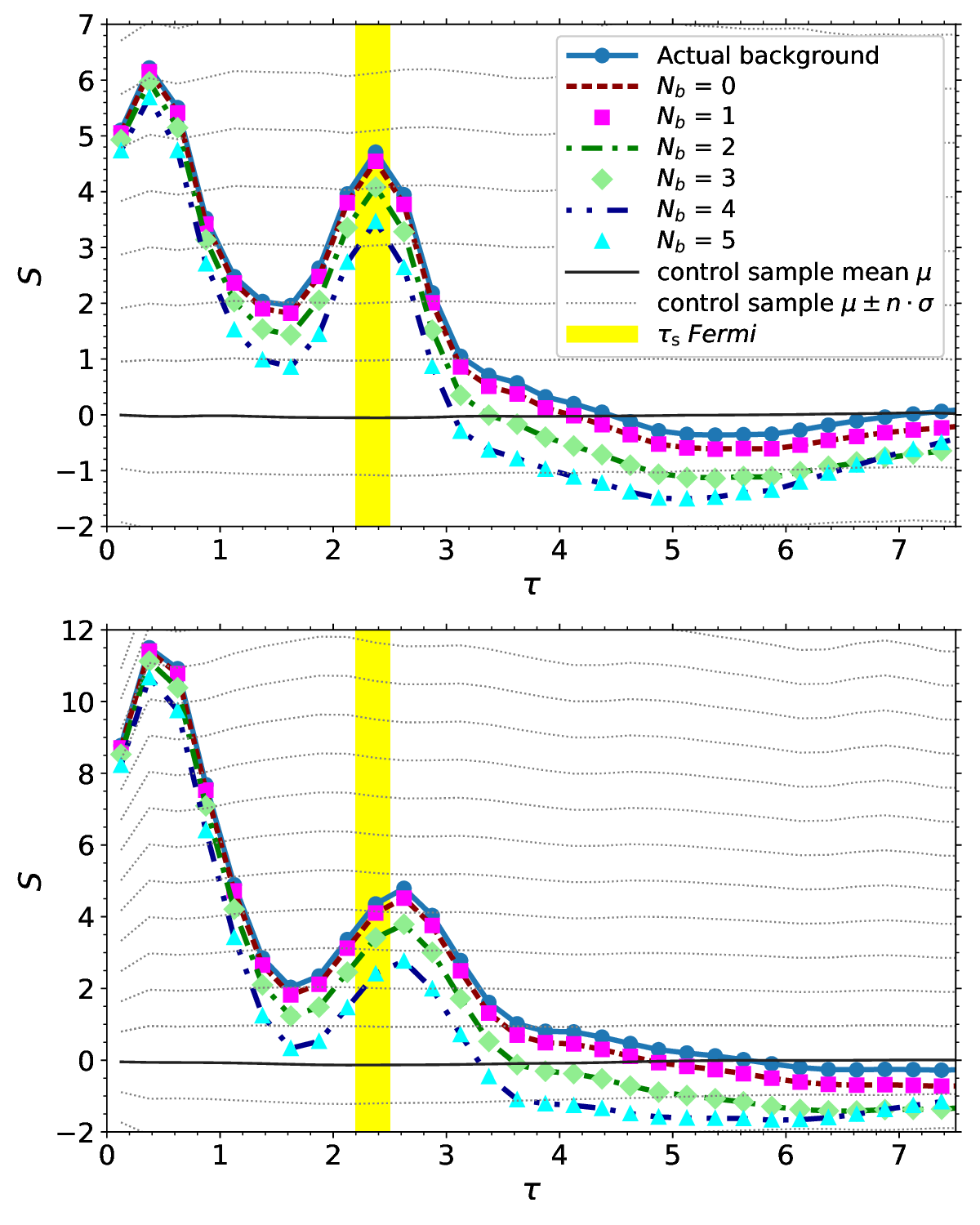}
    \caption{
    Sensitivity of the excess significance $S(\tau)$ to the order $N_b$ of the background polynomial fit according to control samples.
    Results shown for the high-frequency (top panel) and the co-added low-frequency (bottom) channels in cleared data, after injecting the nominal central and {\planar} virial model signals (best-fitting parameters in Table \ref{tab:fitting_virial}; top row for each band), fitting the result at orders (see legend) $N_b = 0$ (red dashed curve), $1$ (magenta squares), $2$ (green dot-dashed curve), $3$ (light green diamonds), $4$ (blue double dot-dashed curve), or $5$ (cyan triangles), and radial binning.
    Due to radial binning, results for any odd choice of $N_b$ (symbols) are similar, albeit not identical, to those of $N_b-1$ (curves).
    For comparison, $S(\tau)$ is shown (blue disks with solid line to guide the eye) after removing the actual background, measured before injection; equivalently, this curve represents the stacked and binned model, with no background.
    Other notations are as in Fig.~\ref{fig:band7_sig}.
    }
    \label{fig:model_bg_order}
\end{figure}

\begin{bfigure*}
    \centering
    \includegraphics[width=\textwidth,trim={0 0.5cm 0 0},clip]{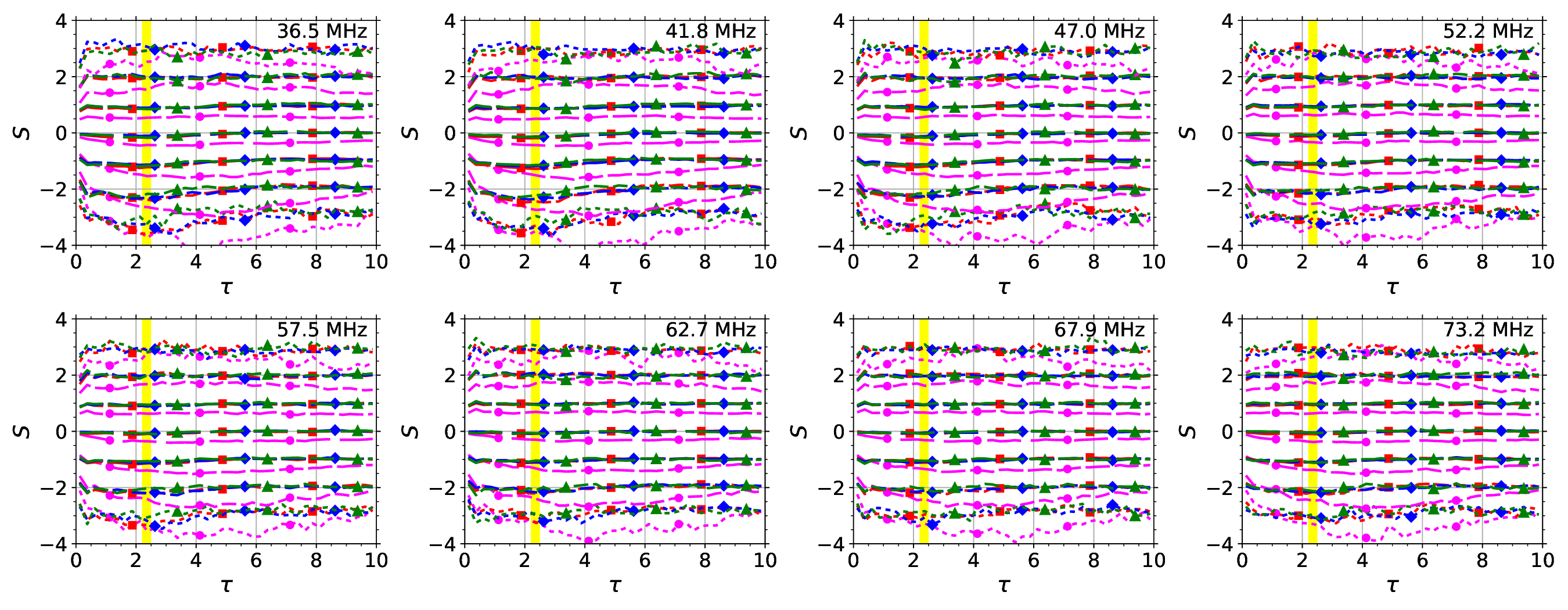}
	\addtocounter{figure}{+1}
    \caption{
    Statistics of data stacked around the control cluster sample, after removing a polynomial background of order $N_b=1$ (magenta lines with disks), $N_b=2$ (red lines with squares), $N_b=3$ (blue lines with diamonds), or $N_b=4$ (green lines with triangles), and calibrating the noise level using $\bar{\eta}$, for each frequency channel (see legend) using the \cleandata.
    Beam co-addition curves are shown with the same notations as in Fig.~\ref{fig:mock}.
    }
    \label{fig:clbgtest}
\end{bfigure*}

\begin{figure}
    \centering
    \includegraphics[width=0.45\textwidth,trim={0 0.5cm 0 0},clip]{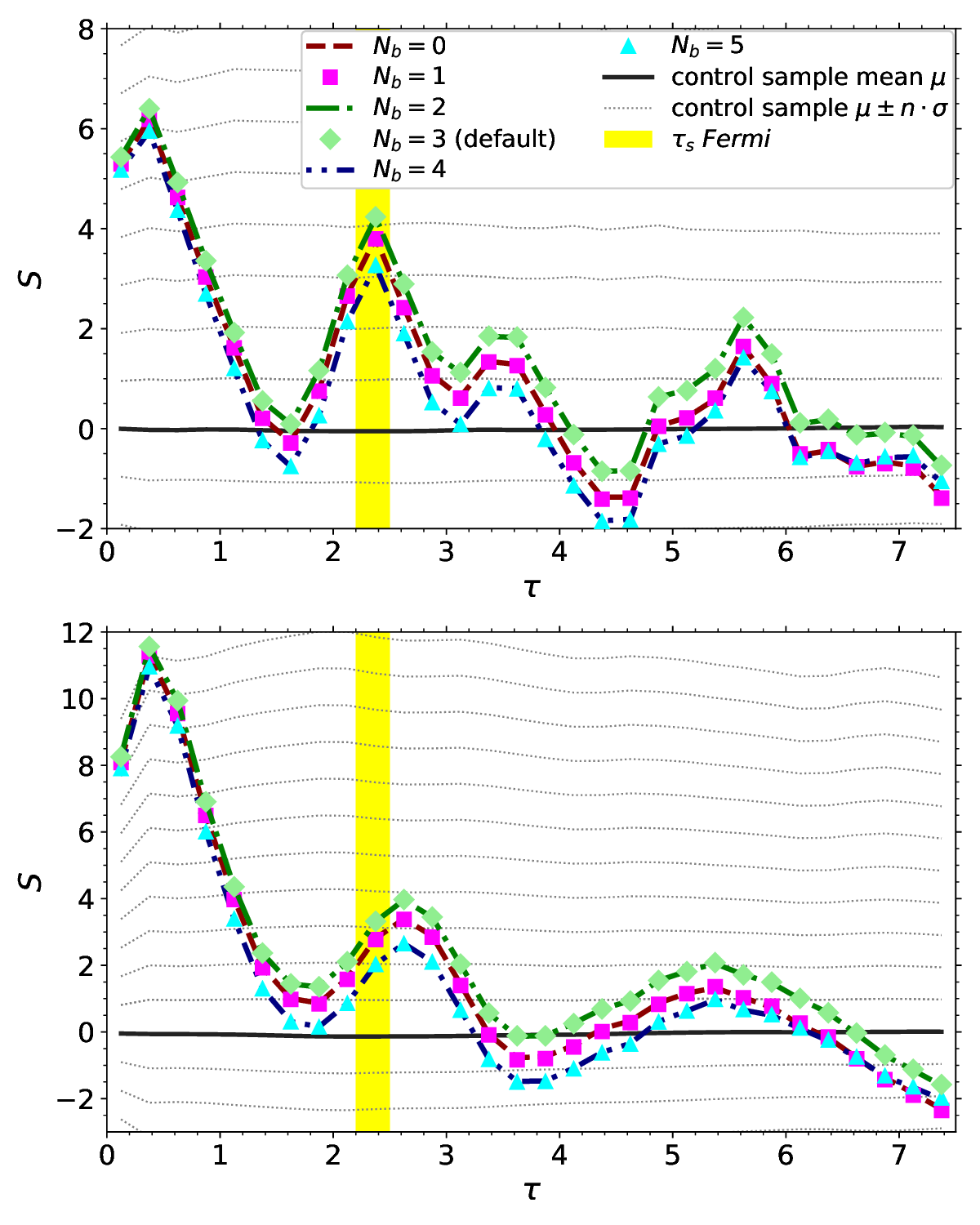}
	\addtocounter{figure}{-2}
    \caption{
    Same as Fig.~\ref{fig:model_bg_order}, but for the real MCXC sample.
    }
    \label{fig:BGorder}
\end{figure}

An accurate estimate of the background, essential in order to study the weak virial excess signal, can be obtained on the relevant scales by approximating the background as a polynomial in the two sky coordinates (see \S\ref{sec:T0}).
We test polynomials of different orders $N_b$, determined from data within different outer radii $\theta_b$, using Monte-Carlo simulations of mock clusters (see \S\ref{subsec:controlsamples}).
The sensitivity of our analysis to these two parameters is then quantified.

\begin{figure}
    \centering
    \includegraphics[width=0.45\textwidth,trim={0 0.5cm 0 0},clip]{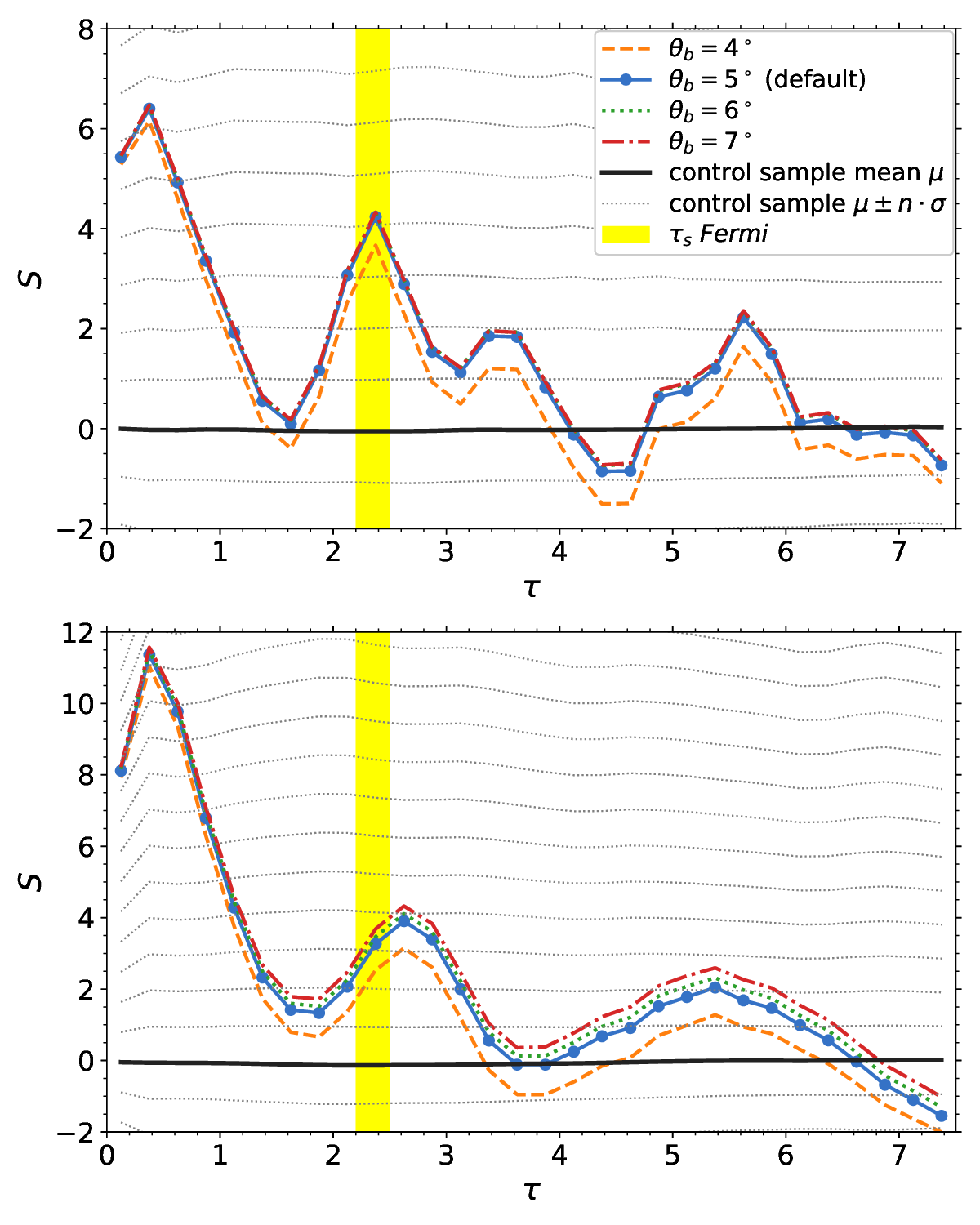}
	\addtocounter{figure}{+1}
    \caption{
    Same as Fig.~\ref{fig:BGorder}, but for variations in the outer angular radius $\theta_b$ of the fit.
    Results shown (see legend) for $\theta_b = 4\dgr$ (dashed blue curve), $5\dgr$ (solid orange), $6\dgr$ (dotted green), and $7\dgr$ (dot-dashed red).
    }
    \label{fig:theta_b}
\end{figure}

\addtocounter{section}{+1}
\begin{table*}
	\caption{
	Best fit for joint central and virial shock modelling for the cleared data.
	} 
    \centering
\hspace{-0.3cm}
\begin{adjustbox}{width=1\textwidth}
\begin{tabular}{cccccccccccccc}
Channel & Model & Weights
& $F_{73}$ & $\alpha_e$ & $\zeta$ & $\tau_{\rm cut}$ & $\tau_s$ & $\dot{m}\xi_e\xi_B^{\frac{2+p}{4}}$ & $p$ & $\chi^2$ & $\DF$ & TS ($\sigma$) & $\chi^2$ Eq. \\
(1) & (2) & (3) & (4) & (5) & (6) & (7) & (8) & $[10^{-4}]$ (9) & (10) & (11) & (12) & (13) & (14) \\ \hline
\multirow{11}{*}{high} & \multirow{7}{*}{{\planar}} & \multirow{3}{*}{bm} & $1.63^{+0.23}_{-0.20}$ & \multirow{3}{*}{—} & \textbf{0.9} & \textbf{0.9} & $2.45^{+0.11}_{-0.09}$  & $2.85^{+0.66}_{-0.65}$ & \multirow{3}{*}{\textbf{2.0}} & 17.2 & 15 & 18.5 (3.9$\sigma$) & \eqref{eq:chi2} \\
 &  &  & $1.63^{+0.19}_{-0.19}$ &  & \textbf{0.9} & \textbf{0.9} & \textbf{2.4} & $2.82^{+0.67}_{-0.67}$ &  & 17.5 & 16 & 18.2 (4.3$\sigma$) & \eqref{eq:chi2} \\
 & & & $1.54^{+0.18}_{-0.17}$ &  & \textbf{1.3} & \textbf{1.2} & $2.45^{+0.11}_{-0.11}$ & $3.09^{+0.67}_{-0.67}$ &  & 10.9 & 15 & 20.6 (4.1$\sigma$) & \eqref{eq:chi2} \\ \cline{3-14}
 &  & \multirow{3}{*}{cl} & $1.51^{+0.23}_{-0.20}$ & \multirow{3}{*}{—} & \textbf{0.9} & \textbf{0.9} & $2.44^{+0.11}_{-0.09}$  & $2.79^{+0.70}_{-0.64}$ & \multirow{3}{*}{\textbf{2.0}} & 16.9 & 15 & 17.7 (3.8$\sigma$) & \eqref{eq:chi2} \\
 &  &  & $1.51^{+0.18}_{-0.18}$ &  & \textbf{0.9} & \textbf{0.9} & \textbf{2.4} & $2.80^{+0.67}_{-0.67}$ &  & 16.9 & 16 & 17.7 (4.2$\sigma$) & \eqref{eq:chi2} \\
 &  &  & $1.44^{+0.17}_{-0.16}$ &  & \textbf{1.3} & \textbf{1.2} & $2.44^{+0.12}_{-0.12}$ & $2.99^{+0.67}_{-0.67}$ &  & 11.0 & 15 & 19.1 (4.0$\sigma$) & \eqref{eq:chi2} \\ \cline{3-14}
 & & — & $1.53^{+0.16}_{-0.16}$ & — & \textbf{1.3} & \textbf{1.2} & $2.43^{+0.08}_{-0.09}$ & $1.45^{+0.25}_{-0.25}$ & \textbf{2.0} & 1054.1 & 876 & 33.2 (5.4$\sigma$) & \eqref{eq:chi2_0} \\ \cline{2-14}
 & \multirow{4}{*}{shell} & \multirow{2}{*}{bm} & $1.73^{+0.14}_{-0.16}$ & \multirow{2}{*}{—} & \textbf{0.9} & \textbf{0.9} & $4.04^{+0.21}_{-0.23}$ & $4.88^{+1.19}_{-1.16}$ & \multirow{2}{*}{\textbf{2.0}} & 16.7 & 15 & 19.0 (4.0$\sigma$) & \eqref{eq:chi2} \\
 &  &  & $1.47^{+0.21}_{-0.20}$ &  & \textbf{1.3} & \textbf{1.2} & $2.78^{+0.37}_{-0.19}$ & $2.88^{+0.89}_{-0.85}$ &  & 20.0 & 15 & 11.4 (2.9$\sigma$) & \eqref{eq:chi2} \\ \cline{3-14}
 &  &
bm\,$^{\dagger}$
 & $1.65^{+0.23}_{-0.23}$ & — & \textbf{0.9} & \textbf{0.9} & $2.78^{+0.31}_{-0.21}$ & $2.29^{+1.17}_{-0.95}$ & \textbf{2.0} & 13.2 & 7 & 7.6 (2.3$\sigma$) & \eqref{eq:chi2} \\  \cline{3-14}
 & & — & $1.42^{+0.17}_{-0.17}$ & — & \textbf{1.3} & \textbf{1.2} & $2.72^{+0.13}_{-0.13}$ & $1.77^{+0.30}_{-0.27}$ & \textbf{2.0} & 1046.1 & 876 & 41.1 (6.1$\sigma$) & \eqref{eq:chi2_0} \\ \hline
\multirow{10}{*}{low} & \multirow{5}{*}{{\planar}} & \multirow{2}{*}{bm} & $2.01^{+0.25}_{-0.20}$ & $1.20^{+0.27}_{-0.26}$ & \textbf{1.2} & \textbf{1.6} & $2.63^{+0.09}_{-0.10}$  & $1.95^{+0.36}_{-0.35}$ & \multirow{2}{*}{\textbf{2.0}} & 137.1 & 134 & 30.5 (5.2$\sigma$) & \eqref{eq:chi2} \\
&  &  & $1.77^{+0.16}_{-0.14}$ & $1.23^{+0.29}_{-0.23}$ & \textbf{1.2} & \textbf{1.3} & $2.57^{+0.10}_{-0.10}$ & $2.03^{+0.36}_{-0.36}$ &  & 138.0 & 134 & 28.5 (5.0$\sigma$) & \eqref{eq:chi2} \\ \cline{3-14}
 &  & \multirow{2}{*}{cl} & $1.88^{+0.22}_{-0.23}$ & $1.26^{+0.25}_{-0.26}$ & \textbf{1.2} & \textbf{1.6} & $2.62^{+0.10}_{-0.10}$ & $1.83^{+0.36}_{-0.36}$ & \multirow{2}{*}{\textbf{2.0}} & 142.9 & 134 & 26.0 (4.7$\sigma$) & \eqref{eq:chi2} \\
 &  & & $1.68^{+0.14}_{-0.13}$ & $1.28^{+0.14}_{-0.20}$ & \textbf{1.2} & \textbf{1.3} & $2.57^{+0.11}_{-0.11}$ & $1.92^{+0.36}_{-0.36}$ & & 137.7 & 134 & 24.8 (4.6$\sigma$) & \eqref{eq:chi2} \\ \cline{3-14}
 & & — & $1.70^{+0.13}_{-0.13}$ & $1.22^{+0.16}_{-0.26}$ & \textbf{1.1} & \textbf{1.3} & $2.56^{+0.13}_{-0.13}$ & $0.84^{+0.15}_{-0.13}$ & \textbf{2.0} & 8674.5 & 6154 & 37.5 (5.8$\sigma$) & \eqref{eq:chi2_0} \\ \cline{2-14}
 & \multirow{3}{*}{shell} & \multirow{2}{*}{bm} & $1.95^{+0.25}_{-0.23}$ & $1.14^{+0.28}_{-0.26}$ & \textbf{1.2} & \textbf{1.6} & $2.99^{+0.13}_{-0.13}$ & $2.31^{+0.49}_{-0.47}$ & \multirow{2}{*}{\textbf{2.0}} & 162.0 & 134 & 24.0 (4.5$\sigma$) & \eqref{eq:chi2} \\
 & & & $1.91^{+0.24}_{-0.24}$ & $1.21^{+0.26}_{-0.28}$ & \textbf{1.2} & \textbf{1.3} & $2.94^{+0.14}_{-0.14}$ & $2.44^{+0.48}_{-0.48}$ & & 149.1 & 134 & 25.6 (4.7$\sigma$) & \eqref{eq:chi2} \\ \cline{3-14}
 & & — & $1.69^{+0.14}_{-0.18}$ & $1.30^{+0.17}_{-0.28}$ & \textbf{1.1} & \textbf{1.3} & $3.14^{+0.06}_{-0.09}$ & $1.58^{+0.19}_{-0.20}$ & \textbf{2.0} & 8540.0 & 6154 & 70.0 (8.1$\sigma$) & \eqref{eq:chi2_0} \\ \cline{2-14}
 & \multirow{2}{*}{\planar} & bm & \textbf{2.01} & \textbf{1.20} & \textbf{1.2} & \textbf{1.6} & $2.62^{+0.10}_{-0.10}$  & $2.27^{+0.71}_{-0.50}$ & $2.02^{+0.17}_{-0.16}$  & 137.1 & 133 & 29.7 (4.8$\sigma$) & \eqref{eq:chi2} \\ \cline{3-14}
 &  & — & \textbf{1.70} & \textbf{1.22} & \textbf{1.1} & \textbf{1.3} & $2.58^{+0.13}_{-0.13}$  & $1.01^{+0.15}_{-0.16}$ & $2.02^{+0.10}_{-0.10}$  & 8667.1 & 6153 & 39.4 (5.7$\sigma$) & \eqref{eq:chi2_0} \\ \cline{2-14}
 & \multirow{2}{*}{shell} & bm & \textbf{1.95} & \textbf{1.14} & \textbf{1.2} & \textbf{1.6} & $3.00^{+0.13}_{-0.13}$ & $2.73^{+0.63}_{-0.59}$ & $2.04^{+0.12}_{-0.11}$ & 160.0 & 133 & 23.8 (4.2$\sigma$) & \eqref{eq:chi2} \\  \cline{3-14}
 &  & — & \textbf{1.69} & \textbf{1.30} & \textbf{1.1} & \textbf{1.3} & $3.24^{+0.17}_{-0.17}$ & $1.77^{+0.28}_{-0.24}$ & $2.01^{+0.11}_{-0.11}$ & 8552.3 & 6153 & 69.5 (7.8$\sigma$) & \eqref{eq:chi2_0} \\  \hline
\end{tabular}
\end{adjustbox}
    \label{tab:fitting_virial_detail}
\begin{tablenotes}
\item
    {\bf Columns:}
    (1) The high-frequency (high) or the seven co-added low-frequency (low) channels; (2) Shock model; (3) Stacking method;
    (4) Flux of the central emission normalised to the 73MHz frequency, in units of $10^{-23}$ erg s$^{-1}$ cm$^{-2}$ Hz$^{-1}$;
    (5) Spectral index of the extended central emission;
    (6) Slope of spatial distribution for the extended component of the central emission [Eq.~(\ref{eq:central_emision_point})];
    (7) Cutoff radius for the extended component of the central emission [Eq.~(\ref{eq:central_emision_point})];
    (8) Shock radius normalised to $R_{500}$;
    (9) Normalisation, in $10^{-4}$ units;
    (10) Injected spectral index of CREs;
    (11) $\chi^2$ values of the fit (without the $\mathcal{C}$ correction; see \S\ref{subsec:Fitting_procedure});
    (12) Number of degrees of freedom;
    (13) TS value (equivalent significance);
    (14) The $\chi^2$ equation.
    The values in \textbf{boldface} are fixed parameters. The prior of shock position $\tau_s = 2.4$ is from \RK.
    In the low-frequency channels, the uncertainties in columns (4), (5), (8) and (9) are corrected using Eq.~(\ref{eq:uncertainty_corr}), and the values of TS are computed using $\mathcal{C}\chi^2$. \\
    $^{\dagger}$ --- using $\tau_{\rm fit}=3$.
\end{tablenotes}
\end{table*}

Different choices of $N_b$ are illustrated using the control samples in Fig.~\ref{fig:model_bg_order}, and using the real MCXC sample in Fig.~\ref{fig:BGorder}.
The figures show the radially-binned significance profiles $S(\tau)$ for cleared data in the high (top panel) and co-added low (bottom) frequency channels, using nominal parameters including $\theta_b=5\dgr$.
The contribution of odd-order terms in the polynomial is integrated out by the radial binning, so the results for any odd choice of $N_b$ are similar, although not identical, to those of $N_b-1$.

Figure \ref{fig:model_bg_order} illustrates the $S(\tau)$ profile obtained after injecting our nominal, best-fitting central and virial planar signals (see Table \ref{tab:fitting_virial_detail}; first row for each frequency range) to random points in the relevant LWA sky.
Namely, for each choice of $N_b$ (see legend), the data around each mock cluster is fitted after injection as a polynomial, and the fit is removed to obtain $S(\tau)$, using the same pipeline applied to the real sample.
The figure also shows (blue disks with solid lines to guide the eye) the $S(\tau)$ profile obtained after removing the real LWA background of the control sample, which was measured before injection. This profile is simply the stacked model, without any LWA data background, so no fitting is needed.

As the figure shows, for $N_b\in\{0,1\}$, the $S(\tau)$ profile is very close to this stacked model, indicating that the real background on relevant scales is fairly uniform.
For larger $N_b$, the background fit incorporates an increasing fraction of the virial excess.
Consequently, removing a background of higher $N_b$ progressively lowers $S(\tau)$.
As the figure shows, $S(\tau)$ outside the central $\tau\sim 1$ is thus diminished approximately by a constant.
The central signal is weakly affected by the choice of $N_b$ because the latter is too sharp and compact to be captured by a low-order polynomial.

In the presence of the Galactic foreground and extended foreground and background sources, it is desirable to include a curvature term in the background fit, so orders $N_b<2$ are insufficient.
Moreover, we find that while $S(\tau)$ approximately follows a normal, normalised distribution for $N_b\geq 2$, this is no longer true for $N_b\leq 1$.
Figure \ref{fig:clbgtest} shows that the $S(\tau)$ distribution for $N_b = 1$ (magenta disks and dashed curve) is not symmetric, does not match the normal confidence levels, and shows strong variations with $\tau$ throughout the $0<\tau<10$ range.

Therefore, orders $N_b=2$ and $N_b = 3$, which are very similar to each other, are the lowest orders possible for the analysis.
As increasing $N_b$ shifts more of the virial excess into the background, we designate $N_b=3$ as our nominal choice.
Comparing the blue disks and green diamonds in Fig.~\ref{fig:model_bg_order} indicates that had we known the true background, we could have reported a somewhat more significant virial excess.

Figure \ref{fig:BGorder} shows the results for the real, MCXC sample, with polynomial orders ranging from $N_b=0$ to $N_b=5$.
The results are very similar to those anticipated using the control sample.
As expected, the virial excess is found to be more significant for orders $N_b=2$ and 3 than it is for higher orders.
Repeating the least-squares procedure for different orders, we find that most model parameters are not sensitive to the choice on $N_b$.
However, the normalisation of the virial excess does diminish with an increasing $N_b$; conversely, it should be slightly higher if we could use the real background.
In particular, the nominal, best-fitting normalisation $\dot{m}\xi_e\xi_B^{(2+p)/4}$ of the virial excess at order $N_b=3$ is higher than found for $N_b\geq4$.
For example, for $N_b=4$, the nominal $\dot{m}\xi_e\xi_B^{(2+p)/4}$ is $\sim20\%$ lower than its $N_b=3$ counterpart in the result \eqref{eq:best_xi}, \ie for the {\planar} model with nominal $\chi^2$ in the high-frequency channel.

Figure \ref{fig:theta_b} shows the effect of changing the radius $\theta_b$ used to estimate the background polynomial.
We find that increasing $\theta_b$ beyond the nominal $5\dgr$ does not have a significant effect on the results.
However, a smaller, $\theta_b=4\dgr$ region
is not sufficiently large for a good handle on the background, and is strongly affected by the central excess, thus diminishing the $S(\tau\gtrsim 1)$ profile.

\addtocounter{section}{-1}
\section{Control samples and best-fit results}
\label{append:describe_controlsample}

Our nominal analysis is based on the cleared data of the high-frequency channel.
This choice combines a high resolution with the removal of point-source contamination using the modified CLEAN algorithm.
The corresponding control sample, shown in the top left panel of Fig.~\ref{fig:mock},
approximately follows a normal distribution, at least out to the accessible $\pm3\sigma$ confidence level.
Good results are obtained for these nominal data even with no calibration at all, as evident from $\bar{\eta}\sim0.9$ (see Table \ref{tab:maps_sum}) being close to unity in this channel.
The well-behaved control data support the application of the estimates
(\ref{eq:photon_add}) and (\ref{eq:cluster_add})
to the nominal, real sample.

\addtocounter{section}{+1}
\begin{table*}
	\caption{Linear regression with different corrections to correlations among low-frequency channels.
	}
    \centering
\begin{tabular}{ccccccccccccc}
Model & Weights & $F_{73}$ & $\alpha$ & $\zeta$ & $\tau_{\rm cut}$ & $\tau_s$ & $\dot{m}\xi_e\xi_B^{\frac{2+p}{4}}$  & $p$ & $\chi^2$ & $\DF$ & $\chi^2$ Eq. & $\MyChiCorr$ \\
(1) & (2) & (3) & (4) & (5) & (6) & (7) & $[10^{-4}]$(8) & (9) & (10) & (11) & (12) & (13) \\ \hline
\multirow{6}{*}{\planar} & \multirow{3}{*}{bm} & \multirow{3}{*}{\textbf{1.98}} & \multirow{3}{*}{\textbf{1.25}} & \multirow{3}{*}{\textbf{1.2}} & \multirow{3}{*}{\textbf{1.6}} & $2.61^{+0.05}_{-0.05}$  & $1.99^{+0.18}_{-0.19}$ & \multirow{3}{*}{\textbf{2.0}} & \multirow{2}{*}{128.9} & \multirow{2}{*}{134} & ({\ref{eq:chi2}}) & 1 \\
 &  &  &  &  &  & $2.61^{+0.09}_{-0.10}$ & $1.99^{+0.36}_{-0.37}$ &  &  &  & ({\ref{eq:chi2}}) & 0.25 \\
 &  &  &  &  &  & $2.63^{+0.10}_{-0.09}$ & $2.01^{+0.38}_{-0.36}$ &  & 10.9 & 14 & ({\ref{eq:chi2_2}}) & --- \\ \cline{2-13}
 & \multirow{3}{*}{cl} & \multirow{3}{*}{\textbf{1.83}} & \multirow{3}{*}{\textbf{1.29}} & \multirow{3}{*}{\textbf{1.2}} & \multirow{3}{*}{\textbf{1.6}} & $2.61^{+0.05}_{-0.06}$ & $1.85^{+0.19}_{-0.19}$ & \multirow{3}{*}{\textbf{2.0}} & \multirow{2}{*}{131.9} & \multirow{2}{*}{134} & ({\ref{eq:chi2}}) & 1 \\
 &  &  &  &  &  & $2.61^{+0.10}_{-0.11}$ & $1.85^{+0.37}_{-0.36}$ &  &  &  & ({\ref{eq:chi2}}) & 0.25 \\
 &  &  &  &  &  & $2.63^{+0.10}_{-0.10}$ & $1.91^{+0.38}_{-0.37}$ &  & 11.4 & 14 & ({\ref{eq:chi2_2}}) & --- \\ \cline{1-13}
 \multirow{6}{*}{shell} & \multirow{3}{*}{bm} & \multirow{3}{*}{\textbf{1.87}} & \multirow{3}{*}{\textbf{1.18}} & \multirow{3}{*}{\textbf{1.2}} & \multirow{3}{*}{\textbf{1.6}} & $2.97^{+0.07}_{-0.07}$ & $2.40^{+0.26}_{-0.23}$ & \multirow{3}{*}{\textbf{2.0}} & \multirow{2}{*}{150.2} & \multirow{2}{*}{134} & ({\ref{eq:chi2}}) & 1 \\
 &  &  &  &  &  & $2.97^{+0.13}_{-0.13}$ & $2.40^{+0.50}_{-0.47}$ &  &  &  & ({\ref{eq:chi2}}) & 0.25 \\
 &  &  &  &  &  & $2.99^{+0.13}_{-0.14}$ & $2.37^{+0.46}_{-0.47}$ &  & 16.6 & 14 & ({\ref{eq:chi2_2}}) & --- \\ \cline{2-13}
 & \multirow{3}{*}{cl} & \multirow{3}{*}{\textbf{1.75}} & \multirow{3}{*}{\textbf{1.23}} & \multirow{3}{*}{\textbf{1.2}} & \multirow{3}{*}{\textbf{1.6}} & $2.98^{+0.07}_{-0.08}$ & $2.22^{+0.25}_{-0.24}$ & \multirow{3}{*}{\textbf{2.0}} & \multirow{2}{*}{149.9} & \multirow{2}{*}{134} & ({\ref{eq:chi2}}) & 1 \\
 &  &  &  &  &  & $2.98^{+0.14}_{-0.15}$ & $2.22^{+0.49}_{-0.48}$ &  &  &  & ({\ref{eq:chi2}}) & 0.25 \\
 &  &  &  &  &  & $2.99^{+0.14}_{-0.15}$ & $2.19^{+0.48}_{-0.45}$ &  & 16.2 & 14 & ({\ref{eq:chi2_2}}) & --- \\ \hline
\end{tabular}
    \label{tab:fitting_virial_chi2_check}
\begin{tablenotes}
\item
    {\bf Columns:}
    (1) Shock model; (2) Stacking method;
    (3) Flux of the central emission normalised to the 73MHz frequency, in units of $10^{-23}$ erg s$^{-1}$ cm$^{-2}$ Hz$^{-1}$;
    (4) Spectral index of the extended central emission;
    (5) Slope of spatial distribution for the extended component of the central emission [Eq.~(\ref{eq:central_emision_point})];
    (6) Cutoff radius for the extended component of the central emission [Eq.~(\ref{eq:central_emision_point})];
    (7) Shock radius normalised to $R_{500}$;
    (8) Normalisation, in $10^{-4}$ units;
    (9) Injected spectral index of CREs;
    (10) $\chi^2$ values of the fit (without the $\mathcal{C}$ correction; see \S\ref{subsec:Fitting_procedure});
    (11) Number of degrees of freedom;
    (12) The equation used to compute $\chi^2$; and
    (13) The correlation correction factor $\MyChiCorr$ multiplying $\chi^2$ when using Eq.~\eqref{eq:chi2}.
    The values in \textbf{boldface} are fixed parameters, and the uncertainties are corrected using Eq.~(\ref{eq:uncertainty_corr}).
\end{tablenotes}
\end{table*}
\addtocounter{section}{-1}

As pointed out above, the restored data (bottom panels in Fig.~\ref{fig:mock}) show a somewhat skewed distribution, due to the effect of the restored point sources.
In the high-frequency channel, the distribution of $S$ is fairly independent of $\tau$ even in the restored data.
We find the
confidence intervals here to be consistent with a normalised skew normal distribution \citep[\eg][]{10.1093/biomet/63.1.201},
$S\propto e^{-x^2/2} \left[1 + {\rm erf}\left(\alpha_s x/2\right)\right]$,
where erf is the error function.
The best-fitting skew parameter $\alpha_s\simeq 0.12$ is obtained for the log-histogram in the $\tau > 0.5$ range.
One could
in principle correct the skewed distribution to a normal one.
Without such a correction, Eqs.~(\ref{eq:photon_add}) and (\ref{eq:cluster_add})  can still be used, but small, $\lesssim 10\%$ errors are introduced in the restored-data $S$.
When plotting $S$, in addition to the nominal values based on these equations, we also present
the confidence levels extrapolated from the control samples assuming a normal distribution.
Such an extrapolation provides a fairly accurate, $|\Delta S|\lesssim0.2$ description of the cleared data, but has only $|\Delta S/S|\lesssim10\%$ accuracy for the restored data due to the skewness.

As \append{\S\ref{append:controlsample}} shows, individual low-frequency channels show a similar behaviour: an approximately symmetric, $\tau$-independent, normal distribution of $S$ for the cleared data, and a somewhat asymmetric, approximately skew-normal distribution for the restored data,
with $\alpha_s$ in the range $\sim0.08$--$0.12$ for different channels.
In the lower frequencies, where the resolution generally deteriorates, some variations with $\tau$
become more apparent.
Similar results are found after the co-addition
of low-frequency channels in Eq.~\eqref{eq:band_stacking}, as shown in the right panels of Fig.~\ref{fig:mock}.
Here, after $\bar{\eta}$ is calibrated in each channel, we calibrate also the correlation parameter $\psi$, by demanding that the standard deviation of $S$ in the control samples be unity, after channel co-addition and averaging over $\tau$.
The results
show an approximately symmetric, normal and normalised distribution of $S$ for the cleared data, and a slightly skew-normal distribution in the restored data
with $\alpha_s\simeq 0.11$.

With the $\bar{\eta}$ and $\psi$ parameters calibrated above (see Table \ref{tab:maps_sum}), we use the same
method to substantiate the standard deviations of the stacked brightness $\Delta I$, in its different co-addition variants.
Namely, we test if the control samples are consistent with the stacked brightness in Eqs.~\eqref{eq:excess_Intensity_bm}, \eqref{eq:excess_Intensity_cl}, and \eqref{eq:Intensity_sum} having the uncertainties estimated in Eqs.~\eqref{eq:err_Intensity_bm}, \eqref{eq:err_Intensity_cl}, and \eqref{eq:err_Intensity_sum}, respectively, and with the same calibration.
To do so, we compare the distribution of $\Delta I$ among the control samples to a normal distribution with the standard deviation $\sigma_I$ given by the above calibrated equations. We find a good agreement, within $\lesssim2\%$, out to the accessible, $\pm3\sigma$ confidence level for the cleared data.

Table \ref{tab:fitting_virial} shows the best-fit results, based on the calibrated control samples, for our nominal models. Table \ref{tab:fitting_virial_detail} provides results for a range of model variants, demonstrating the robustness of the results.

\section{Calibrating noise and correlation factors}
\label{append:controlsample}

Although the LWA antenna noise is approximately known in each channel, a more accurate determination of the noise level and any inter-correlations between channels is necessary in order to correctly determine the significance level of any excess signal.
We use the statistical distribution of the control samples, first to calibrate the noise level in each channel, and then to estimate the correlations between channels, as shown in \S\ref{subsec:controlsamples}.

\begin{figure*}
    \centering
    \includegraphics[width=\textwidth,trim={0 0.5cm 0 0},clip]{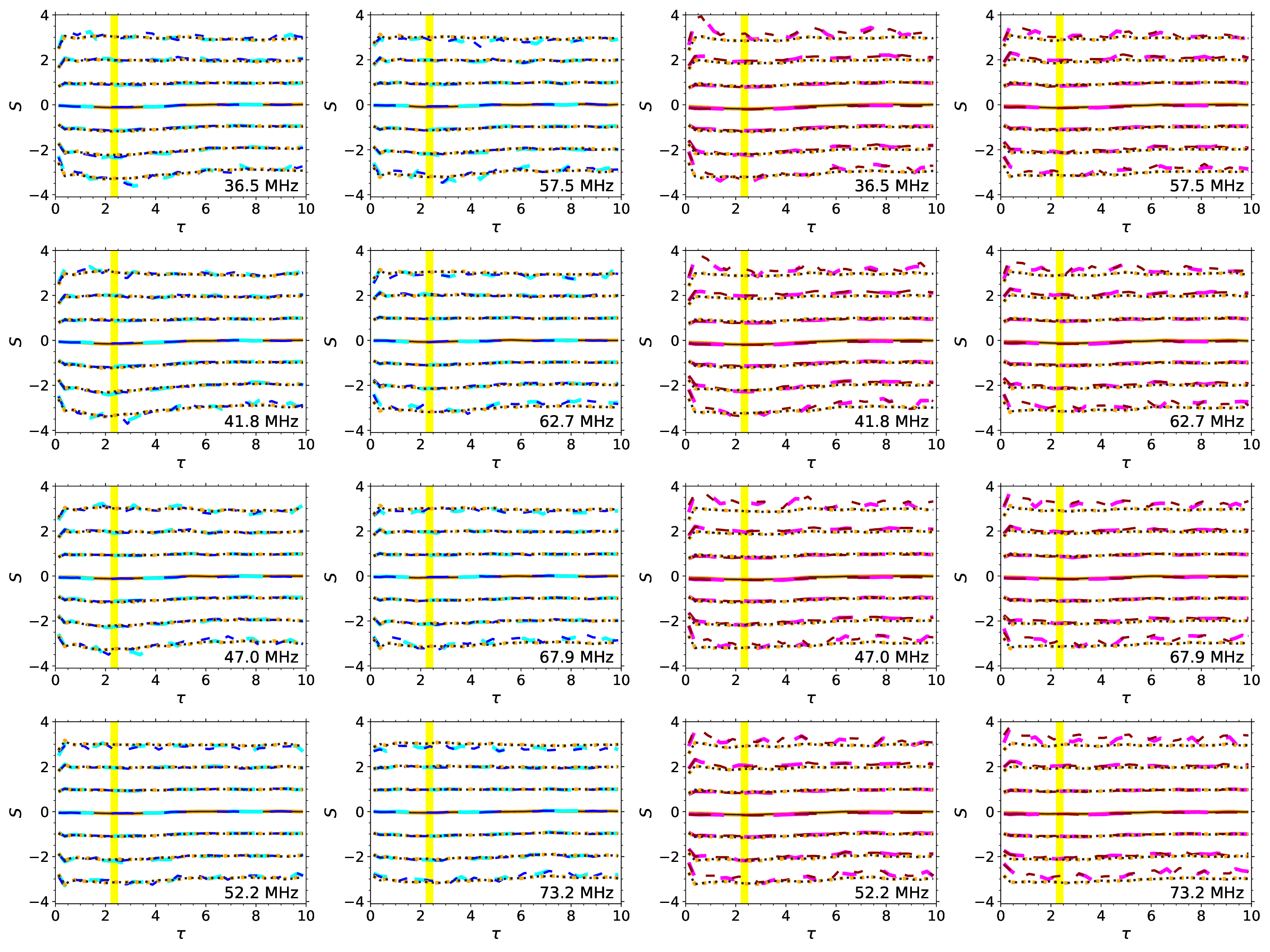}
    \caption{
    Same as Fig.~\ref{fig:mock}, for the cleared (left two columns) and restored (right two columns) data, but for each low-frequency channel separately.
    }
    \label{fig:uncorr_cl}
\end{figure*}

Figure \ref{fig:uncorr_cl} demonstrates the control sample distribution in the same manner as Fig.~\ref{fig:mock}, but for each channel separately.
The figure shows that after calibrating the noise levels, good agreement is obtained between the nominal confidence level $S$ (values on the $y$-axis), the corresponding median and $68\%$, $95\%$, and $99.7\%$ containment levels (dashed curves), and the corresponding normal-distribution significance levels (the mean offset by integer multiples of the standard deviation; dotted curves).
The figure adopts the same notations as Fig.~\ref{fig:mock}, but for simplicity, only includes beam co-addition results.

The noise calibration factors $\eta(\nu)$
are properties of the telescope, and so, should not depend on the analysis details.
Indeed, the figure shows that the factors needed to obtain the above agreement do not vary significantly with $\tau$ (except at small radii, where the statistics
is poor).
We thus adopt the $\tau$-averaged value
$\bar{\eta}(\Myfr)$
in each channel, as summarised in Table \ref{tab:maps_sum}.

As another illustration, we examine the statistical distribution of control samples after the same noise calibration, with different choices of $N_b$. Figure \ref{fig:clbgtest} demonstrates the confidence level estimates for $N_b = 1$, $2$, $3$, and $4$, for each frequency channel, using the \cleandata.
We find similar distributions of $S(\tau)$ for $2\leq N_b\leq 4$, agreeing with a normal, normalised distribution across the explored $\tau$ range, when using the same calibration factors.
For $N_b = 1$, the $S(\tau)$ distribution becomes skewed and is neither normal nor normalised; however, as mentioned in \append{\S\ref{append:backgroundremove}}, this results from the low-order fit and is unrelated to the noise calibration.
More generally, we find that the same noise calibration works for all reasonable analysis variations.

Our procedure of effectively correcting the linear regression for correlations among the low-frequency channels is described in \S\ref{sec:StackedQuantities} and \S\ref{subsec:Fitting_procedure}.
Table \ref{tab:fitting_virial_chi2_check} demonstrates how replacing $\chi^2$ by a corrected $\MyChiCorr \chi^2$ in the nominal (cluster but not channel co-added) least-squares Eq.~\eqref{eq:chi2} effectively offsets these correlations.
Namely, the table illustrates how applying $\MyChiCorr\simeq 0.25$ to Eq.~\eqref{eq:chi2} yields the same confidence intervals as those obtained with the channel co-added Eq.~\eqref{eq:chi2_2}, \ie by fitting the data after co-adding channels with the correlation correction in Eq.~(\ref{eq:band_stacking}).
Moreover, introducing this $\MyChiCorr$ correction is shown in \S\ref{subsec:Fitting_procedure} to be equivalent to a generalised Eq.~(\ref{eq:band_stacking}) with the same parameter $\psi$.
The best-fitting values are unaffected, of course, by this correction.

\section{Central excess}
\label{append:CentralResults}

For the central emission, $\chi^2$ is computed using the radial range $0\le \tau \le 1.5$.
Notice that a fit to the extended component cutoff parameter $\tau_{\rm cut}$ may still fall outside
the fitting region, as $\tau>1.5$ emission can contribute to the $\tau<1.5$ signal after  PSF convolution.
We verify that the results are not sensitive to the precise $\tau$ range used for the fitting, including larger ranges reaching $0\le \tau \le 2$.

The best-fitting models for the central emission are derived and studied using
the control samples. The results are presented in \S\ref{append:centre_restored} and Table \ref{tab:fitting_centre} based on the raw  (restored without NVSS point-source masking) and in  \S\ref{append:centre_cleared} and Table \ref{tab:fitting_centre_cleared} based on the cleared data.

\subsection{Restored data}
\label{append:centre_restored}

We first model the central signal in the raw data.
To simplify this modelling, we use an average FWHM value for the restoring beam in each channel, in the declination range $20\dgr<\delta<60\dgr$ relevant to the (41 out of 44) clusters in our sample.
One way to quantify this FWHM is by fitting the raw data to the convolution of a  circular Gaussian with many (say, 300) bright ($F_{\rm 1.4GHz} > 1$ Jy) and compact (major-axis $<0\dgr.25$) known (NVSS) sources.
The resulting FWHM values (see Table \ref{tab:maps_sum}) are fairly constant in this declination range, showing for example $\lesssim3\%$
differences between $20\dgr<\delta<40\dgr$ and $40\dgr<\delta<60\dgr$ fits.
It is challenging to fit the central cluster emission
with our oversimplified model, composed of only two, point-like and extended,
components, each with a
different power-law spectrum.

The central signal is reasonably fitted in the seven low-frequency channels by the Gaussian-convolved central point-source model ($A_e = 0$), with a $\chi^2$ per DOF value of $\chi_n^2(\tau \le 1.5)\simeq 2.8$ or equivalently $\MyChiCorr\chi_n^2(\tau \le 1.5)\simeq 1.3$. However, such a point-source model does not provide a good fit in the high-frequency channel, where $\chi_n^2(\tau \le 1.5)\simeq 12.2$.
Modelling point and extended components simultaneously provides a good fit both in the high-frequency channel, with $\chi_n^2(\tau \le 1.5)\simeq0.8$, and at low frequencies, where $\MyChiCorr\chi_n^2(\tau \le 1.5)\simeq 0.6$; see best-fit parameters in Table \ref{tab:fitting_centre}.
Here, we assume that the extended component is bright and compact enough to be mostly picked-up by CLEAN, and hence convolve both point-like and extended components with the Gaussian kernel.
This procedure is somewhat inaccurate, because
$\sim1/4$ of the total raw central excess is not picked up by CLEAN (see \S\ref{append:centre_cleared}).
At both high and low frequencies,
we find that the point-like and extended components make equal contributions to the raw data, \ie $F_e\simeq 0.5F_{73}$.
When excluding the cluster MCXCJ2338.4+2700, whose centre harbours bright NVSS point-sources, including an $F_{\rm 1.4GHz} \simeq 2.8$ Jy source, the contribution of the extended component increases to $80\%$.

\begin{table*}
	\caption{Best-fit results for the central emission in the raw (restored, without NVSS point-source masking) data.
	} 
    \centering
\begin{adjustbox}{width=1\textwidth}
\begin{tabular}{cccccccccccc}
Channel & Central source model & Weights & $F_{73}$ & $F_e/F_{73}$ & $\alpha_e$ & $\alpha_p$ & $\zeta$ & $\tau_{\rm cut}$ & $\chi^2$ & $\DF$ & $\chi^2$ Eq. \\
(1) & (2) & (3) & (4) & (5) & (6) & (7) & (8) & (9) & (10) & (11) & (12) \\ \hline
\multirow{2}{*}{high} & \multirow{2}{*}{\parbox{2cm}{(point+extended)\\$\times$Gaussian}} & bm & $6.58^{+0.31}_{-0.29}$ & $0.48^{+0.09}_{-0.09}$ & — & — & $0.79^{+0.47}_{-0.65}$ & $0.91^{+0.20}_{-0.17}$ & 1.3 & \multirow{2}{*}{2} & \multirow{2}{*}{\eqref{eq:chi2}} \\
& & cl & $5.88^{+0.47}_{-0.47}$ & $0.49^{+0.07}_{-0.08}$ & — & —  & $0.20^{+0.63}_{-0.77}$ & $0.89^{+0.19}_{-0.16}$  & 1.1 &  & \\
\multirow{2}{*}{low}  & \multirow{2}{*}{\parbox{2cm}{(point+extended)\\$\times$Gaussian}} & bm & $9.53^{+0.33}_{-0.34}$ & $0.49^{+0.04}_{-0.04}$ & $0.76^{+0.09}_{-0.08}$ & $1.16^{+0.10}_{-0.10}$  & $1.18^{+0.11}_{-0.10}$  & $1.79^{+0.13}_{-0.13}$  & 30.1 & \multirow{2}{*}{36} & \multirow{2}{*}{\eqref{eq:chi2}} \\
& & cl & $8.15^{+0.34}_{-0.37}$ & $0.46^{+0.04}_{-0.04}$ & $0.95^{+0.15}_{-0.17}$ & $1.22^{+0.13}_{-0.14}$  & $1.22^{+0.13}_{-0.14}$ & $2.06^{+0.18}_{-1.06}$ & 20.6    &   & \\
\end{tabular}
\end{adjustbox}
    \label{tab:fitting_centre}
\begin{tablenotes}
\item
    {\bf Columns:}
    (1) The high-frequency (high) or the seven co-added low-frequency (low) channels;
    (2) Central source model, specifying point or extended components in Eq.~(\ref{eq:central_emision_point}) with Gaussian or PSF convolution;
    (3) Stacking method: beam co-addition ('bm' for brevity) or cluster co-addition ('cl' for brevity);
    (4) Flux of the central emission at 73 MHz
    in units of $10^{-23}$ erg s$^{-1}$ cm$^{-2}$ Hz$^{-1}$;
    (5) Ratio between the extended component flux and total flux;
    (6) Spectral index $\alpha=-d\ln I_\nu/d\ln \nu$ of the central emission for the extended component;
    (7) Spectral index of the central emission for the point source;
    (8) Radial power law of the extended central emission in Eq.~(\ref{eq:central_emision_point});
    (9) Normalised cutoff radius for the extended
    central emission in Eq.~(\ref{eq:central_emision_point});
    (10) $\chi^2$ value of the best fit, uncorrected for correlations;
    (11) Number of degrees of freedom in the fit;
    (12) The $\chi^2$ equation.
    In the low-frequency channels, the uncertainties are corrected using Eq.~(\ref{eq:uncertainty_corr}).
\end{tablenotes}
\end{table*}

\begin{figure*}
    \centering
    \includegraphics[width=0.8\textwidth,trim={0 0.5cm 0 0},clip]{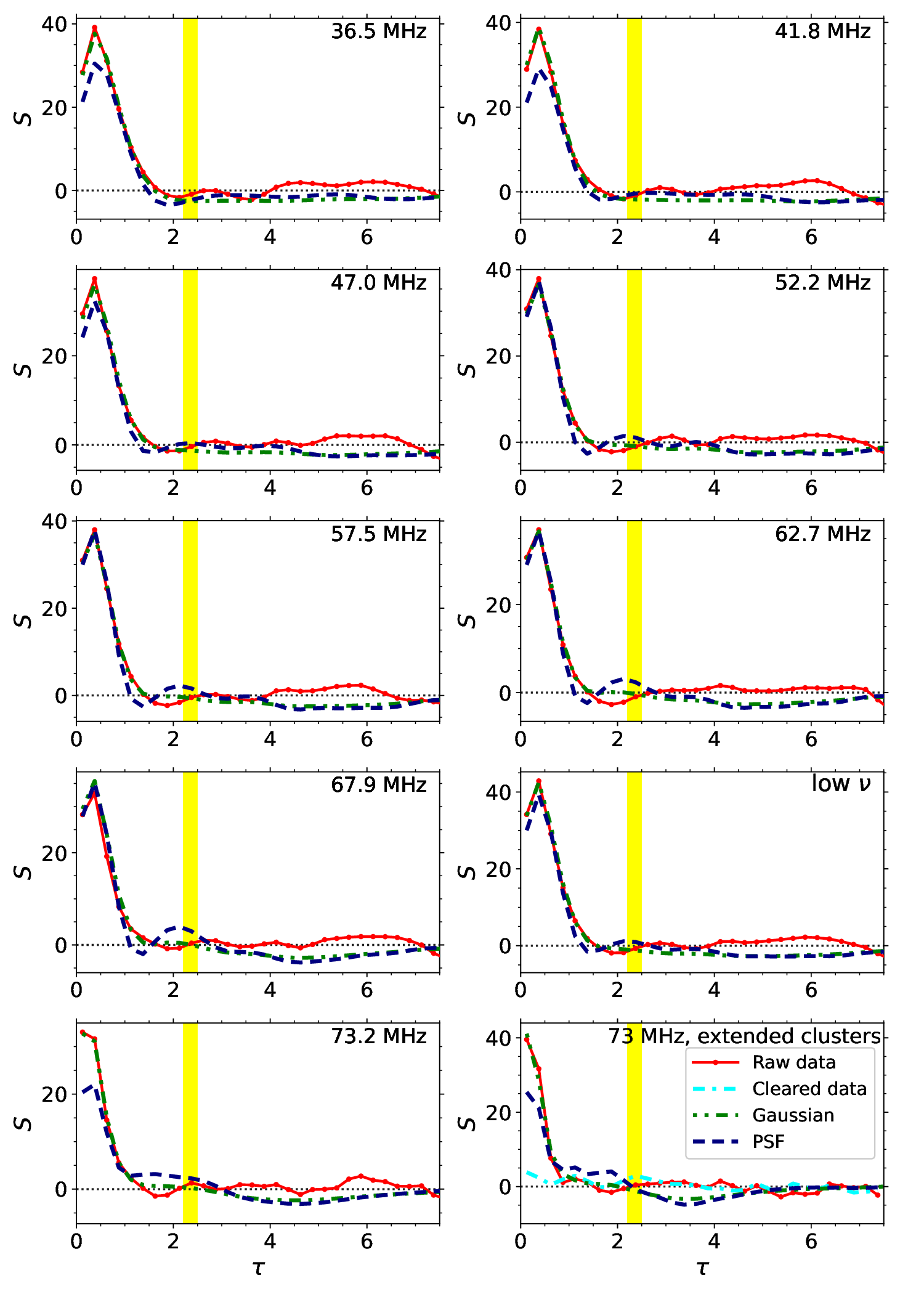}
    \caption{
    Best-fit to the raw (restored without NVSS point-source masking) data using the central
    model only,
    shown in each individual channel (labelled frequencies) and stacked over the seven low-frequency channels (labelled ``low $\nu$'').
    The seven low-frequency channels are jointly fit with six free parameters ($F_p$, $F_e$, $\alpha_p$, $\alpha_e$, $\zeta$ and $\tau_{\rm cut}$), while the high-frequency channel is fitted separately and without spectral parameters.
    Each panel shows the beam co-added significance profiles of the raw data
    (solid red curve), the best-fit model with a Gaussian restoring beam
    (using Eq.~\eqref{eq:chi2} for $\chi^2$; double dot-dashed green) and the same model convolved with the PSF (dashed blue).
    In the 73 MHz channel (bottom-left panel), PSF convolution yields a flat, $1\lesssim\tau\lesssim2.5$ excess due to the more compact clusters; the bottom right panel shows that this feature does not reach the virial radius for the larger clusters
    (shown for the eight clusters of $\theta_{500} > 0\dgrdot3$), which alone still show a virial excess in cleared data (cyan dot-dashed curve).
    The virial shock range of \emph{Fermi} is highlighted (notations as in Fig.~\ref{fig:band7_sig}).
    }
    \label{fig:fake_center_restored}
\end{figure*}

Figure \ref{fig:fake_center_restored} shows that the central signal (for the 44-cluster sample; solid red) is fairly well fit by this  Gaussian-convolved two-component model in each channel separately
(double-dot-dashed green).
The best-fit spectral index is
\begin{equation}
  \alpha_p =
  1.16^{+0.10}_{-0.10}~(1.22^{+0.13}_{-0.14})
    \label{eq:central_alpha_point}
\end{equation}
for the point-like component, and \begin{equation}
  \alpha_e =
  0.76^{+0.09}_{-0.08}~(0.95^{+0.15}_{-0.17})
    \label{eq:central_alpha_extend}
\end{equation}
for the extended component, using beam (cluster) co-addition.
Differences between beam and cluster co-addition
might be due to the aforementioned cluster MCXCJ2338.4+2700.
Indeed, if we exclude this cluster, the spectral indices become  $\alpha_p = 1.20^{+0.20}_{-0.20}~(1.13^{+0.16}_{-0.16})$ and a more consistent  $\alpha_e = 1.07^{+0.29}_{-0.14}~(1.05^{+0.18}_{-0.13})$ for beam (cluster) co-addition.
The spectra slightly softens when excluding this source, which itself shows an $\alpha \simeq 0.75 \pm 0.5$ spectrum based on comparing its catalogued GMRT \citep[Giant Metrewave Radio Telescope;][]{2017A&A...598A..78I} and NVSS fluxes.

These spectra are
sensitive not only to the co-addition method, but also to the choice of channels.
Fitting the four lowest-frequency channels, focusing for example on beam co-addition,
gives a slightly
harder $\alpha_p = 1.05^{+0.09}_{-0.09}$
but a similar $\alpha_e=0.82^{+0.27}_{-0.26}$;
using the three higher frequency channels gives
a softer
$\alpha_e =0.93^{+0.13}_{-0.14}$
and a much softer $\alpha_p=2.14^{+0.20}_{-0.20}$.
Overall, we find spectral indices in the range $1.0\lesssim\alpha_p\lesssim2.2$ and $0.7\lesssim\alpha_e\lesssim1.1$ for our best-fit $\zeta$ and $\tau_{\rm cut}$ values.
We cannot determine the spectrum more precisely, due to our weak constraints on $\zeta$ and $\tau_{\rm cut}$ (see Table \ref{tab:fitting_centre}) and to the sensitivity of the results to assumptions.

As expected,
modelling the data with central components alone
does not provide a good fit to the data beyond $\tau\simeq 2$, and in particular, does not reproduce the virial excess.
Nevertheless, although the sidelobe pattern induced by the strong central signal should ideally be entirely removed by CLEAN, some residual may remain in the data due to PSF inaccuracies, noise, and faint sources below the CLEAN threshold.
The worst-case putative residual artefacts can be crudely estimated by a PSF convolution of the best-fit to the raw data (itself obtained by convolving the model with the Gaussian restoring beam),
shown as dashed curves in Fig.~\ref{fig:fake_center_restored}.
Such a
PSF convolution leads to a significantly worse fit to the raw data,
especially in the $1<\tau<2$ region, suggesting that CLEAN has been rather successful in removing the full PSF-convolved central signal.
For example,
our best fit to the central region, convolved with the PSF, gives $\chi^2_n \simeq 134.5$
for the high-frequency channel and
$\MyChiCorr\chi^2_n \simeq 7.1$
for the low-frequency channels in the $\tau \le 1.5$ region.

Even a putative failure of CLEAN only at large angular separations, corresponding to $\tau>2$, does not provide a good explanation for the virial excess.
In terms of amplitude alone,
this scenario would require CLEAN to remove essentially all of the sidelobe flux inside $\tau<2$, but little to none of it beyond $\tau>2$,
and even that would not suffice to explain the excess.
In terms of the angular position of the ripples, the $\tau$-dependence does not match the data either,
except for a few chance combinations of individual clusters in specific channels (see Fig.~\ref{fig:fake_center_restored}).
Quantitatively,
the PSF-convolved models derived based on the $\tau \le 1.5$ data provide a poor fit in the $2<\tau<5$ region, with
$\chi^2_n(2<\tau<5) \simeq 10.2$ ($\MyChiCorr\chi^2_n \simeq 3.1$)
for the high (low) frequency channels.
This should be compared to
$\chi^2_n\simeq 5.9$ ($\MyChiCorr\chi^2_n \simeq2.9$) with the Gaussian convolution.
As the figure shows, neither a PSF nor a Gaussian convolution of the model shows a visually reasonable fit in the $2<\tau<5$ range.
These conclusions are not sensitive to the precise range of fitted radii.

Indeed, while putative PSF sidelobe residuals of the central signal could, in principle, generate some excess rings, we find that they cannot mimic the virial excess: the first ripple does not reach the virial radius,  while the second ripple is both too weak and typically too extended to reproduce the virial signal.
Consider the low-frequency channels first.
As Fig.~\ref{fig:fake_center_restored} shows, in the two lowest-frequency channels, the PSF residuals are negligible.
Higher-frequency channels show stronger residuals, but misaligned with the virial excess of most or all clusters.
In the highest of these channels, $67.9\MHz$, the first ripple is expected to lie, for different clusters, at $1.3\lesssim\tau\lesssim2.5$ for different clusters, while the second ripple lies at $2.0\lesssim\tau\lesssim4.0$; the higher $\tau$ values are more relevant due to more numerous compact clusters.
As the figure
shows, the co-added result gives a first ripple peaked at $\tau\sim2.2$, slightly inside the low-frequency virial excess, and a very weak second ripple at $\tau\sim3.7$, far outside the virial signal.
In any case, the mismatch between the restored data and the PSF-convolved central excess (compare solid and dashed curves in the figure) shows that sidelobes are removed quite accurately by CLEAN.

For the high-frequency channel, PSF-convolution shows a peculiar flat excess in the $1\lesssim\tau\lesssim2.5$ range. This excess can be traced to a positive average of the PSF outside the main beam in this channel.
As mentioned above, our results indicate that CLEAN efficiently removed this residual.
But even if this were not the case, one can show that this residual cannot mimic the virial excess. For example, the bottom-right panel of Fig.~\ref{fig:fake_center_restored} shows that this flat residual does not even reach the virial radius for the more extended ($\theta_{500} > 0\dgrdot3$) clusters in our sample.

\subsection{Cleared data}
\label{append:centre_cleared}

Fitting the \clean\ data with both components of the central emission model, we find that the point-like component makes a negligible contribution, indicating an efficient CLEAN.
Fitting the point-like component alone fails to reproduce the central excess in the cleared data (compare the solid blue and dot-dashed green curves in Fig.~\ref{fig:fake_center_clear}), with fairly large $\chi^2_n(\tau \le 1.5) \simeq 3.5$ for the high-frequency channel and $\MyChiCorr\chi^2_n(\tau \le 1.5) \simeq 1.2$ for the co-added low-frequency channels.
In contrast, the extended emission model nicely reproduces the central
significance profiles (compare the solid blue and double dot-dashed purple curves in the figure), with
$\chi^2_n(\tau \le 1.5) \simeq 0.2$ for the high-frequency channel and $\MyChiCorr\chi^2_n(\tau \le 1.5) \simeq 0.1$ for the co-added low-frequency channels.

\addtocounter{figure}{+1}
\begin{figure}
    \centering
    \includegraphics[width=0.45\textwidth,trim={0 0.4cm 0 0},clip]{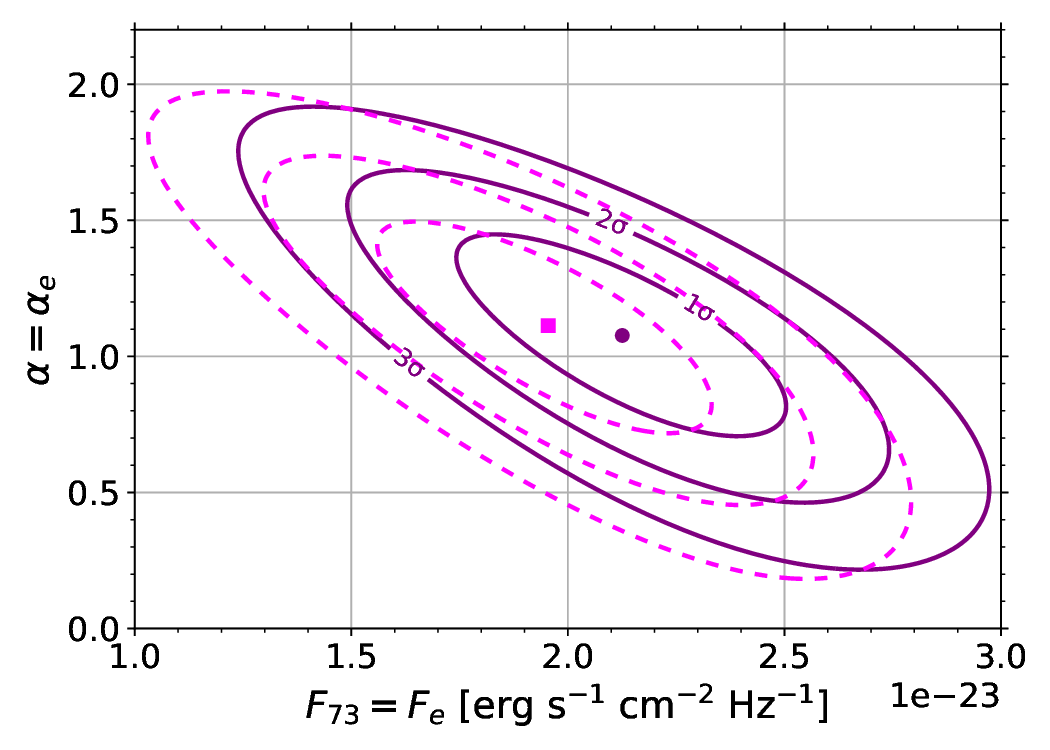}
    \caption{
    Best fit (symbols and $1\sigma$ to $3\sigma$ confidence contours) for central extended-source parameters
    $F_{e}$ (at $73\MHz$) and $\alpha_e$, for the low-frequency \cleandata.
    Results are shown for nominal analysis with beam (circle with purple solid contours) and cluster (square with magenta dashed contours) co-addition.
    }
    \label{fig:alpha_clear}
\end{figure}
\addtocounter{figure}{-2}

\begin{figure*}
    \centering
    \includegraphics[width=\textwidth,trim={0 0.5cm 0 0},clip]{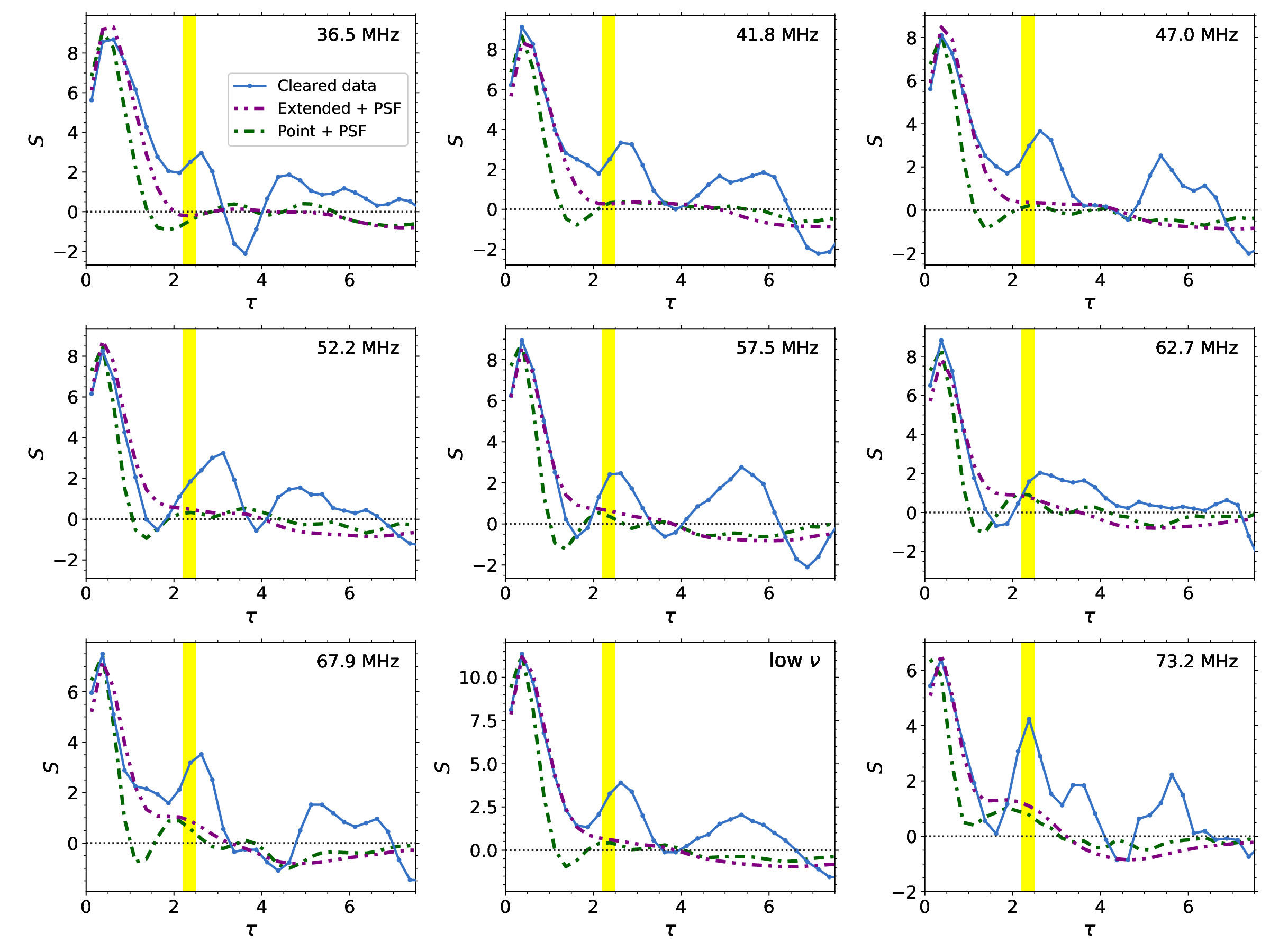}
    \caption{
    Same as Fig.~\ref{fig:fake_center_restored} but for \cleandata.
    The model (\ref{eq:central_emision_point}) is used with a single parameter $F_{73}$ for the high-frequency channel, while the seven low frequencies are jointly fit with the same two parameters, $F_{73}$ and $\alpha$. The parameters $\zeta$ and $\tau_{\rm cut}$ are fixed.
    Each panel shows the beam co-added significance profiles of the data (solid blue curve) and of the PSF-convolved point (dot-dashed green) and extended (double dot-dashed purple) central emission models.
    }
    \label{fig:fake_center_clear}
\end{figure*}

With this extended component alone, we find that the cleared data has a total central flux ($F_{73}$) which is about a quarter of its raw-data counterpart (compare Tables \ref{tab:fitting_centre} and \ref{tab:fitting_centre_cleared}).
The spectral index of the extended source for the cleared data is
\begin{equation} \label{eq:alpha_ext}
\alpha_{\ext} = 1.08^{+0.24}_{-0.24} \quad (1.11^{+0.25}_{-0.26})
\end{equation}
for beam (cluster) co-addition; see Fig.~\ref{fig:alpha_clear}.
This spectrum is consistent with the backward estimate of the cleared-central emission in \S\ref{subsec:spec_lwa}.

\begin{table*}
	\caption{Best-fit results for the central emission in the cleared data.} 
    \centering
\begin{adjustbox}{width=1\textwidth}
\begin{tabular}{cccccccccc}
Channel                     & Central source model                     & Weights & $F_{73}$               & $\alpha_e$               & $\zeta$      & $\tau_{\rm cut}$ & $\chi^2$ & $\DF$ & $\chi^2$ Eq.        \\
(1)                       & (2)                         & (3)                                      & (4)      & (5)                    & (6)                    & (7)          & (8)          & (9)      & (10)         \\ \hline
\multirow{2}{*}{high } & \multirow{2}{*}{extended $\times$ PSF}          & bm       & $1.83^{+0.24}_{-0.22}$ & —                      & $0.94^{+0.39}_{-0.52}$  & $0.95^{+0.32}_{-0.21}$ & 0.7      & 3  & \eqref{eq:chi2} \\
& & —  & $1.60^{+0.22}_{-0.21}$ & — & $1.27^{+0.35}_{-0.41}$ & $1.16^{+0.54}_{-0.35}$ & 442.1 & 261 & \eqref{eq:chi2_0}\\
\multirow{2}{*}{low }  & \multirow{2}{*}{extended $\times$ PSF}          & bm       & $2.13^{+0.25}_{-0.25}$ & $1.08^{+0.24}_{-0.24}$  & $1.24^{+0.14}_{-0.19}$  & $1.61^{+0.37}_{-0.37}$  & 19.0     & 38 & \eqref{eq:chi2} \\
& & — & $1.57^{+0.20}_{-0.18}$ & $1.13^{+0.28}_{-0.27}$ & $1.12^{+0.21}_{-0.21}$ & $1.32^{+0.40}_{-0.30}$ & 3222.8 & 1844 & \eqref{eq:chi2_0} \\ \hline
\multirow{2}{*}{high } & \multirow{2}{*}{extended $\times$ PSF}          & bm       & $1.87^{+0.18}_{-0.18}$ & —                      & \textbf{0.9} & \textbf{0.9} & 0.9      & 5 & \eqref{eq:chi2} \\
& & — & $1.65^{+0.17}_{-0.17}$ & — & \textbf{1.3} & \textbf{1.2} & 442.2 & 263 & \eqref{eq:chi2_0} \\
\multirow{2}{*}{low }  & \multirow{2}{*}{extended $\times$ PSF}          & bm       & $2.21^{+0.22}_{-0.22}$ & $1.01^{+0.22}_{-0.22}$ & \textbf{1.2} & \textbf{1.6} & 19.5     & 40 & \eqref{eq:chi2} \\
& & — & $1.50^{+0.15}_{-0.14}$ & $1.13^{+0.27}_{-0.26}$ & \textbf{1.1} & \textbf{1.3} & 3223.8 & 1846 & \eqref{eq:chi2_0} \\ \hline
\end{tabular}
\end{adjustbox}
    \label{tab:fitting_centre_cleared}
\begin{tablenotes}
\item
    {\bf Columns:}
    (1) The high-frequency (high) or the seven co-added low-frequency (low) channels;
    (2) Central source model, specifying point or extended components in Eq.~(\ref{eq:central_emision_point}) with Gaussian or PSF convolution;
    (3) Stacking method: beam co-addition ('bm' for brevity) or cluster co-addition ('cl' for brevity);
    (4) Flux of the central emission at 73 MHz
    in units of $10^{-23}$ erg s$^{-1}$ cm$^{-2}$ Hz$^{-1}$;
    (5) Spectral index $\alpha=-d\ln I_\nu/d\ln \nu$ of the central emission;
    (6) Radial power law of the extended central emission in Eq.~(\ref{eq:central_emision_point});
    (7) Normalised cutoff radius for the extended
    central emission in Eq.~(\ref{eq:central_emision_point});
    (8) $\chi^2$ value of the best fit, uncorrected for correlations;
    (9) Number of degrees of freedom in the fit;
    (10) The $\chi^2$ equation.
    Values in \textbf{boldface} are fixed, not fit, parameters.
    In the low-frequency channels, the uncertainties are corrected using Eq.~(\ref{eq:uncertainty_corr}).
\end{tablenotes}
\end{table*}

The $\zeta$ and $\tau_{\rm cut}$ values obtained using the cleared data are
broadly consistent with their raw-data counterparts,
but the large statistical uncertainties in these two parameters
and the different values obtained for high vs. low frequencies indicate that the model is too crude (see Table \ref{tab:fitting_centre_cleared}). However, the best-fitting results for $\alpha_e$ are not sensitive to the choice of $\zeta$ and $\tau_{\rm cut}$.
For simplicity, we fix $\zeta = 0.9$ (1.2) and $\tau_{\rm cut}=0.9$ (1.6) for the high (low) frequency data when simultaneously modelling the central and virial excess signals in \S\ref{sec:virialemission}.
Adopting the standard $\chi^2$ of Eq.~\eqref{eq:chi2_0} instead of our nominal Eq.~\eqref{eq:chi2} gives more consistent best-fitting values for $\zeta$ and $\tau_{\rm cut}$ among high and low frequencies (see Table \ref{tab:fitting_centre_cleared}), without changing the spectrum significantly; namely, giving $\alpha_e = 1.13^{+0.28}_{-0.27}$ consistent with Eq.~\eqref{eq:alpha_ext}.

The resulting model, convolved with the PSF before fitting to the data, shows (see Fig.~\ref{fig:fake_center_clear}) some minor sidelobe ripples, but again, these artefacts cannot account for the virial access in terms of neither amplitude nor position.
This conclusion pertains to each of the channels separately, and to both extended and point-like models.
We conclude that the extended central excess in the \cleandata\ has a very minor effect near the virial radius.
This strengthens our conclusion that any central excess sidelobes missed by CLEAN cannot mimic the virial excess.

\section{More tests ruling out virial PSF ripples}
\label{append:psf}

\begin{bfigure}
    \centering
    \includegraphics[width=0.45\textwidth,trim={0 0.5cm 0 0},clip]{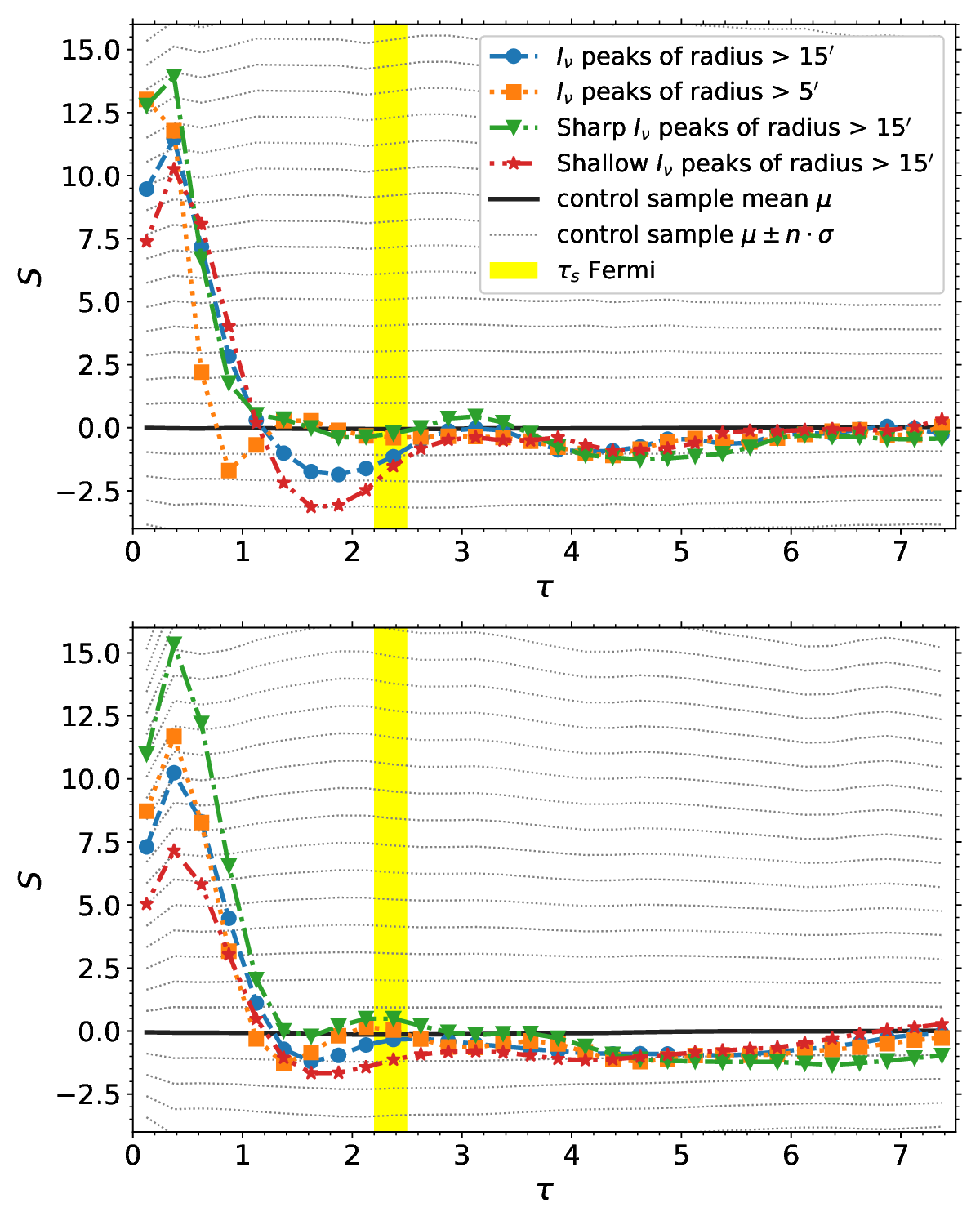}
    \caption{
    Excess significance profiles of the control samples purposely chosen around random local brightness maxima, for the high (top panel) and co-added low (bottom) frequency channels, using the cleared data.
    Shown are results using the smoothing scale $15\arcmin$ (blue disks with dashed lines to guide the eye), $5\arcmin$ (orange squares with dotted lines to guide the eye), $15\arcmin$ with sharp peaks (green triangles with dot-dashed lines to guide the eye), $15\arcmin$ with shallow peaks (red stars with double dot-dashed lines to guide the eye).
    Other notations are as in Fig.~\ref{fig:band7_sig}.
    }
    \label{fig:local_max}
\end{bfigure}

\begin{bfigure}
    \centering
    \includegraphics[width=0.45\textwidth,trim={0 0.5cm 0 0},clip]{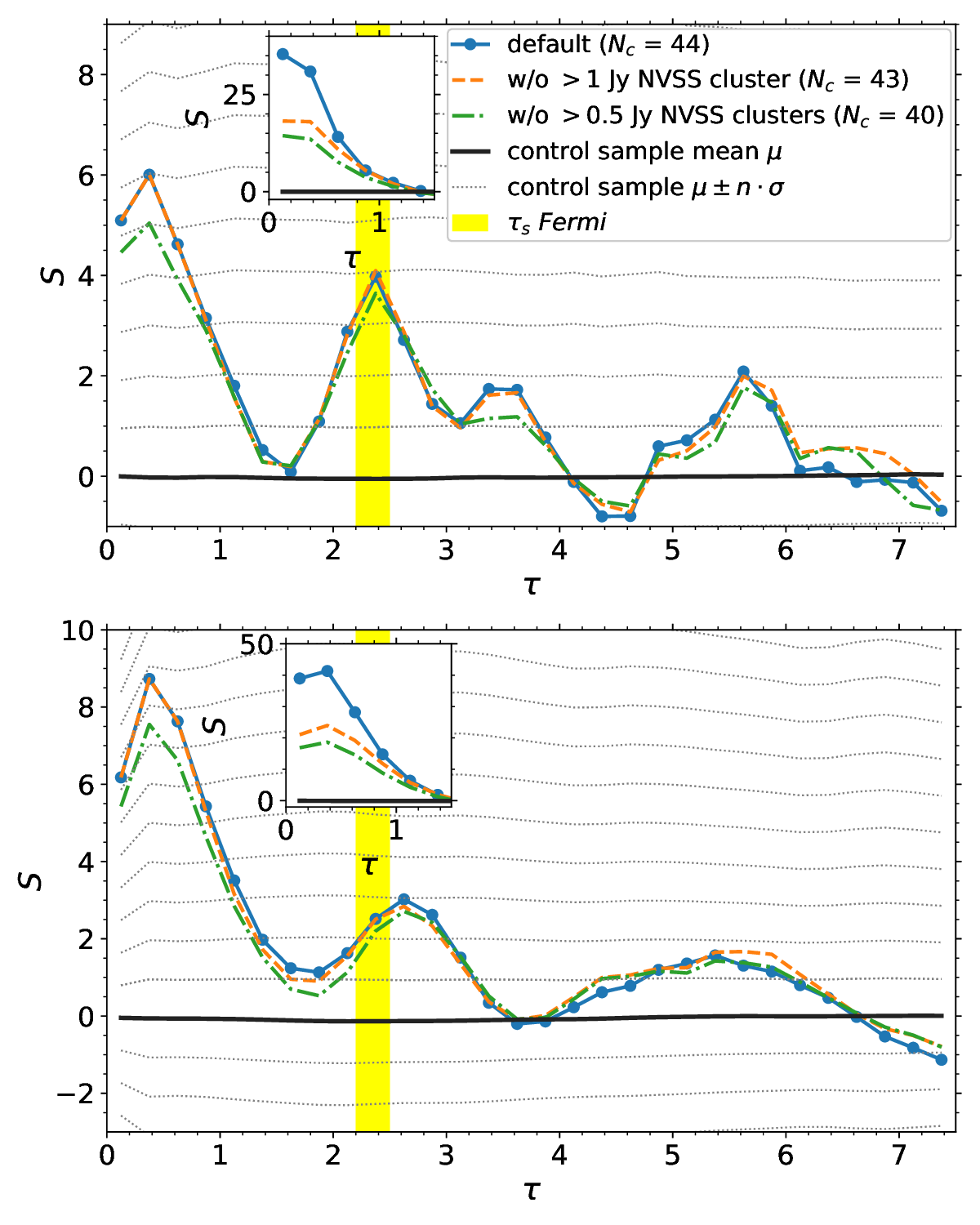}
    \caption{
    Excess significance
    in the high (top panel) and co-added low (bottom) frequency channels,
    before (blue disks with solid line to guide the eye) and after (curves) removing the one cluster with $F_{\rm 1.4GHz} > 1$ Jy (dashed orange), or the four clusters with $F_{\rm 1.4GHz} > 0.5$ Jy (dot-dashed green), NVSS sources within $0\dgrdot2$ from their centres.
    Other notations are as in
    Fig.~\ref{fig:band7_sig}.
    While removing these clusters from the nominal sample substantially lowers the central excess in the restored, unmasked data (insets),
    the effect on the nominal virial excess in the cleared, masked data (main figures) is minute.
    }
    \label{fig:43c_test}
\end{bfigure}

As the centres of clusters are often radio bright and the stacked virial signal assumes the shape of a concentric ring, it is important to test if the virial signature could be an artefact arising from PSF sidelobes of the central emission.
Several tests were outlined in \S\ref{subsec:sidelobe} and reviewed in \S\ref{sec:summary}, indicating that the virial signal is not such an artefact.

Here, we present two such tests. In the first, we attempt to mimic the virial ring using PSF sidelobes, by stacking bright sources or bright regions in the sky that are unrelated to clusters.
In the second, we selectively exclude from our sample clusters with a particularly strong central excess, to see if this may have any effect on the virial excess.
Both tests, like the others, summarised in \S\ref{sec:summary}, do not support a putative PSF sidelobe origin for the virial excess.

\subsection{PSF sidelobes fail to mimic virial excess}
The procedure is identical to that used for control clusters in \S\ref{subsec:controlsamples}, but instead of stacking random regions in the allowed part of the sky (\S\ref{subsec:cluster}), here we stack local brightness peaks in the restored data, around which one may find sidelobe emission that was not fully removed by CLEAN.
We test different prescriptions for the scale and compactness of these peaks, and examine the cleared-data results using the same nominal stacking and analysis pipelines described in \S\ref{sec:analysis}.

Figure \ref{fig:local_max} demonstrates the resulting high-frequency (top panel) and low-frequency (bottom) significance profiles, for different choices of brightness peak samples.
These include peaks in the LWA sky on small (smoothed on radius $5\arcmin$) or medium (radius $\sim\theta_{500}\sim 15\arcmin$) scales, after large-scale sky structures (obtained by smoothing with a Gaussian of radius $40\arcmin$) were removed.
We also examine sharp vs. shallow peaks on $15\arcmin$ scales, defined by comparing the maximal gradient at the smoothing radius to its ensemble median.

As the figure illustrates, in all our attempts, we were unable to obtain a significant excess at scales corresponding to cluster peripheries, even though we
choose a control sample with
a stacked central signal stronger than in the real sample. Some ripples can be seen, but in addition to being weak, they are also misaligned with the real virial excess.

\subsection{Virial excess unaffected by changing central excess}

If the virial excess were a sidelobe artefact of emission from the centres of clusters, then lowering the latter would diminish the former.
Such a scenario is readily tested, as the central excess is dominated by a few clusters harbouring bright NVSS sources near their centres.
Figure \ref{fig:43c_test} demonstrates that if one or more of these clusters are excluded from our nominal cluster sample, the
raw (restored and unmasked; see inset) central excess is nearly halved, while the nominal (cleared and masked) virial excess is barely affected.
This test alone proves that the virial signal does not depend on most, if not all, of the central emission.

In particular, only one cluster harbours $F_{\rm 1.4 GHz} > 1$ Jy NVSS sources.
This cluster alone contributes $\sim 50\%$ ($\sim 45\%$) of the central excess at high (low) frequencies, as two very bright, $\gtrsim 2.4$ Jy sources are found within $0\dgrdot2$ of its centre.
As the figure shows, removing this cluster alone dramatically lowers the
raw central excess, with no significant effect on the nominal virial excess.
Further excluding clusters harbouring $F_{\rm 1.4 GHz} > 0.5$ Jy NVSS sources within $0\dgrdot2$ from their centres removes three additional clusters from the sample.
The remaining, 40 cluster sample shows
a raw
central excess fainter by a factor of $\gtrsim 2.3$
than in the nominal sample, yet the virial excess is still barely modified.

\section{Sensitivity and consistency tests}
\label{append:sensi}

\begin{bfigure}
    \centering
    \includegraphics[width=0.45\textwidth,trim={0 0.5cm 0 0},clip]{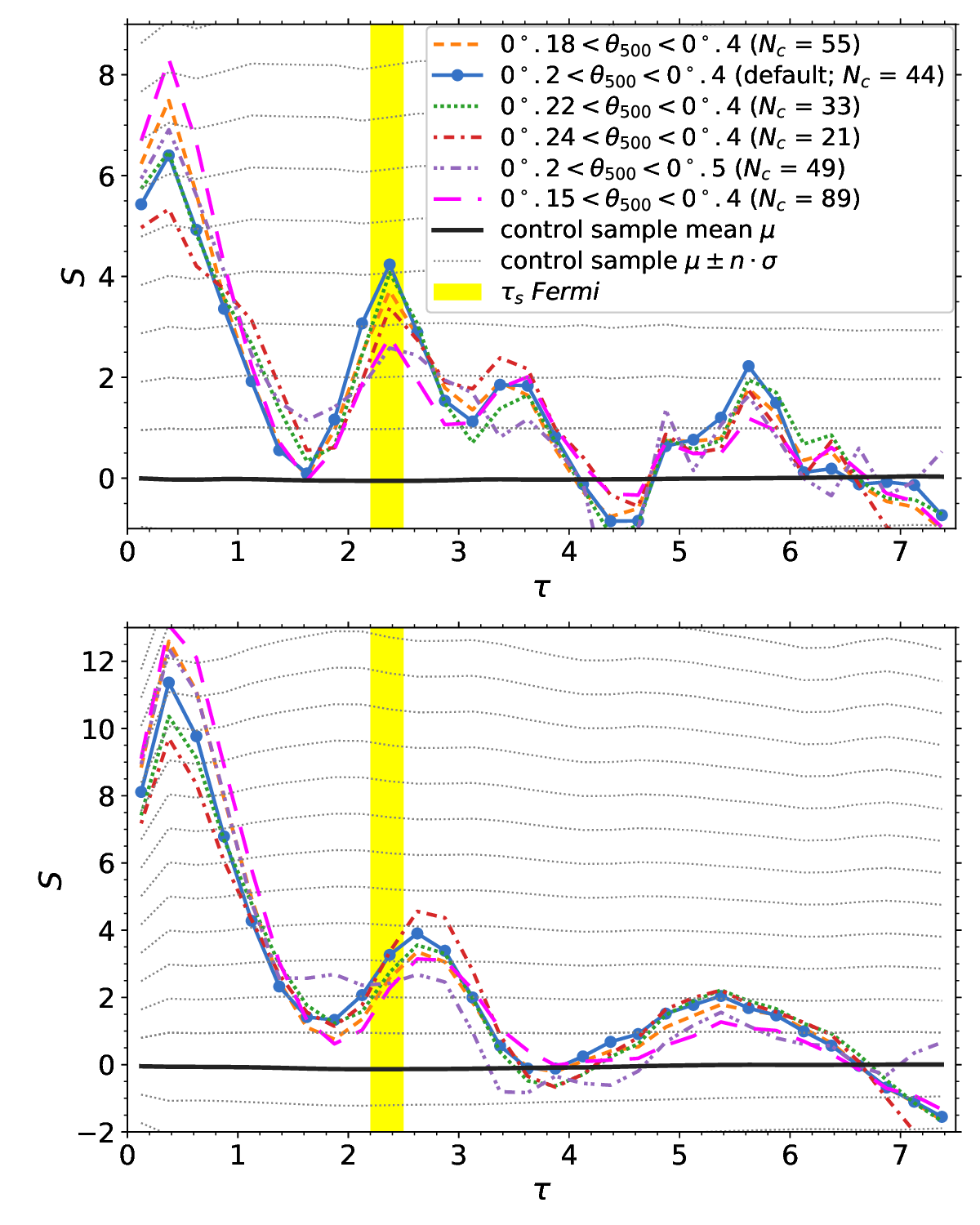}
    \caption{
    Significance profiles with various $\theta_{500}$ selections: $0\dgr.18 < \theta_{500} < 0\dgr.4$ (dashed orange curve), $0\dgr.2 < \theta_{500} < 0\dgr.4$ (default; solid blue), $0\dgr.22 < \theta_{500} < 0\dgr.4$ (dotted green), $0\dgr.24 < \theta_{500} < 0\dgr.4$ (dot-dashed red), $0\dgr.2 < \theta_{500} < 0\dgr.5$ (double dot-dashed purple), $0\dgr.15 < \theta_{500} < 0\dgr.4$ (long dashed magenta), for the high-frequency (top penal) and low-frequency data (bottom), using the \cleandata. The number of clusters ($N_c$) in each sample is shown in the legend.
    Other notations are as in
    Fig.~\ref{fig:band7_sig}.
    }
    \label{fig:theta_500_test}
\end{bfigure}

\begin{bfigure}
    \centering
    \includegraphics[width=0.45\textwidth,trim={0 0.5cm 0 0},clip]{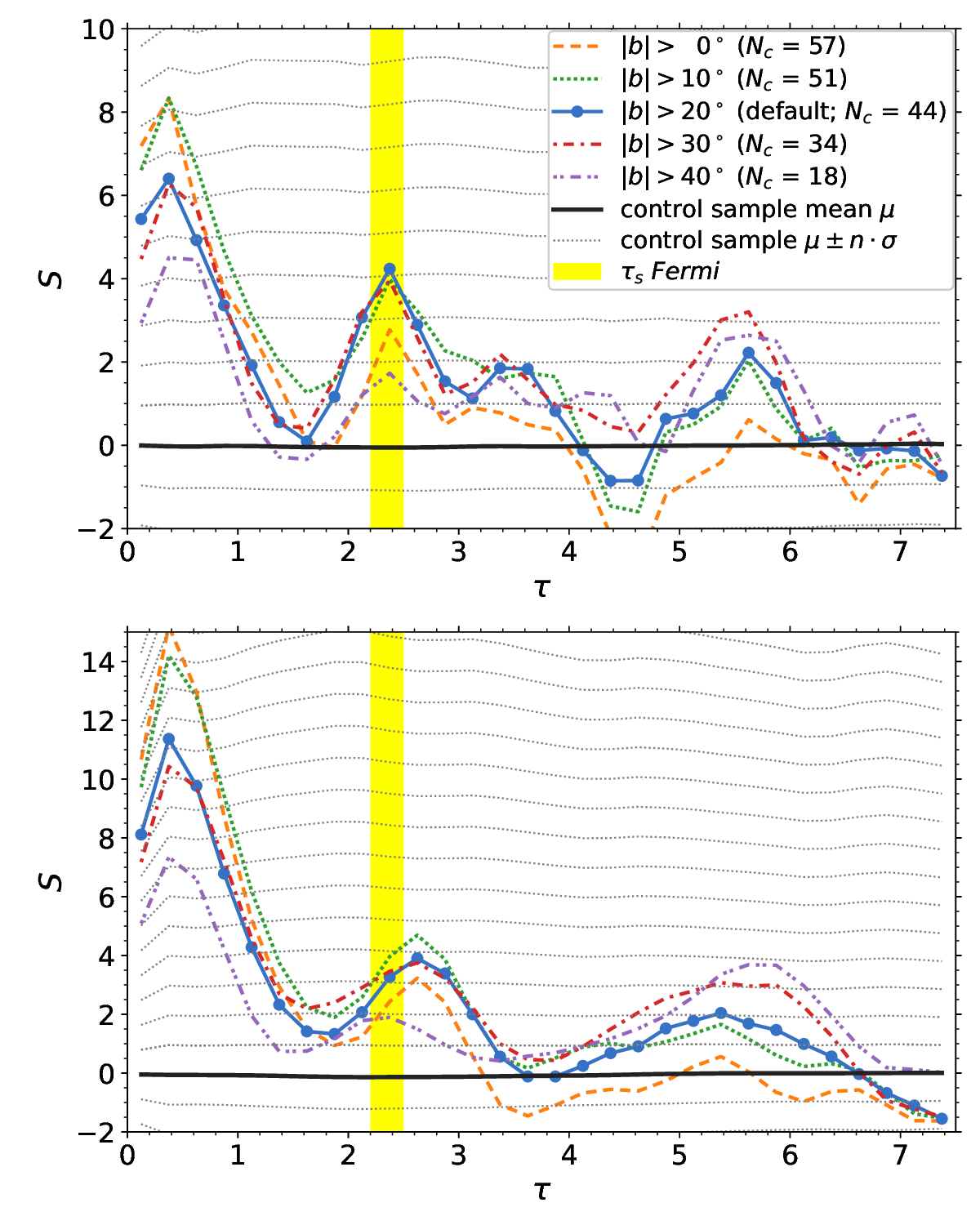}
    \caption{Same as Fig.~\ref{fig:theta_500_test}, but varying the latitude  cut:
    $|b| > 0\dgr$ (dashed orange curve), $|b| > 10\dgr$ (dotted green), $|b| > 20\dgr$ (default; solid blue), $|b| > 30\dgr$ (dot-dashed red), and $|b| > 40\dgr$ (double dot-dashed purple).
    }
    \label{fig:b_test}
\end{bfigure}

We apply sensitivity and consistency tests to each stage of the analysis, in order to examine its robustness and identify optimal control parameter values.
Below we provide tests of parameters controlling the
cluster selection (range of angular size $\theta_{500}$, the cut on galactic latitude $|b|$, and the cut on the declination $\delta$), the data analysis (the cut on the flux $F_{\rm 1.4GHz}$ of NVSS point sources and the masking radius around them), the and stacking procedure (choice of radial bin size $\Delta\tau$).

For simplicity, in the following figures, each parameter is varied individually, leaving the remaining parameters nominal:
galaxy clusters with angular dimensions in the range $0\dgr.2 \leq \theta_{500} < 0\dgr.4$,
galactic latitudes $|b| > 20\dgr$,
and declination $\delta > 20\dgr$,
after masking the data within $0\dgrdot3$ from NVSS point sources of flux exceeding 1 Jy,
removing a background polynomial model of order $N_b = 3$ estimated within $\theta_b=5\dgr$,
and radially binning the results with resolution $\Delta\tau=0.25$.
Each figure shows both high (top panel) and co-added low (bottom) frequency channel results, but focuses for brevity on the beam co-added, cleared data.

Figure \ref{fig:theta_500_test} demonstrates variations in the angular range $\theta_{500}$ of
clusters in our sample. Alternative ranges shown are $0\dgrdot18 < \theta_{500} < 0\dgrdot4$ (leaving $N_c=55$ clusters instead of the nominal $N_c=44$), $0\dgrdot22 < \theta_{500} < 0\dgrdot4$ ($N_c=33$), $0\dgrdot24 < \theta_{500} < 0\dgrdot4$ ($N_c=21$), $0\dgrdot2 < \theta_{500} < 0\dgrdot5$ ($N_c=49$), and $0\dgrdot15 < \theta_{500} < 0\dgrdot4$ ($N_c=89$); see label.
As the figure shows, the virial signal is not sensitive to small changes near the nominal range.
Lowering the minimal $\theta_{500}$ admits more clusters at marginal or insufficient resolution; therefore, a substantially smaller $\theta_{500}$ washes out the excess.
Raising the maximal $\theta_{500}$ admits highly extended clusters, thus contaminating the virial signal due to two effects: (i) substantial foreground and background structures on these scales are introduced to the stacking; and (ii) the extended clusters are bright, and thus offset the background removal themselves.

Figure \ref{fig:b_test} shows the effect of varying the cut on $|b|$.
Alternative cuts shown in the figure are $|b| > 0\dgr$ ($N_c=57$), $|b| > 10\dgr$ ($N_c=51$), $|b| > 30\dgr$ ($N_c=34$), and $|b| > 40\dgr$ ($N_c=18$); see label.
As the figure shows, raising the cut from the nominal $20\dgr$ to $30\dgr$ does not significantly change the virial excess, in spite of the smaller sample size, suggesting that Galactic contamination is not entirely negligible at $20\dgr<b<30\dgr$.
The signal starts dropping if the $|b|$ cut is raised further, due to the smaller $N_c$.
Lowering the latitude cut to $10\dgr$ also does not significantly change the virial excess, but does increase the noise.
Lowering the cut further does diminish the virial signal, although it remains detectable even with no latitude cut.

\begin{figure}
    \begin{center}
    \includegraphics[width=0.45\textwidth,trim={0 0.5cm 0 0},clip]{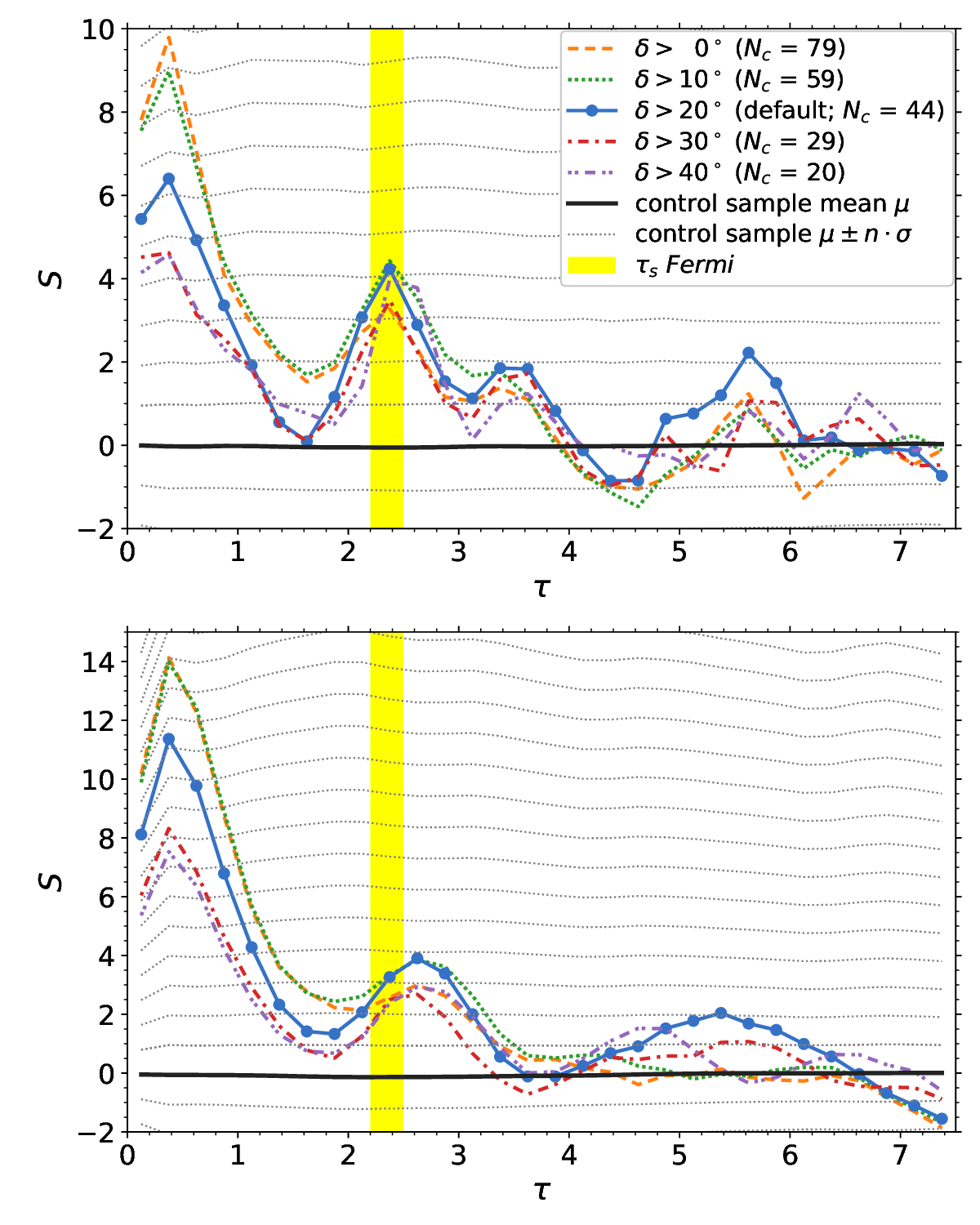}
    \end{center}
    \caption{
    Same as Fig.~\ref{fig:theta_500_test}, but varying the declination cut: $\delta > 0\dgr$ (dashed orange curve), $\delta > 10\dgr$ (dotted green), $\delta > 20\dgr$ (default; solid blue), $\delta > 30\dgr$ (dot-dashed red), and $\delta > 40\dgr$ (double dot-dashed purple).
    }
    \label{fig:dec_test}
\end{figure}

Figure \ref{fig:dec_test} demonstrates variations in the minimal declination threshold of the cluster sample.
Alternative thresholds shown are $\delta > 0\dgr$ ($N_c=79$), $\delta > 10\dgr$ ($N_c=59$), $\delta > 30\dgr$ ($N_c=29$),  and $\delta > 40\dgr$ ($N_c=16$); see label.
As shown in the figure, the virial excess is not sensitive to the choice of the cut. A high $\delta$ cut leaves fewer clusters in the sample, but these clusters have better PSF properties.

\begin{figure}
    \centering
    \includegraphics[width=0.45\textwidth,trim={0 0.5cm 0 0},clip]{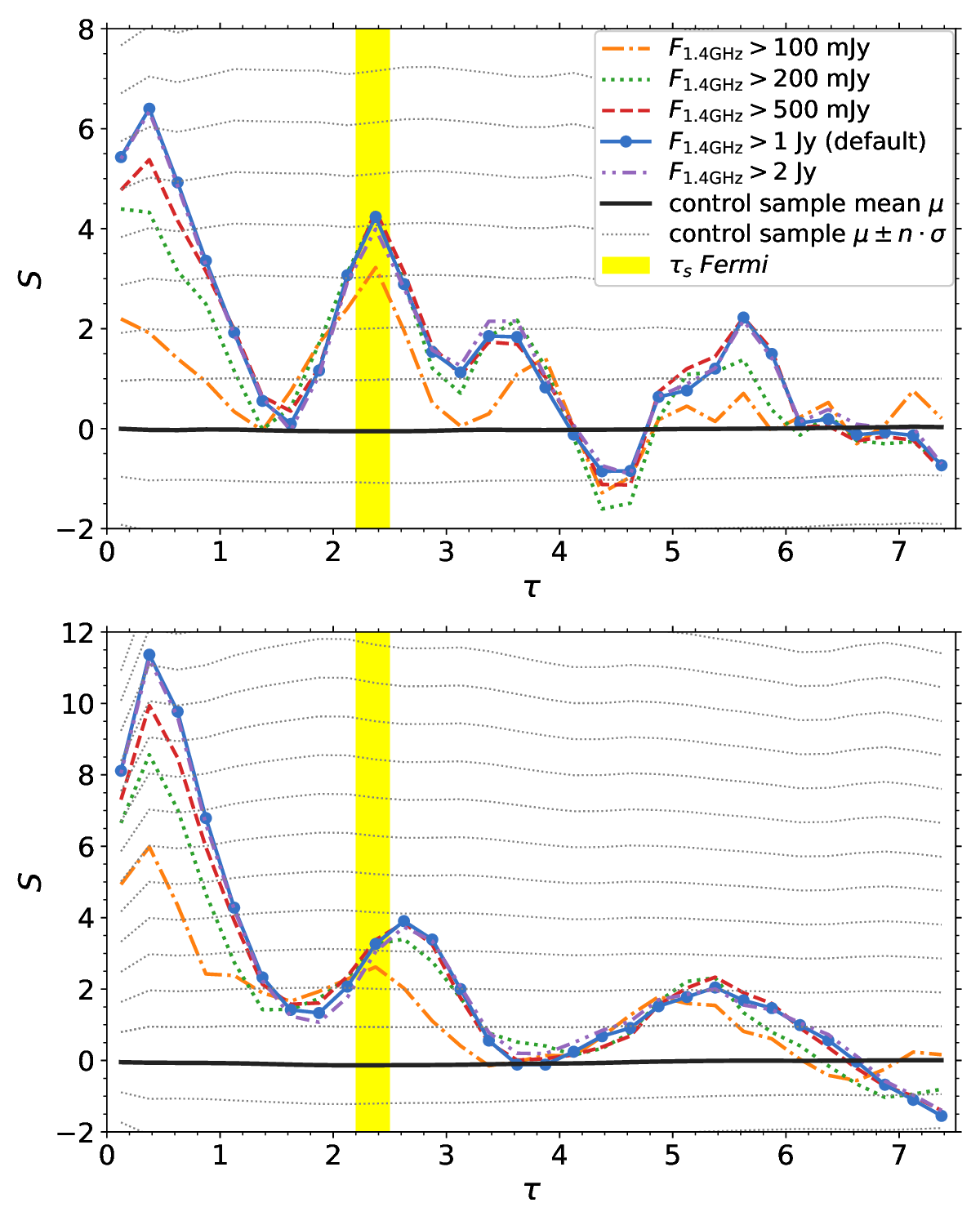}
    \caption{
    Same as Fig.~\ref{fig:theta_500_test}, but varying the point-source flux cut:
    $F_{\rm 1.4GHz} > 100$ mJy (dot-dashed orange curve), $F_{\rm 1.4GHz} > 200$ mJy (dotted green), $F_{\rm 1.4GHz} > 500$ mJy (dashed red), $F_{\rm 1.4GHz} > 1$ Jy (default; solid blue), and $F_{\rm 1.4GHz} > 2$ Jy (double dot-dashed purple).
    }
    \label{fig:psc_test}
\end{figure}

Figure \ref{fig:psc_test} demonstrates the effect of varying the cut on the NVSS point-source flux, in the range 100 mJy to 2 Jy.
Masking pixels within $0\dgr.3$ around NVSS point sources with $F_{\rm 1.4 GHz} > 200$ mJy or $F_{\rm 1.4 GHz} > 500$ mJy does not significantly affect the virial shock signal, but it does diminish the central signal.
The approximate independence of the virial excess upon the NVSS flux cut indicates either that CLEAN has successfully removed point sources down to the $F_{\rm 1.4 GHz} \sim 200$ mJy level within a radius $0\dgrdot3$, or that point sources neither contaminate the virial excess nor contribute to it;
the change in central excess would favour the latter.
The virial excess is still present when the cut is lowered further, to $100$ mJy, but becomes fainter, probably due to the excessive masking of pixels: here, only $\sim 60\%$ of the initial pixels in the virial bin remain unmasked, compared to $\sim 85\%$ ($\sim99\%$) for the $F_{\rm 1.4 GHz} = 200$ (1000) mJy cut.

\begin{figure}
    \centering
    \includegraphics[width=0.45\textwidth,trim={0 0.5cm 0 0},clip]{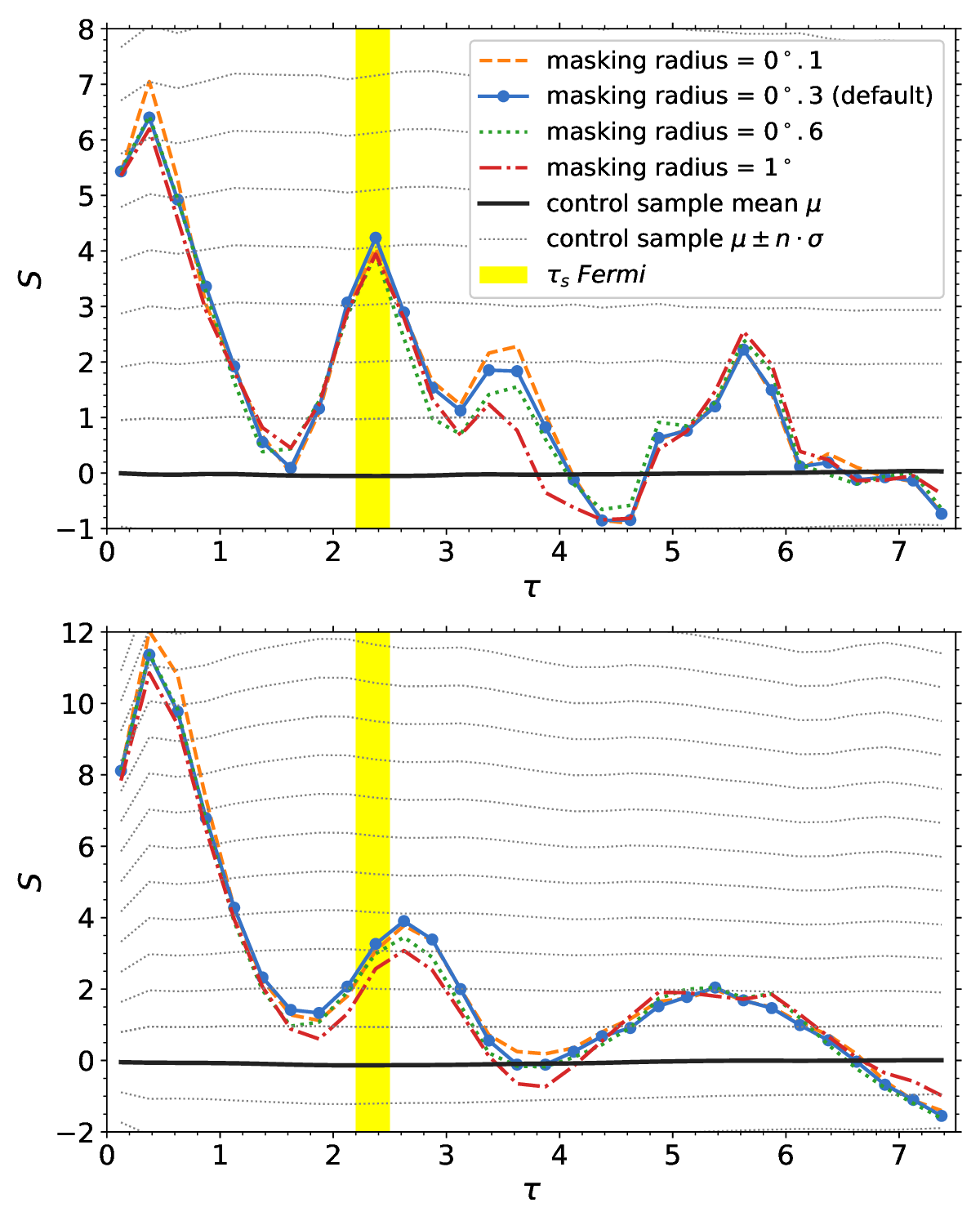}
    \caption{
    Same as Fig.~\ref{fig:theta_500_test}, but varying the point source masking radius:
    $0\dgrdot1$ (dashed orange curve), $0\dgrdot3$ (default; solid blue), $0\dgrdot6$ (dotted green), and $1\dgrdot$ (dash-dotted red).
    }
    \label{fig:psc_maskinglength_test}
\end{figure}

Figure \ref{fig:psc_maskinglength_test} demonstrates the dependence of the results upon the radius within which HEALPix pixels are masked around NVSS point sources. Varying the masking radius in the range $0\dgrdot1$ to $0\dgrdot6$ has a negligible effect on the results, but further raising the masking radius removes enough pixels to start diminishing the virial excess, especially at low frequencies.

\begin{figure}
    \centering
    \includegraphics[width=0.45\textwidth,trim={0 0.5cm 0 0},clip]{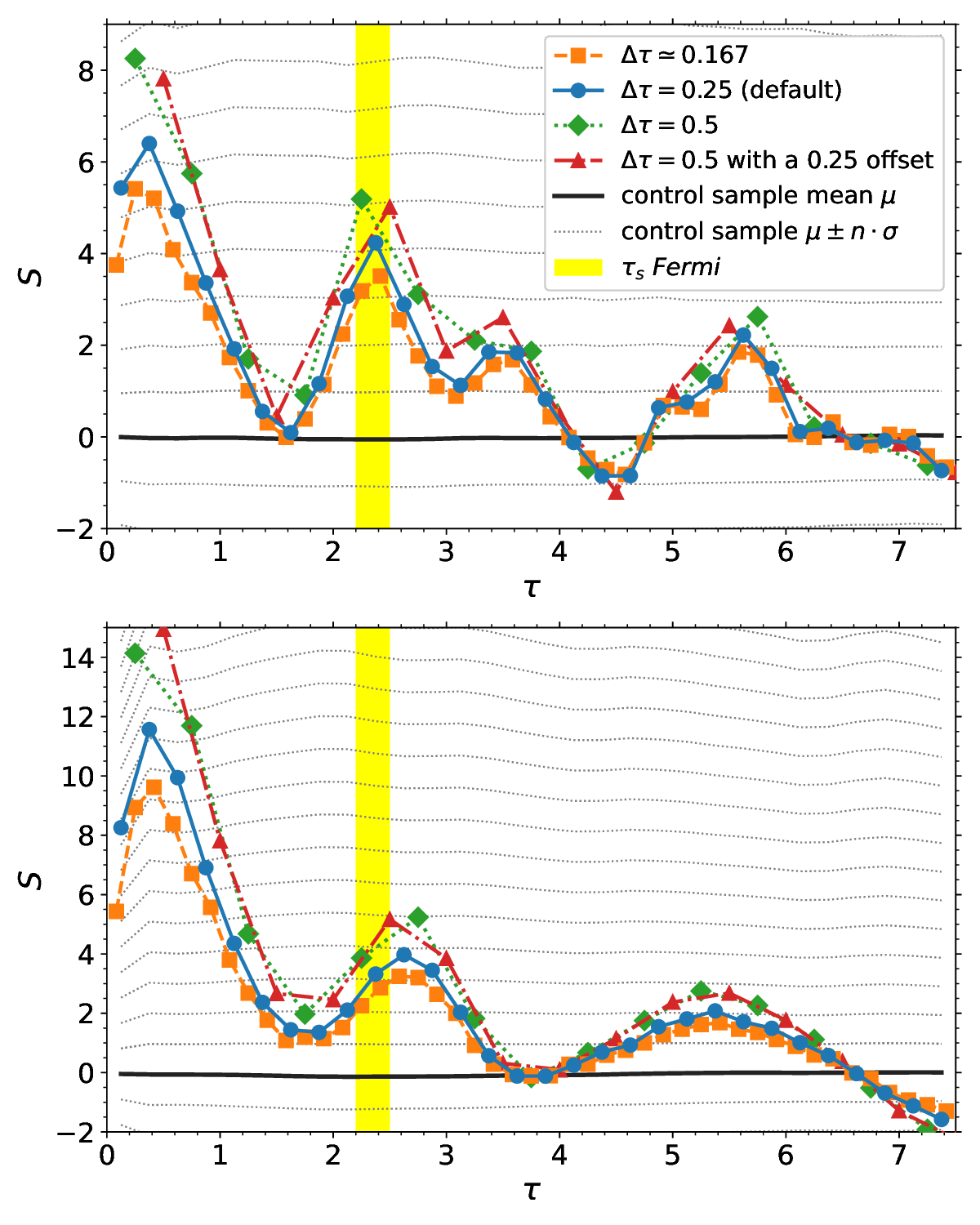}
    \caption{Same as Fig.~\ref{fig:theta_500_test}, but varying the radial bin size: $\Delta\tau \simeq 0.167$ (dashed orange curve), $\Delta\tau = 0.25$ (default: solid blue), and $\Delta\tau = 0.5$ (dotted green). A bin size $\Delta\tau = 0.5$ with a $0.25$ offset from the origin (dot-dashed red) is also shown.
    }
    \label{fig:bins}
\end{figure}

Figure \ref{fig:bins} demonstrates variations in the radial bin size, from $\Delta\tau \simeq 0.167$ (six bins inside per $\tau$) to $\Delta\tau=0.5$ (two bins per $\tau$).
The $S(\tau)$ profiles of different resolutions are consistent with each other, and show that both central and virial excess signals are extended, as their significance increases as the bins are enlarged beyond the beam size.
The peak of the excess slightly shifts between high and low frequencies at nominal and low resolutions, but only between adjacent radial bins, so we cannot substantiate any frequency dependence.

\begin{figure}
    \centering
    \DrawFig{\includegraphics[width=0.45\textwidth,trim={0 0.5cm 0 0},clip]{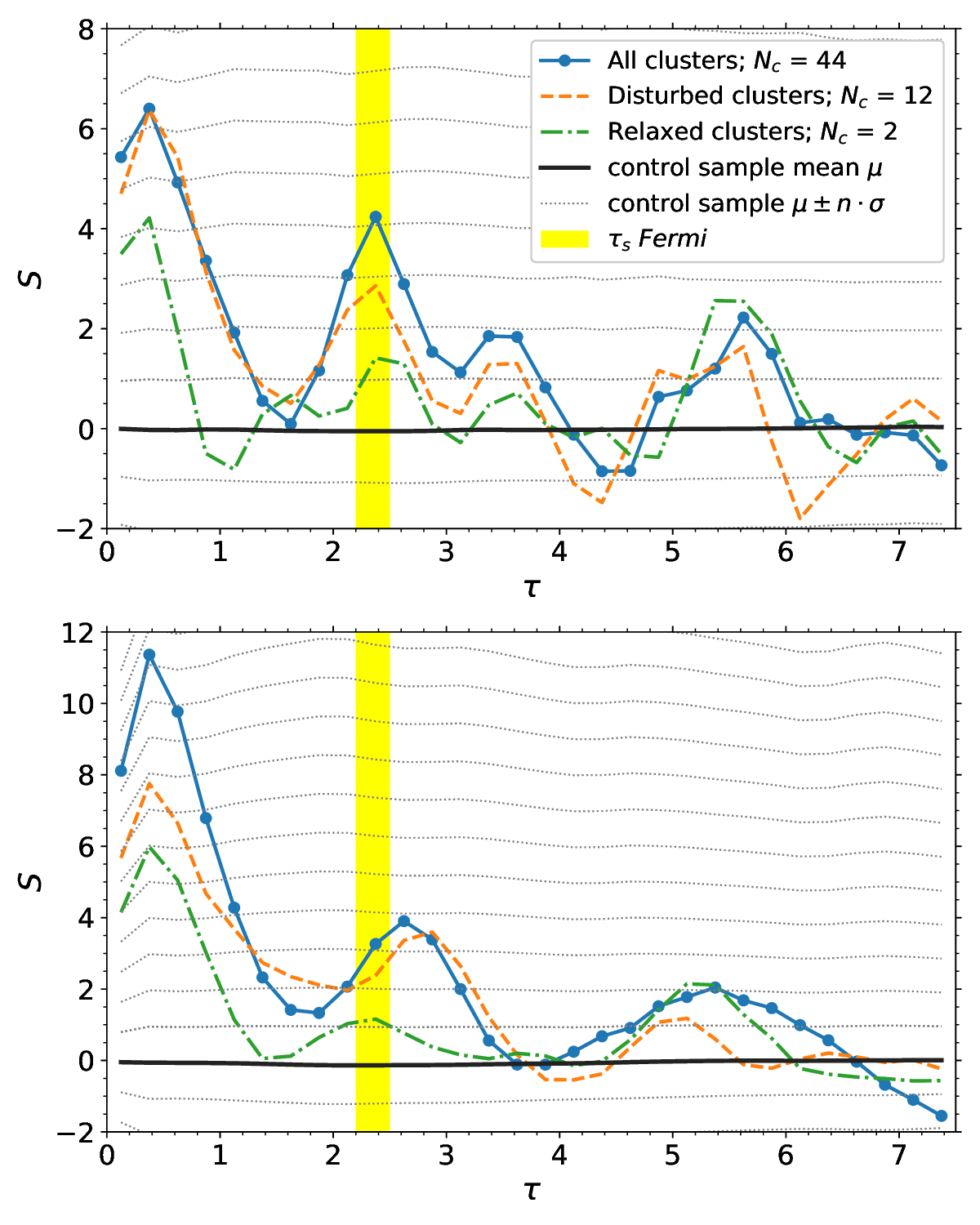}}
    \caption{
	Same as Fig.~\ref{fig:theta_500_test}, but showing, in addition to the nominal (solid blue curves) sample, also sub-samples consisting of the 12 disturbed (dashed orange) and the two relaxed (dot-dashed green) clusters. See Table \ref{tab:MCXC_sample} for cluster classification.
    }
    \label{fig:disturbed}
\end{figure}

Out of the 44 clusters in our sample, 12 can be presently classified as perturbed (due to long cooling times, an absent cool-core, or an identified merger), and two as relaxed. 
One expects the peripheral, virial excess to be fairly indifferent to the dynamical state of the cluster, whereas the central signal should be modified, for example become more extended, in perturbed clusters. 
To examine the effect of the dynamical state of the cluster, Fig.~\ref{fig:disturbed} shows the radial significance profiles obtained separately for the 12 disturbed clusters and for the two relaxed clusters. 
Both sub-samples show central and virial signals, but due to the small sample sizes, we cannot identify any significant differences between sub-samples.

\section{Virial shock Model}
\label{append:model}

We generalise the virial shock model of {\RK} for radio emission, using the same underlying assumptions but incorporating also the magnetic field and the resulting synchrotron emission, recovering Eq.~(\ref{eq:shock_prop}) used in \S\ref{sec:virialemission}.

To determine the synchrotron luminosity of the virial shock, we first derive an approximation for the accretion rate through the shock and its downstream conditions.
The thermal plasma is assumed neutral, with a particle number density following
an isothermal $\beta$-model,
\begin{align}
    n(r) = n_{0} \left[ 1 + \left(\frac{r}{r_c}\right)^2 \right]^{-3\beta/2} \coma
    \label{eq:beta_model}
\end{align}
where $n_{0}$ is the central
number density, $r_c$ is the core radius, and $\beta$ is the slope parameter.
Approximating the cluster as in hydrostatic equilibrium, the total (gravitating) mass inside a radius $r$ can be written as
\begin{equation}
    M(r) \simeq \frac{3 \beta k_B T r}{G \bar{m}} \left(1 + \frac{r_c^2}{r^2} \right)^{-1} \coma
    \label{eq:hydro_eq_mass}
\end{equation}
where $G$ is Newton’s constant.
As the typical core radius $r_c \sim 0.1 R_{500}$ is much smaller than the virial shock position at $r\gtrsim 2 R_{500}$, to study the distribution near the shock one can neglect the core and approximate $n(r) \simeq n_{0} (r/r_c)^{-3 \beta}$.
As $r \gg r_c$, the mass can be approximated as
\begin{equation}
    M(r) \simeq \frac{3 \beta k_B T}{G \bar{m}}r \fin
    \label{eq:hydro_eq_appr_mass}
\end{equation}
This approximation yields the enclosing radius
\begin{equation}
    R_\delta = \left[\frac{9 \beta k_B T}{4 \pi \rho_c(z) \delta G \bar{m}}\right]^{1/2}
    \label{eq:hydro_eq_appr_mass}
\end{equation}
and the mass-temperature relation
\begin{align} \label{eq:MassTemperatureRelation}
M_\delta = \frac{9}{2 \sqrt{\pi \rho_c(z) \delta}} \left( \frac{\beta k_B T}{G \bar{m}}\right)^{3/2} \coma
\end{align}
where
$\delta$ is the over-density parameter defining a radius $R_\delta$ and an enclosed mass $M_\delta$, such that the mean enclosed mass density  $M_\delta / [ (4/3) \pi R_\delta^3 ]$.
The critical mass density is defined as $\rho_c(z) \equiv \rho_0 \mathcal{H}^2$, where $\rho_0 = 3 H_0^2 / (8 \pi G)$
is the present critical mass density, $\mathcal{H} \equiv H(z)/H_0 \simeq [(1-\Omega_m) + (1+z)^3\Omega_m]^{1/2}$ describes the evolution of the Hubble constant, and $\Omega_m$ is the matter fraction of the Universe.

The MCXC catalogue provides $M_{500}$ for all clusters, but $T$ for only some of them, so we use the mass--temperature relation (\ref{eq:MassTemperatureRelation}) to estimate $T_d$, the downstream temperature.
The downstream particle number density,
\begin{equation}
    n_d = (1-\beta) \frac{f_b \rho_c(z)}{\bar{m}} \delta_{\sh} \coma
    \label{eq:down_par_density}
\end{equation}
is estimated by assuming that the total baryon mass $f_b M_{\sh}$ is given by the spatial integral of the isothermal $\beta$-profile (\ref{eq:beta_model}),
and quantities with subscript $\sh$ are evaluated at the shock radius.
The magnetic field can then be parameterised by assuming that magnetic energy is a fraction $\xi_B$ of the downstream thermal energy,
\begin{equation}
    \frac{B^2}{8 \pi} \equiv \xi_B \frac{3}{2} n_d k_B T_d \,,
    \label{eq:xiB_def}
\end{equation}
where we assumed that the adiabatic index is $5/3$.

Assuming that a fraction $\xi_e$ of the thermal energy density downstream of the shock is deposited in CREs of spectral index $p$, solving for the steady-state CRE distribution yields \citep[\eg][]{KeshetEtAl03}
\begin{equation}\label{eq:CRESpectrum}
\frac{dN_e}{d\gamma} \simeq \myK \gamma^{-(p+1)} \coma
\end{equation}
assumed to prevail over Lorentz factors $\gmin<\gamma<\gmax$.
The effective value of $\gmin$ near or below the cooling break has little effect on our results, so is taken for simplicity as unity.
The maximal CRE energy is determined by the balance between acceleration time and Compton cooling,
giving
\citep[\eg][]{KeshetEtAl03,ReissKeshet18}
\begin{equation}
    \gmax \simeq 8.2 \times 10^{7} \left( \xi_{B1} \delta_{100} \right)^{1/4} M_{14}^{1/2} \coma
\end{equation}
where
$\delta_{100}\equiv 100\delta_{\sh}\simeq 500\tau_{\sh}^{-2}$,
$\xi_{B1}\equiv \xi_B/0.01$, and we took $\beta=2/3$. For simplicity, we adopt  $\gmax = 10^8$.
The volume-integrated normalisation in Eq.~(\ref{eq:CRESpectrum}) is given by
\begin{equation} \label{eq:myK}
    \myK=\frac{9 k_B T \xi_e}{8(p-1)c u_{cmb} \sigma_T}\frac{f_b \dot{M_{\sh}}}{\bar{m}}\times
    \begin{cases}
    \frac{1}{\ln(\gmax/\gmin)} & \mbox{for $p=2$\,;} \\ \\
    \frac{p-2}{\gminP{2-p} - \gmaxP{2-p}} & \mbox{for $p \neq 2$\,,}
    \end{cases}
\end{equation}
where $\sigma_T$ is the Thomson cross section,
and we assumed that CRE cooling is dominated by inverse-Compton emission off CMB photons.
To estimate $f_b \dot{M_{\sh}} / \bar{m}$, \ie the number accretion rate of gas particles through $R_{\sh}$,
we parameterise the accretion rate as proportional to $M_{\sh}$ by introducing the dimensionless accretion parameter
\begin{equation}
    \dot{m} \equiv \frac{\dot{M_{\sh}}}{M_{\sh} H(z) }\fin
    \label{eq:accretion_rate}
\end{equation}

For the steady state CRE spectral index $p+1$, the luminosity of synchrotron radiation is \citep[\eg][]{RybickiLightman86}
\begin{equation}
L_\nu = \alpha_e h\nu\,\myK \Phi q(p)\left( \frac{3\nu_B}{\nu}\right)^{1+\frac{p}{2}} {\rm \ erg\ s^{-1}} \coma
\label{eq:symRL}
\end{equation}
where $\alpha_e=e^2/(\hbar c)$ is the fine-structure constant, $h=2\pi\hbar$ is Planck's constant, $e$ is the electron charge, $m_e$ is the electron mass, $\nu_B=e B/(2\pi m_e c)$ is the cyclotron frequency,
$\Gamma(y)$ is the gamma function,
and the $p$-dependent numerical factor
\begin{equation}
q(p)\equiv \frac{ \,\Gamma\left( \frac{p}{4} + \frac{11}{6}\right) \Gamma\left( \frac{p}{4} + \frac{1}{6}\right)}{(p+2)\sqrt{3}}\,.
\end{equation}
The dependence $\Phi \equiv \sin^{(p+2)/2}{\phi}$ upon the pitch angle $\phi$
is isotropically averaged to give
\begin{equation}
 \langle\Phi\rangle = \frac{\sqrt{\pi}\,\Gamma\left(\frac{p}{4}+\frac{3}{2}\right)}{2 \Gamma\left(\frac{p}{4}+2\right)}  \fin
\label{eq:pitch_angle}
\end{equation}

The specific energy flux can now be estimated as
\begin{equation}
    \label{eq:FluxGeneral}
    F_\nu = \frac{L_\nu}{4 \pi d^2_L} =  \frac{\theta_\delta^2 L_\nu}{4 \pi (1+z)^4 R_\delta^2}\coma
\end{equation}
where $d_L$ is the luminosity distance.
The specific brightness is then given by
\begin{equation} \label{eq:Inu}
I_\nu = F_\nu f(\theta) = \frac{F_\nu {\tilde{f}}(\tilde{\theta})}{\theta_{\sh}^2}
= \frac{L_\nu \tilde{f}(\tilde{\theta})}{4 \pi^2 (1+z)^4 R_{\sh}^2}\coma
\end{equation}
where $\tilde{\theta}\equiv \theta/\theta_{\sh}$. The functions $f(\theta)$ or equivalently $\tilde{f}(\tilde{\theta})=\theta_{\sh}^2f(\theta)$ describe the brightness distribution across the cluster, and are normalised to unity,
\begin{equation}
\int f(\theta)\,d\Omega = 2\pi \int_0^{\infty} \tilde{f}(\tilde{\theta}) \tilde{\theta} \,d\tilde{\theta} = 1 \fin
\end{equation}
In the limit of emission in a thin ring in the plane of the sky, one has $\tilde{f}=\delta(\tilde{\theta}-1)/2\pi$.
The function $\tilde{f}(\tilde{\theta})$ was derived for the projection of a finite shell with an evolving CRE spectrum in \citet{2018ApJ...869...53K}; in the thin shell limit, the result reduces to $\tilde{f}=(1-\tilde{\theta}^2)^{-1/2}\Theta(1-\tilde{\theta})/(2\pi)$.

For completeness, we combine Eqs. (\ref{eq:hydro_eq_appr_mass})--(\ref{eq:FluxGeneral}) to derive the flux in an explicit form.
In the strong shock limit $p=2$, we find
\begin{equation}
    \begin{aligned}
    \nu F_\nu & \simeq \frac{1-\beta}{12\beta^2}\left(\frac{\pi}{6}\right)^{\frac{1}{3}} \frac{(f_b G)^2\rho_0^{\frac{7}{3}}H_0\mathcal{H}^{\frac{17}{3}}}{u_{cmb} \ln(\gmax/\gmin)}\delta_{\sh}^{\frac{1}{2}}\dot{m}\xi_e\xi_B \frac{M_\delta^{\frac{5}{3}} \theta_\delta^2 \delta^{\frac{11}{6}}}{(1+z)^4} \\
    & \simeq  4.6\times 10^{-11} \frac{M_{14}^{\frac{5}{3}} \theta_{0.2}^2 \mathcal{H}^{\frac{17}{3}} \delta_{100}^{\frac{1}{2}} \dot{m}\xi_e\xi_B }{(1+z)^4 \ln(\gmax/\gmin)}
    {\rm \ erg\ s^{-1}\ cm^{-2}}
    \coma
    \end{aligned}
\label{eq:shock}
\end{equation}
where we defined $\delta_{100}\equiv \delta_{\sh}/100$ and $\theta_{0.2}\equiv\theta_{500}/0\dgrdot2$, and in the second line we took the typical $\beta=2/3$.
For an arbitrary $p\neq 2$, we obtain
\begin{equation}
    \begin{aligned}
    \nu F_\nu & \simeq
    \frac{(1-\beta)^{\frac{2+p}{4}}(p-2)\Gamma\left(\frac{p}{4}+\frac{3}{2}\right)\Gamma\left(\frac{1}{6}+\frac{p}{4}\right)\Gamma\left(\frac{11}{6}+\frac{p}{4}\right)} {(\beta/3)^{\frac{6+p}{4}}2^{\frac{34-p}{6}}\left(\frac{\pi}{3}\right)^{\frac{p-1}{6}}(p+2)(p-1)\Gamma\left(\frac{p}{4}+2\right)} \\
    & \quad \times  \frac{\left(\frac{e}{m_e c}\right)^{\frac{p-2}{2}}(f_b G)^{\frac{6+p}{4}}\rho_0^{\frac{5+p}{3}}H_0\mathcal{H}^{\frac{13+2p}{3}}}{u_{cmb} \left(\gmin^{2-p}-\gmax^{2-p}\right)} \delta_{\sh}^{\frac{p}{4}}\dot{m}\xi_e\xi_B^{\frac{p+2}{4}}\\
    & \quad \times  \frac{M_\delta^{\frac{8+p}{6}} \theta_\delta^2 \delta^{\frac{20+p}{12}}}{(1+z)^4} \nu^{-\frac{p-2}{2}} \\
    & \simeq  7.4\times 10^{-12}
    \frac{(p-2)\Gamma\left(\frac{p}{4}+\frac{3}{2}\right)\Gamma\left(\frac{1}{6}+\frac{p}{4}\right)\Gamma\left(\frac{11}{6}+\frac{p}{4}\right)} {0.22^p (p+2)(p-1)\Gamma\left(\frac{p}{4}+2\right)} \\
    & \quad \times \frac{M_{14}^{\frac{8+p}{6}} \theta_{0.2}^2 \mathcal{H}^{\frac{13+2p}{3}} \delta_{100}^{\frac{p}{4}}\dot{m}\xi_e\xi_B^{\frac{2+p}{4}}}{(1+z)^4 \left(\gmin^{2-p}-\gmax^{2-p}\right)\nu^{\frac{p-2}{2}}}
    {\rm \ erg\ s^{-1}\ cm^{-2}}
    \coma
    \end{aligned}
\label{eq:shock}
\end{equation}
again taking $\beta=2/3$ in the second equality.

The last result can be combined with Eq.~\eqref{eq:Inu} in the form
\begin{equation}
    \begin{aligned}
    \nu I_\nu & \simeq 2.1\times 10^{-9}
    \frac{(p-2)\Gamma\left(\frac{p}{4}+\frac{3}{2}\right)\Gamma\left(\frac{1}{6}+\frac{p}{4}\right)\Gamma\left(\frac{11}{6}+\frac{p}{4}\right)}
    {0.047^p (p+2)(p-1)\Gamma\left(\frac{p}{4}+2\right)} \\
    & \quad \times \frac{M_{14}^{\frac{8+p}{6}} \mathcal{H}^{\frac{13+2p}{3}} \tau_{\sh}^{-\frac{p+4}{2}}\dot{m}\xi_e\xi_B^{\frac{2+p}{4}}\tilde{f}(\tilde{\theta})}
    {(1+z)^4 \left(\gmin^{2-p}-\gmax^{2-p}\right)\nu^{\frac{p-2}{2}}}
    \erg\se^{-1}\cm^{-2}\sr^{-1}
    \coma
    \end{aligned}
\label{eq:shock_nuInu}
\end{equation}
which is used in Eq.~\eqref{eq:shock_prop}.
In the strong shock limit, this result reduces to
\begin{equation}
    \nu I_\nu  \simeq 2.9\times 10^{-7}
    \frac{M_{14}^{\frac{5}{3}} \mathcal{H}^{\frac{17}{3}} \tau_{\sh}^{-3}\dot{m}\xi_e\xi_B\tilde{f}(\tilde{\theta})}
    {(1+z)^4 \ln(\gmax/\gmin)}
    \erg\se^{-1}\cm^{-2}\sr^{-1}
    \fin
    \label{eq:StrongShock_nuInu}
\end{equation}

\section{Mutual effects of central and virial modelling}
\label{append:mutual_effects}

For better accuracy, the results in \S\ref{subsec:model_high_nu} and \ref{subsec:model_low_nu} were obtained by simultaneously modelling the central and virial signals. However, as the spatial overlap between the two signals is limited, our qualitative conclusions are robust to the analysis details. In particular, modelling each signal separately yields similar results, with only a small enhancement in the individual signal normalisations.

For example, one can approximately neglect the effect of the central signal at large radii beyond, say, $\tau=1.5$.
We thus fit the virial shock model
(in the strong shock limit $p = 2$) alone, by focusing only on data outside $\tau\simeq 1.5$. The high-frequency nominal planar model gives
\begin{equation}
    \tau_s = 2.40^{+0.08}_{-0.08} \coma
\end{equation}
consistent with Eq.~\eqref{eq:best_tau_s}, and
\begin{equation}
    \dot{m}\xi_e\xi_B = 3.99^{+0.62}_{-0.62} \times 10^{-4}\coma
\end{equation}
$\sim 40\%$ higher than Eq.~\eqref{eq:best_xi},
with $\mbox{TS}\simeq 41.5$ ($6.1\sigma$ for $\DF=2$).
The same model in the low-frequency channels yields
\begin{equation}
    \tau_s \simeq 2.56^{+0.09}_{-0.09} \coma
\end{equation}
consistent with Eq.~\eqref{eq:best_tau_s_lowf}, and
\begin{equation}
    \dot{m}\xi_e\xi_B = 2.35^{+0.36}_{-0.36} \times 10^{-4}\coma
\end{equation}
$\sim 20 \%$ higher than Eq.~\eqref{eq:best_xi_lowf}, with $\mbox{TS}\simeq 42.4$ ($6.2\sigma$ for $\DF=2$).
Similar results are obtained for the shell model, with normalisation enhancements varying between $20\%$ and $40\%$ for different model variants.

An alternative, more accurate possibility is to freeze the central model parameters to the values inferred in \S\ref{append:CentralResults}, and then model only the virial signal.
The resulting parameter values are consistent with those inferred above, within the statistical uncertainties.
In particular, in the high-frequency channel, $\tau_s$ shifts by $\lesssim 2 \%$ ($5\%$) and $\dot{m}\xi_e\xi_B$ decreases by $\lesssim 20 \%$ (10\%) for the {\planar} (shell) model.
In low-frequency channels, $\tau_s$ shifts by $\lesssim 1 \%$ ($1\%$) and $\dot{m}\xi_e\xi_B$ diminishes by $\lesssim 10 \%$ ($25 \%$) for the planar (shell) model.

The opposite effect, of the virial excess biasing the central model parameters, can be inferred for the cleared data by comparing the best-fitting parameters obtained with (Table \ref{tab:fitting_virial_detail}) and without (Table \ref{tab:fitting_centre}) the virial shock modelling.
When modelling the central and virial signals simultaneously, we obtain central model parameters comparable to those obtained while neglecting the virial signal in \S\ref{append:CentralResults}, with central flux normalisations smaller than those of Table \ref{tab:fitting_centre} by $\lesssim 10\%$ ($15\%$) for the {\planar} and (shell) model in both high and low frequencies, and a central spectrum
\begin{equation}
    \alpha_e = 1.20^{+0.27}_{-0.26} ~ (1.14^{+0.28}_{-0.26})
\end{equation}
consistent with Eq.~\eqref{eq:alpha_ext}.

\section{Shock parameter degeneracy}
\label{append:decouple_xib}

In the limit of a strong shock, where $p=2$, the radio signal provides a measurement of $\dot{m}\xi_e\xi_B$.
The degeneracy between the three terms in this product can be broken under some assumptions on the gas profile.
If the parameters of the isothermal $\beta$-model are known, and the model can be extrapolated (with some correction) to the cluster periphery, then the density $n_d$ and temperature $T_d$ downstream of the virial shock can be inferred. The downstream velocity is then
\begin{equation}
    v_d \simeq \left(\frac{k_B T_d}{3 \bar{m} }\right)^{1/2} \coma
\end{equation}
so the accretion parameter can be estimated from the definition \eqref{eq:accretion_rate},
\begin{equation}
    \dot{m} \simeq \frac{4 \pi R_{\sh}^2 \bar{m} n_d v_d}{f_b M_{\sh} H_0 \mathcal{H}} \coma
    \label{eq:accretion_rate1}
\end{equation}
facilitating a measurement of the product $\xi_e\xi_B$.
A similar approach was used when combining the galaxy distribution around A2319 \citep{HurierEtAl19} with the product $\dot{m}\xi_e$ inferred from the leptonic signal, giving $\xi_e\simeq 0.5\%$ and $\dot{m}\simeq 1.1$ separately \citep{2018ApJ...869...53K}.

One way to break the degeneracy between $\xi_e$ and $\xi_B$ is to assume equipartition, \ie $\xi_e=\xi_B$.
An alternative with no additional assumption (in the one-zone model) is to use the $\dot{m}\xi_e\simeq 0.6\%$ estimates based on modelling the leptonic signals from the virial shock \citep{2017ApJ...845...24K, 2018ApJ...869...53K, ReissKeshet18}.
After this degeneracy is broken, giving a measure of $\xi_B$, one may equivalently estimate $B$, as $n_d$ and $T_d$ were already determined; see Eq.~(\ref{eq:xiB_def}).
For clusters with available $\beta$-model parameters, typically estimated from the X-ray distribution at $r<R_{500}$, we extrapolate the model to the virial shock radius but assume that by $r_{\sh}$, both temperature and density are diminished by an additional factor $f_\beta\simeq 1/3$.
Alternatively, and without a $\beta$-model, we may estimate $B$ by assuming a $\beta=2/3$ model and substituting $T_d$ using  Eq.~(\ref{eq:MassTemperatureRelation}) and $n_d$ using Eq.~(\ref{eq:down_par_density}) (no $f_\beta$ correction needed), in which we invoke hydrostatic equilibrium and assume $r \gg r_c$.

\section{Cluster sample}
\label{append:MCXC_clusters_list}
The clusters in our sample are listed in Table \ref{tab:MCXC_sample}, based on the MCXC catalogue supplemented by the estimated $\theta_{500}$ values and the dynamical state of the cluster (when available).

\begin{table*}
	\caption{Cluster sample.}
    \centering
	\label{tab:MCXC_sample}
	\begin{tabular}{llcccccccc}
		MCXC Name        & Other Name      &    $b$ &    $l$ &   $M_{500}$ &   $R_{500}$ &   $\theta_{500}$ &   $L_X$ &    $z$ & Dyn. state \\
		(1)        & (2)      & (3) & (4) &  (5) &  (6) &   (7) &  (8) &  (9) & (10) \\
		\hline
		 MCXCJ1204.1+2020 & NGC4066        &  77.25 & 242.53 &        0.32 &        0.48 &             0.26 &   4.880 & 0.0252 & \\
		 MCXCJ2336.5+2108 & A2626          & -38.44 & 100.45 &        1.81 &        0.84 &             0.21 &  86.889 & 0.0565 & R$^a$ \\
		 MCXCJ1109.7+2145 & A1177          &  66.28 & 220.43 &        0.46 &        0.54 &             0.24 &   9.028 & 0.0319 &  \\
		 MCXCJ1122.3+2419 & HCG51          &  69.76 & 215.92 &        0.25 &        0.44 &             0.24 &   3.302 & 0.0258 & D$^b$ \\
		 MCXCJ0838.1+2506 & CGCG120-014    &  33.73 & 199.58 &        0.33 &        0.48 &             0.23 &   5.181 & 0.0286 &  \\
		 MCXCJ0036.5+2544 & ZwCl193        & -37.01 & 118.75 &        0.49 &        0.55 &             0.22 &   9.881 & 0.0341 &  \\
		 MCXCJ1348.8+2635 & A1795          &  77.18 &  33.82 &        5.53 &        1.22 &             0.28 & 547.807 & 0.0622 & R$^c$ \\
		 MCXCJ0058.9+2657 & RXJ0058.9+2657 & -35.89 & 124.99 &        0.82 &        0.65 &             0.20 &  23.388 & 0.0451 &  \\
		 MCXCJ2338.4+2700 & A2634          & -33.09 & 103.48 &        1.22 &        0.75 &             0.34 &  44.138 & 0.0309 & D$^d$ \\
		 MCXCJ1522.4+2742 & A2065          &  56.62 &  42.84 &        3.51 &        1.05 &             0.21 & 262.787 & 0.0723 & D$^c$ \\
		 MCXCJ1658.0+2751 & AWM5           &  35.93 &  49.02 &        0.71 &        0.62 &             0.26 &  18.244 & 0.0337 &  \\
		 MCXCJ1359.2+2758 & A1831          &  74.95 &  40.07 &        1.98 &        0.87 &             0.20 & 101.656 & 0.0612 &  \\
		 MCXCJ1206.6+2811 & NGC4104        &  80.02 & 204.27 &        0.42 &        0.52 &             0.26 &   7.641 & 0.0283 &  \\
		 MCXCJ0228.1+2811 & RXJ0228.2+2811 & -30.01 & 147.57 &        0.66 &        0.61 &             0.24 &  16.276 & 0.0353 &  \\
		 MCXCJ1110.7+2842 & A1185          &  67.75 & 202.97 &        0.60 &        0.59 &             0.26 &  13.790 & 0.0314 & D$^a$  \\
		 MCXCJ0828.6+3025 & A0671          &  33.15 & 192.75 &        1.15 &        0.73 &             0.21 &  41.316 & 0.0503 & D$^a$ \\
		 MCXCJ0200.2+3126 & NGC0777        & -29.18 & 139.74 &        0.28 &        0.46 &             0.38 &   4.025 & 0.0168 &  \\
		 MCXCJ0150.7+3305 & A260           & -28.16 & 137.01 &        0.63 &        0.60 &             0.23 &  14.929 & 0.0363 &  \\
		 MCXCJ1320.2+3308 & NGC5098        &  81.35 &  78.68 &        0.52 &        0.56 &             0.22 &  10.908 & 0.0362 &  \\
		 MCXCJ0110.9+3308 & NGC0410        & -29.54 & 127.63 &        0.22 &        0.42 &             0.32 &   2.545 & 0.0177 &  \\
		 MCXCJ0919.8+3345 & A779           &  44.40 & 191.08 &        0.26 &        0.45 &             0.27 &   3.442 & 0.023  & D$^d$ \\
		 MCXCJ0933.4+3403 & UGC05088       &  47.24 & 191.05 &        0.21 &        0.41 &             0.21 &   2.385 & 0.0269 &  \\
		 MCXCJ1334.3+3441 & NGC5223        &  78.09 &  74.98 &        0.25 &        0.44 &             0.25 &   3.318 & 0.024  &  \\
		 MCXCJ1617.4+3456 & NGC6107        &  45.67 &  56.28 &        0.67 &        0.61 &             0.27 &  16.810 & 0.0315 &  \\
		 MCXCJ1740.5+3538 & RXJ1740.5+3539 &  29.07 &  60.60 &        0.91 &        0.68 &             0.22 &  28.023 & 0.0428 &  \\
		 MCXCJ0246.0+3653 & A0376          & -20.55 & 147.11 &        1.61 &        0.81 &             0.24 &  71.362 & 0.0488 & D$^e$ \\
		 MCXCJ1742.8+3900 &                &  29.41 &  64.47 &        0.77 &        0.64 &             0.21 &  21.333 & 0.0423 &  \\
		 MCXCJ1205.2+3920 & RXJ1205.1+3920 &  74.45 & 158.23 &        1.03 &        0.70 &             0.26 &  34.025 & 0.0381 &  \\
		 MCXCJ1627.6+4055 & A2197          &  43.90 &  64.83 &        0.39 &        0.51 &             0.23 &   6.744 & 0.0301 & D$^d$ \\
		 MCXCJ1627.3+4240 & A2192          &  43.93 &  67.27 &        0.38 &        0.51 &             0.22 &   6.551 & 0.0317 &  \\
		 MCXCJ1714.3+4341 & NGC6329        &  35.42 &  68.95 &        0.31 &        0.48 &             0.24 &   4.769 & 0.0276 &  \\
		 MCXCJ1733.0+4345 & IC1262         &  32.07 &  69.52 &        0.86 &        0.66 &             0.30 &  24.990 & 0.0307 &  \\
		 MCXCJ1134.8+4903 & A1314          &  63.56 & 151.79 &        0.46 &        0.54 &             0.22 &   9.025 & 0.0341 & D$^a$ \\
		 MCXCJ0907.8+4936 & VV196          &  42.12 & 169.27 &        0.40 &        0.51 &             0.20 &   7.063 & 0.0352 &  \\
		 MCXCJ1811.0+4954 & ZwCl8338       &  26.71 &  77.72 &        1.35 &        0.77 &             0.22 &  53.405 & 0.0501 &  \\
		 MCXCJ0751.3+5012 & UGC04052       &  29.84 & 168.39 &        0.42 &        0.52 &             0.32 &   7.650 & 0.0228 &  \\
		 MCXCJ1649.2+5325 & ARP330,SHK016  &  39.48 &  81.26 &        0.30 &        0.47 &             0.22 &   4.539 & 0.029  &  \\
		 MCXCJ0740.9+5525 & UGC03957       &  28.93 & 162.22 &        1.29 &        0.76 &             0.31 &  48.840 & 0.034  &  \\
		 MCXCJ0721.3+5547 & A0576          &  26.25 & 161.36 &        1.68 &        0.83 &             0.31 &  75.714 & 0.0381 & D$^{f,\dagger}$ \\
		 MCXCJ1723.3+5658 & NGC6370        &  34.34 &  85.21 &        0.23 &        0.43 &             0.22 &   2.840 & 0.0272 &  \\
		 MCXCJ1715.3+5724 & NGC6338      &  35.40 &  85.80 &        0.87 &        0.67 &             0.33 &  25.372 & 0.0276 & D$^c$ \\
		 MCXCJ1755.8+6236 & NGC6521        &  30.22 &  91.82 &        0.31 &        0.47 &             0.25 &   4.684 & 0.0266 &  \\
		 MCXCJ1736.3+6803 & Zw1745.6+6703  &  32.00 &  98.27 &        0.24 &        0.43 &             0.24 &   3.050 & 0.0248 &  \\
		 MCXCJ1703.8+7838 & A2256          &  31.76 & 111.01 &        4.25 &        1.12 &             0.28 & 354.347 & 0.0581 & D$^g$ \\
		\hline
		\end{tabular}
\begin{tablenotes}
	\item
		{\bf Columns:}
		(1) MCXC cluster name;
		(2) Other cluster name;
		(3) Galactic latitude in degrees;
		(4) Galactic longitude in degrees;
		(5) Mass within $R_{500}$ in $10^{14} \Msun$ units;
		(6) Radius enclosing an overdensity $\delta = 500$ in Mpc;
		(7) Angular scale of $R_{500}$ in degrees;
		(8) X-ray luminosity in the 0.1 - 2.4 keV energy band, in $10^{42}$ erg ${\rm s}^{-1}$, out to the $R_{500}$ radius;
		(9) Redshift;
		(10) Dynamical state. R --- relaxed: sample has a cooling time less than the Hubble time or is classified as a cool-core cluster; D --- disturbed: smaple has a cooling time greater than the Hubble time, is classified as an non cool-core cluster or merger, or has an observed radio relics.
		References:
		$^a$ --- \citet{Stewart1984}; 
		$^b$ --- \citet{Johnson2011};
		$^c$ --- \citet{Lagana2019};
		$^d$ --- \citet{Jones1999}; 
		$^e$ --- \citet{Proust2003}; 
		$^f$ --- \citet{Dupke2007};
		$^g$ --- \citet{Yuan2015}. \\
		$\dagger$: Reported as a cool-core cluster in \citet{Hudson2010}.
	\end{tablenotes}
\end{table*}

\section{2D significance map}
\label{append:2D_sig_maps}
\begin{figure}
    \centering
    \DrawFig{\includegraphics[width=0.5\textwidth,trim=10 40 0 80, clip]{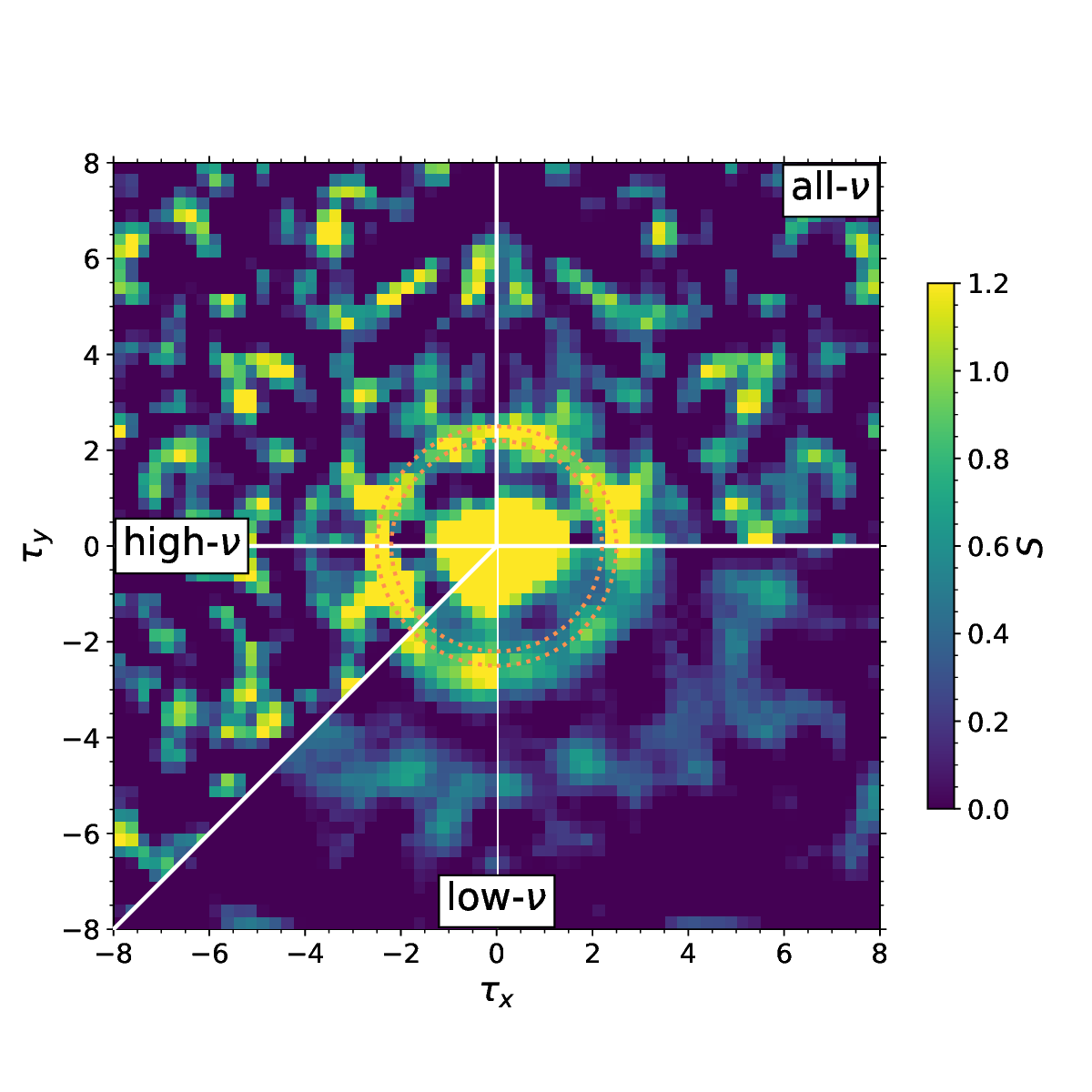}}
    \caption{
	Same as Fig.~\ref{fig:2dsig}, but with a larger $\tau$ range. 
    }
    \label{fig:2dsig_1}
\end{figure}

\begin{figure}
	\centering
	\DrawFig{\includegraphics[width=0.5\textwidth,trim=10 40 0 80, clip]{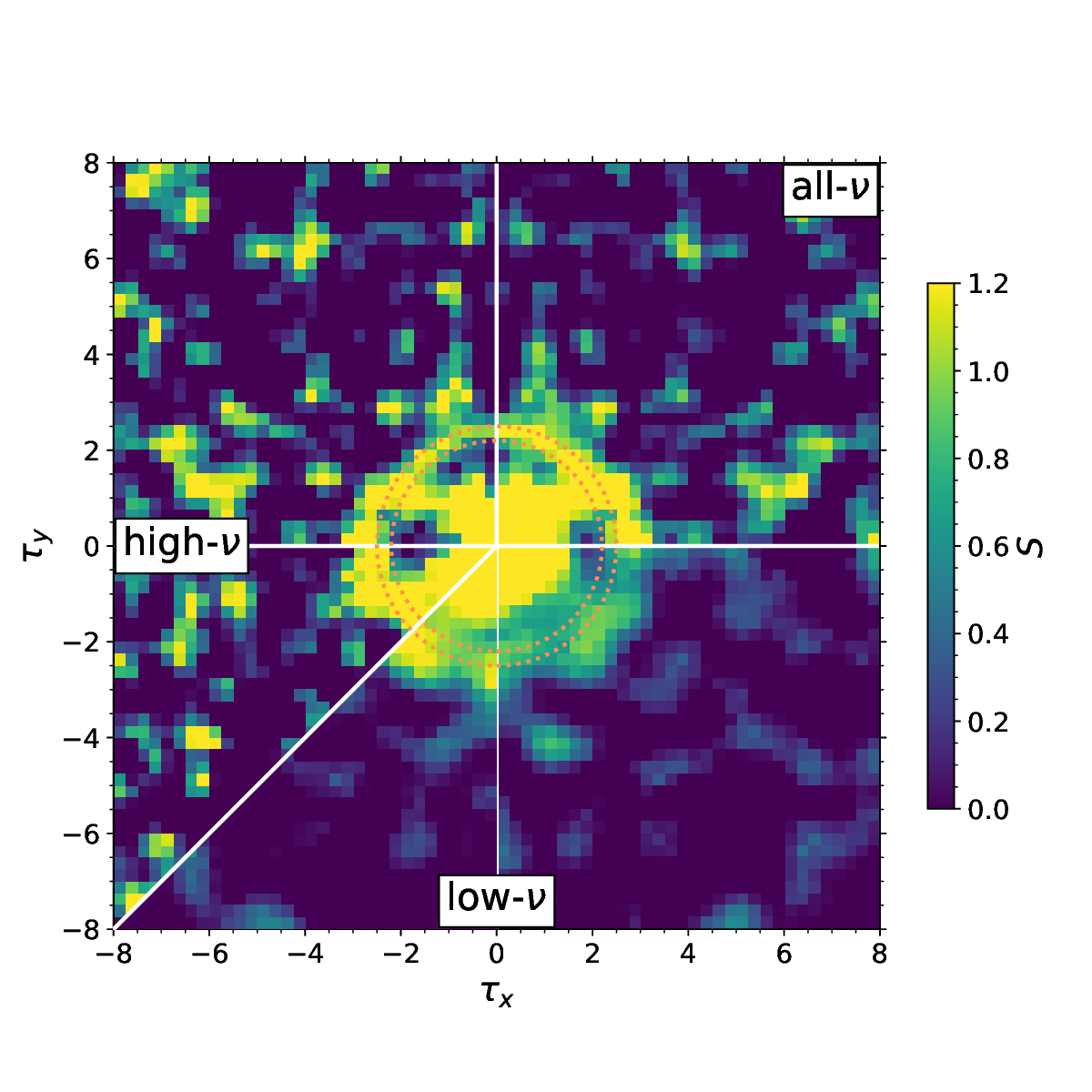}}
	\caption{
	Same as Fig.~\ref{fig:2dsig_1}, but with a more relaxed  declination cut of $\delta > 10\dgr$ (leaving 59 clusters). 
	}
	\label{fig:2dsig_5}
\end{figure}

\begin{figure}
    \centering
	\DrawFig{\includegraphics[width=0.5\textwidth,trim=10 40 0 80, clip]{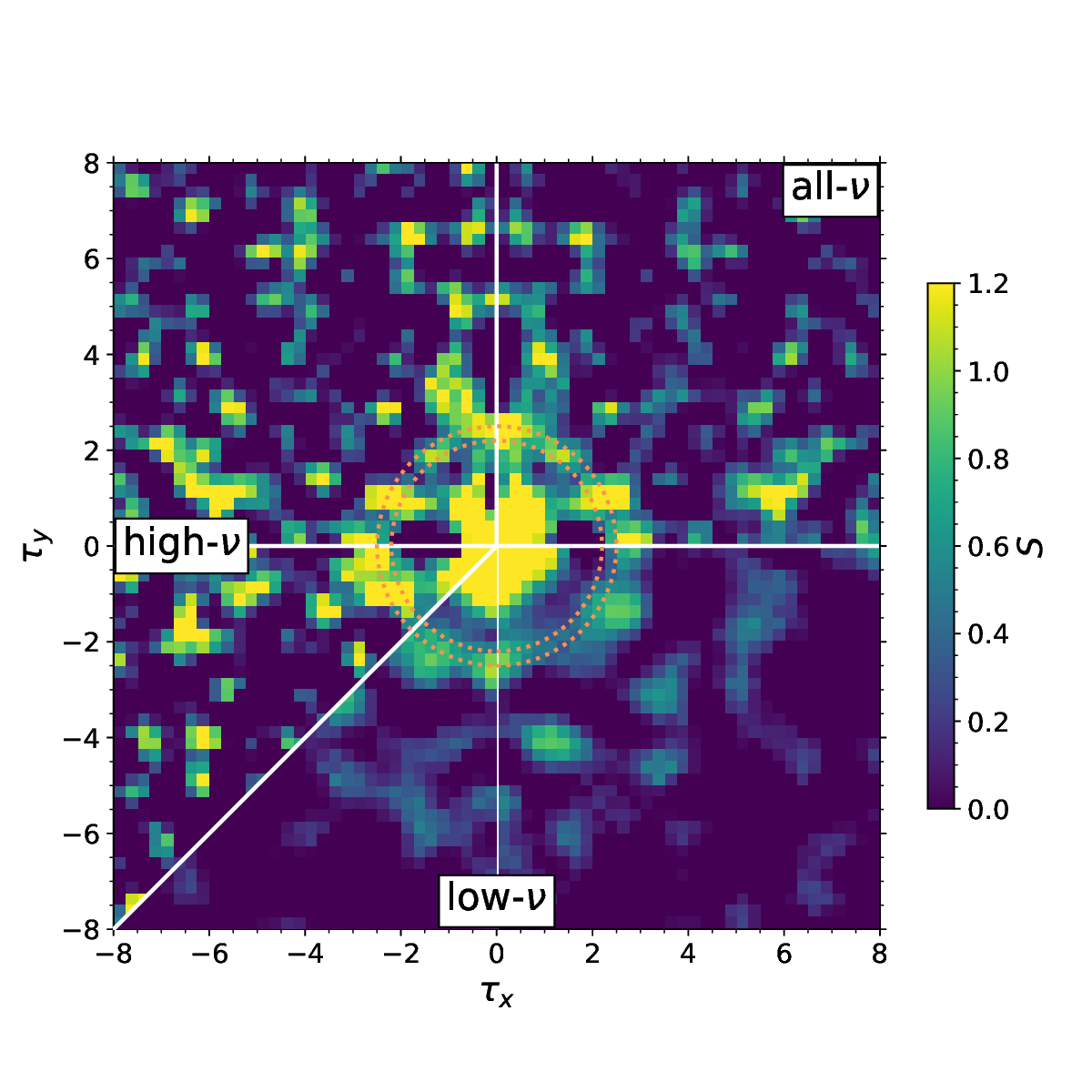}}
    \caption{
	Same as Fig.~\ref{fig:2dsig_1}, but with a more restrictive declination cut, $\delta > 30\dgr$ (leaving 29 clusters). 
    }
\label{fig:2dsig_2}
\end{figure}

Figures \ref{fig:2dsig_1}--\ref{fig:2dsig_2} demonstrate 2D significance maps in a large, $|\tau|<8$ radial range, for different choices of the declination cut used for cluster selection: $\delta > 20\dgr$  (nominal), $10\dgr$, and $30\dgr$.

\end{document}